%% file: science-case.tex
\documentclass[11pt,fleqn]{book} 

\usepackage{tcolorbox}
\usepackage{wrapfig}
\usepackage{url}
\usepackage[utf8]{inputenc}
\usepackage{aas_macros}
\usepackage{pdfpages}
\usepackage{multicol}
\usepackage{setspace}
\usepackage{tcolorbox}

\input{structure}

\begin{document}
\pagenumbering{roman}

\includepdf{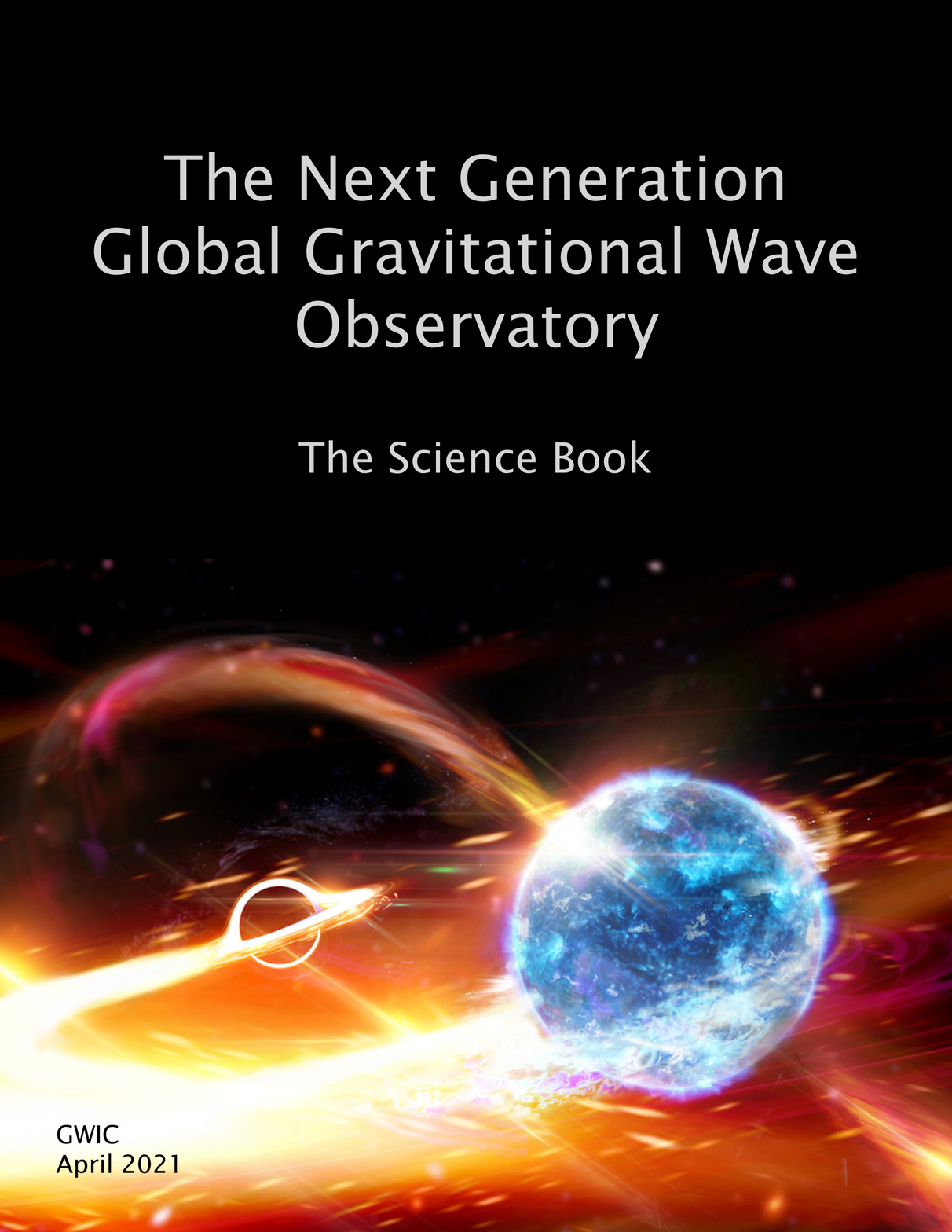}
\input{cover}

\markboth{\sffamily\bfseries Executive Summary}{\sffamily\bfseries Executive Summary}
\input{exec}
\clearpage

\pagenumbering{arabic}
\input{intro} 
\input{extreme-matter} 
\input{bbh} 
\input{cosmology} 
\input{extreme-gravity} 
\input{explosions} 
\input{acronyms}
\clearpage

\bibliographystyle{JHEP}
\chapterimage{bib.jpg} 
\addcontentsline{toc}{chapter}{\color{ocre} Bibliography} 
\small
\begin{spacing}{1.00}
\markboth{\sffamily\bfseries Bibliography}{\sffamily\bfseries Bibliography}
\bibliography{3g}
\vfill
\end{spacing}
\end{document}

%% file: structure.tex
\usepackage[top=2.54cm,bottom=2.54cm,left=2.54cm,right=2.54cm,headsep=10pt,letterpaper]{geometry} 
\usepackage{xcolor} 
\definecolor{ocre}{RGB}{52,177,201} 

\usepackage{authblk}

\usepackage{avant} 
\usepackage{mathptmx} 

\usepackage{microtype} 
\usepackage[utf8]{inputenc} 
\usepackage[T1]{fontenc} 

\usepackage{titlesec} 

\usepackage{graphicx} 
\graphicspath{{Pictures/}{figures/cosmo/}} 

\usepackage{lipsum} 

\usepackage{tikz} 

\usepackage[english]{babel} 

\usepackage{enumitem} 
\setlist{nolistsep} 

\usepackage{booktabs} 

\usepackage{eso-pic} 

\usepackage{titletoc} 

\contentsmargin{0cm} 
\titlecontents{chapter}[1.25cm] 
{\addvspace{15pt}\Large\sffamily\bfseries} 
{\color{ocre!60}\contentslabel[\LARGE\thecontentslabel]{1.25cm}\color{ocre}} 
{}
{\color{ocre!60}\normalsize\sffamily\bfseries\;\titlerule*[.5pc]{.}\;\thecontentspage} 
\titlecontents{section}[1.25cm] 
{\addvspace{5pt}\sffamily\bfseries} 
{\contentslabel[\thecontentslabel]{1.25cm}} 
{}
{\sffamily\hfill\color{black}\thecontentspage} 
[]
\titlecontents{subsection}[1.25cm] 
{\addvspace{1pt}\sffamily\small} 
{\contentslabel[\thecontentslabel]{1.25cm}} 
{}
{\sffamily\;\titlerule*[.5pc]{.}\;\thecontentspage} 
[]

\titlecontents{lsection}[0em] 
{\footnotesize\sffamily} 
{}
{}
{}

\titlecontents{lsubsection}[.5em] 
{\normalfont\footnotesize\sffamily} 
{}
{}
{}

\usepackage{fancyhdr} 

\fancypagestyle{plain}{\fancyhead{}} 

\makeatletter
\renewcommand{\cleardoublepage}{
\clearpage\ifodd\c@page\else
\hbox{}
\vspace*{\fill}
\thispagestyle{empty}
\fi}

\usepackage{amsmath,amsfonts,amssymb,amsthm} 

\newtheoremstyle{ocrenumbox}
{0pt}
{0pt}
{\normalfont}
{}
{\small\bf\sffamily\color{ocre}}
{\;}
{0.25em}
{\small\sffamily\color{ocre}\thmname{#1}\nobreakspace\thmnumber{\@ifnotempty{#1}{}\@upn{#2}}
\thmnote{\nobreakspace\the\thm@notefont\sffamily\bfseries\color{black}---\nobreakspace#3.}} 

\newtheoremstyle{blacknumex}
{5pt}
{5pt}
{\normalfont}
{} 
{\small\bf\sffamily}
{\;}
{0.25em}
{\small\sffamily{\tiny\ensuremath{\blacksquare}}\nobreakspace\thmname{#1}\nobreakspace\thmnumber{\@ifnotempty{#1}{}\@upn{#2}}
\thmnote{\nobreakspace\the\thm@notefont\sffamily\bfseries---\nobreakspace#3.}}

\newtheoremstyle{blacknumbox} 
{0pt}
{0pt}
{\normalfont}
{}
{\small\bf\sffamily}
{\;}
{0.25em}
{\small\sffamily\thmname{#1}\nobreakspace\thmnumber{\@ifnotempty{#1}{}\@upn{#2}}
\thmnote{\nobreakspace\the\thm@notefont\sffamily\bfseries---\nobreakspace#3.}}

\newtheoremstyle{ocrenum}
{5pt}
{5pt}
{\normalfont}
{}
{\small\bf\sffamily\color{ocre}}
{\;}
{0.25em}
{\small\sffamily\color{ocre}\thmname{#1}\nobreakspace\thmnumber{\@ifnotempty{#1}{}\@upn{#2}}
\thmnote{\nobreakspace\the\thm@notefont\sffamily\bfseries\color{black}---\nobreakspace#3.}} 
\makeatother

\newcounter{dummy}
\numberwithin{dummy}{section}
\theoremstyle{ocrenumbox}
\newtheorem{theoremeT}[dummy]{Theorem}

\newtheorem{exerciseT}{Exercise}[chapter]
\theoremstyle{blacknumex}
\newtheorem{exampleT}{Example}[chapter]
\theoremstyle{blacknumbox}

\newtheorem{definitionT}{Definition}[section]
\newtheorem{corollaryT}[dummy]{Corollary}
\theoremstyle{ocrenum}

\RequirePackage[framemethod=default]{mdframed} 

\newmdenv[skipabove=7pt,
skipbelow=7pt,
backgroundcolor=black!5,
linecolor=ocre,
innerleftmargin=5pt,
innerrightmargin=5pt,
innertopmargin=5pt,
leftmargin=0cm,
rightmargin=0cm,
innerbottommargin=5pt]{tBox}

\newmdenv[skipabove=7pt,
skipbelow=7pt,
rightline=false,
leftline=true,
topline=false,
bottomline=false,
backgroundcolor=ocre!10,
linecolor=ocre,
innerleftmargin=5pt,
innerrightmargin=5pt,
innertopmargin=5pt,
innerbottommargin=5pt,
leftmargin=0cm,
rightmargin=0cm,
linewidth=4pt]{eBox}	

\newmdenv[skipabove=7pt,
skipbelow=7pt,
rightline=false,
leftline=true,
topline=false,
bottomline=false,
linecolor=ocre,
innerleftmargin=5pt,
innerrightmargin=5pt,
innertopmargin=0pt,
leftmargin=0cm,
rightmargin=0cm,
linewidth=4pt,
innerbottommargin=0pt]{dBox}	

\newmdenv[skipabove=7pt,
skipbelow=7pt,
rightline=false,
leftline=true,
topline=false,
bottomline=false,
linecolor=gray,
backgroundcolor=black!5,
innerleftmargin=5pt,
innerrightmargin=5pt,
innertopmargin=5pt,
leftmargin=0cm,
rightmargin=0cm,
linewidth=4pt,
innerbottommargin=5pt]{cBox}

\makeatletter
\renewcommand{\@seccntformat}[1]{\llap{\textcolor{ocre}{\csname the#1\endcsname}\hspace{1em}}}
\renewcommand{\section}{\@startsection{section}{1}{\z@}
{-2ex \@plus -1ex \@minus -.2ex}
{1ex \@plus.1ex }
{\normalfont\large\sffamily\bfseries}}
\renewcommand{\subsection}{\@startsection {subsection}{2}{\z@}
{-2ex \@plus -0.1ex \@minus -.2ex}
{0.5ex \@plus.2ex }
{\normalfont\sffamily\bfseries}}
\renewcommand{\subsubsection}{\@startsection {subsubsection}{3}{\z@}
{-2ex \@plus -0.1ex \@minus -.2ex}
{.2ex \@plus.2ex }
{\normalfont\small\sffamily\bfseries}}
\renewcommand\paragraph{\@startsection{paragraph}{4}{\z@}
{-2ex \@plus-.2ex \@minus .2ex}
{.1ex}
{\normalfont\small\sffamily\bfseries}}

\usepackage[hidelinks,backref=page,pagebackref=true,hyperindex=true,colorlinks=false,breaklinks=true,urlcolor= ocre,bookmarks=true,bookmarksopen=false,pdftitle={Title},pdfauthor={Author}]{hyperref}

\newcommand{\thechapterimage}{}
\newcommand{\chapterimage}[1]{\renewcommand{\thechapterimage}{#1}}

\def\thechapter{\arabic{chapter}}
\def\@makechapterhead#1{
\thispagestyle{empty}
{\centering \normalfont\sffamily
\ifnum \c@secnumdepth >\m@ne
\if@mainmatter
\startcontents
\begin{tikzpicture}[remember picture,overlay]
\node at (current page.north west)
{\begin{tikzpicture}[remember picture,overlay]
\node[anchor=north west,inner sep=0pt] at (0,0) {\includegraphics[width=\paperwidth]{\thechapterimage}};
\draw[anchor=west] (2.46cm,-6cm) node [rounded corners=20pt,fill=ocre!10!white,text opacity=10,draw=ocre,draw opacity=1,line width=1.5pt,fill opacity=.6,inner sep=12pt]{\LARGE\sffamily\bfseries\textcolor{black}{\thechapter. #1\strut\makebox[22cm]{}}};
\end{tikzpicture}};
\end{tikzpicture}}
\par\vspace*{130\p@}
\fi
\fi}

\def\@makeschapterhead#1{
\thispagestyle{empty}
{\centering \normalfont\sffamily
\ifnum \c@secnumdepth >\m@ne
\if@mainmatter
\begin{tikzpicture}[remember picture,overlay]
\node at (current page.north west)
{\begin{tikzpicture}[remember picture,overlay]
\node[anchor=north west,inner sep=0pt] at (0,0) {\includegraphics[width=\paperwidth]{\thechapterimage}};
\draw[anchor=west] (2.46cm,-6cm) node [rounded corners=20pt,fill=ocre!10!white,fill opacity=.6,inner sep=12pt,text opacity=1,draw=ocre,draw opacity=1,line width=1.5pt]{\LARGE\sffamily\bfseries\textcolor{black}{#1\strut\makebox[22cm]{}}};
\end{tikzpicture}};
\end{tikzpicture}}
\par\vspace*{120\p@}
\fi
\fi
}
\makeatother

\usepackage{graphicx}
\graphicspath{{./figures/}}
\usepackage{appendix}
\usepackage{latexsym,amsmath,amsfonts,amssymb,booktabs}
\usepackage[font=small]{caption}
\usepackage{slashed,upgreek,amscd,cancel,tensor,color}
\usepackage{adjustbox}
\usepackage[numbers,sort&compress,square]{natbib}
\usepackage{epsfig,latexsym}
\usepackage{url}
\numberwithin{equation}{section}
\usepackage{doi}
\usepackage{subcaption}
\usepackage{mathtools}
\usepackage{upgreek}
\usepackage{stfloats}
\usepackage{afterpage}
\usepackage{multirow}

%% file: cover.tex

\begingroup
\thispagestyle{empty}



\newpage
\phantom{.}


\input{poem}

\newpage


\input{committees}
\vfill
\input{copyright}
\newpage


\input{authors}
\newpage


\chapterimage{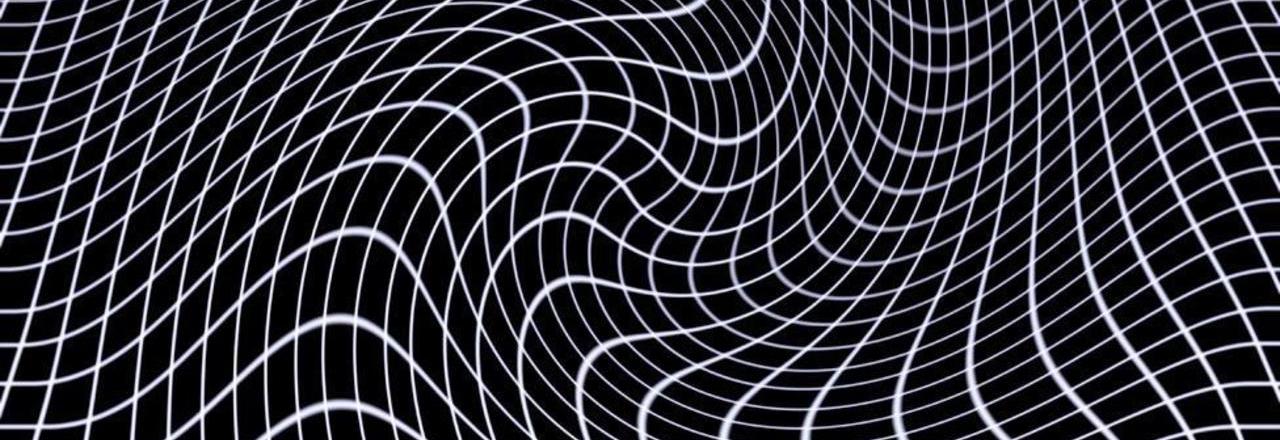} 
\pagestyle{empty} 
\tableofcontents 


\pagestyle{fancy} 

%% file: poem.tex
\phantom{.}
\vfill
\thispagestyle{empty}
\begin{flushright}
\it
Out of the cosmic rife, I just picked me a star\\
another came along, from not so far\\
Thought it would be a real good bet\\
The best is yet to come\\[20pt]

The best is yet to come and may be, it’ll be  fine\\
You think you’ve seen the sun\\
But you ain’t seen two rattle and shine \\[20pt]

A wait till the 3rd-gen’s underway\\
Wait till our feisty stars have met\\
And wait till you see that everyday\\
You ain’t seen nothing yet.   \\[20pt]
\hfill {\rm --- Sanjay Reddy with apologies to Frank Sinatra}

\end{flushright}
\vskip3cm

\noindent Front Cover: Artist's impression of a black hole-neutron star merger, Carl Knoz, OzGrav

%% file: committees.tex
{\small 
\noindent \textbf{SCIENCE BOOK SUBCOMMITTEE} \\[12pt] 
\noindent Vicky Kalogera, Northwestern University, USA (Co-chair) \\
\noindent B.S. Sathyaprakash, Penn State USA and Cardiff University, UK (Co-chair) \\
\noindent Matthew Bailes, Swinburne, Australia \\
\noindent Marie-Anne Bizouard, CNRS \& Observatoire de la Cote d'Azur, France \\
\noindent Alessandra Buonanno, AEI, Potsdam, Germany and University of Maryland, USA \\
\noindent Adam Burrows, Princeton, USA \\
\noindent Monica Colpi, INFN, Italy \\
\noindent Matt Evans, MIT, USA \\
\noindent Stephen Fairhurst, Cardiff University, UK \\
\noindent Stefan Hild, Maastricht University, Netherlands \\
\noindent Mansi M. Kasliwal, Caltech, USA \\
\noindent Luis Lehner, Perimeter Institute, Canada \\
\noindent Ilya Mandel, University of Birmingham, UK \\
\noindent Vuk Mandic, University of Minnesota, USA \\
\noindent Samaya Nissanke, University of Amsterdam, Netherlands \\
\noindent Maria Alessandra Papa, AEI, Hannover, Germany \\
\noindent Sanjay Reddy, University of Washington, USA \\
\noindent Stephan Rosswog, Oskar Klein Centre, Sweden \\
\noindent Chris Van Den Broeck, Nikhef, Netherlands \\
\vskip24pt
\noindent \textbf{STEERING COMMITTEE} \\[12pt] 
\noindent Michele Punturo, INFN Perugia, Italy (Co-chair) \\
\noindent David Reitze, Caltech, USA (Co-chair) \\
\noindent Peter Couvares, Caltech, USA \\
\noindent Stavros Katsanevas, European Gravitational Observatory \\
\noindent Takaaki Kajita, University of Tokyo, Japan \\
\noindent Vicky Kalogera, Northwestern University, USA \\
\noindent Harald Lueck, AEI, Hannover, Germany \\
\noindent David McClelland, Australian National University, Australia \\
\noindent Sheila Rowan, University of Glasgow, UK \\
\noindent Gary Sanders, Caltech, USA \\
\noindent B.S. Sathyaprakash, Penn State University, USA and Cardiff University, UK \\
\noindent David Shoemaker, MIT, USA (Secretary) \\
\noindent Jo van den Brand, Nikhef, Netherlands \\
}

%% file: copyright.tex
{\small 
\noindent \textsc{Gravitational Wave International Committee}

\noindent This document was produced by the GWIC 3G Committee, the GWIC 3G Science Case Team and the
International 3G Science Team Consortium

\noindent \textit{Final release, April 2021} 
}

%% file: authors.tex
{\small 
\noindent \textbf{\large{Additional Authors}} \\

\noindent P.  Ajith,  International Centre for Theoretical Sciences, India \\
\noindent Shreya Anand,  Caltech, USA \\
\noindent Igor  Andreoni,  Caltech, USA \\
\noindent K.G.  Arun,  Chennai Mathematical Institute, India \\
\noindent Enrico  Barausse,  SISSA, Italy \\
\noindent Masha  Baryakhtar,  New York University, USA and University of Washington, USA \\
\noindent Enis  Belgacem,  Utrecht University, The Netherlands \\
\noindent Christopher P.L.  Berry,  Northwestern University, USA \\
\noindent Daniele  Bertacca,  Universit\`a degli Studi di Padova, Italy and INFN Sezione di Padova, Italy \\
\noindent Richard  Brito,  Sapienza University of Rome, Italy \\
\noindent Chiara  Caprini,  Universit\'e Paris-Diderot, France \\
\noindent Katerina  Chatziioannou,  Caltech, USA \\
\noindent Michael  Coughlin,  University of Minnesota, USA \\
\noindent Giulia  Cusin,  University of Geneva, Switzerland \\
\noindent Tim  Dietrich,  University of Potsdam, Germany \\
\noindent Yves  Dirian,  University of Geneva and University of Zuerich, Switzerland \\
\noindent William E.  East,  Perimeter Institute, Canada \\
\noindent Xilong  Fan,  Wuhan University and Hubei University of Education, China \\
\noindent Daniel  Figueroa,  \'Ecole Polytechnique F\'ed\'erale de Lausanne, Switzerland \\
\noindent Stefano  Foffa,  University of Geneva, Switzerland \\
\noindent Archisman  Ghosh,  Ghent University, Belgium \\
\noindent Evan  Hall,  MIT, USA \\
\noindent Jan  Harms,  Gran Sasso Science Institute, Italy \\
\noindent Ian  Harry,  Max Planck Institute for Gravitational Physics, Potsdam, Germany and University of Portsmouth, UK \\
\noindent Tanja  Hinderer,  Utrecht University, The Netherlands \\
\noindent Thomas  Janka,  Max Planck Institute for Astrophysics, Germany \\
\noindent Stephen  Justham,  University of the Chinese Academy of Sciences, China and University of Amsterdam, The Netherlands \\
\noindent Dan  Kasen,  University of California Berkeley, USA \\
\noindent Kei  Kotake,  Fukuoka University, Japan \\
\noindent Geoffrey  Lovelace,  California State University Fullerton, USA \\
\noindent Michele  Maggiore,  University of Geneva, Switzerland \\
\noindent Alberto  Mangiagli,  Astroparticule et Cosmologie, France \\
\noindent Michela  Mapelli,  University of Padova and INFN Padova, Italy \\
\noindent Andrea  Maselli,  Gran Sasso Science Institute, Italy \\
\noindent Andrew  Matas,  Max Planck Institute for Gravitational Physics, Potsdam, Germany \\
\noindent Jess  McIver,  Caltech, USA and the University of British Columbia, Canada \\
\noindent Bronson  Messer,  ORNL and University of Tennessee, USA \\
\noindent Tony  Mezzacappa,  ORNL and University of Tennessee, USA \\
\noindent Cameron  Mills,  Cardiff University, UK \\
\noindent Bernhard  Mueller,  Monash University, Australia \\
\noindent Ewald  M\"uller,  Max Planck Institute for Astrophysics, Germany \\
\noindent Michael  P\"urrer,  Max Planck Institute for Gravitational Physics, Potsdam, Germany \\
\noindent Paolo  Pani,  Sapienza University of Rome, Italy \\
\noindent Geraint  Pratten,  University of Balearic Islands, Spain and University of Birmingham, UK \\
\noindent Tania  Regimbau,  Universit\'e Grenoble Alpes, France \\
\noindent Mairi  Sakellariadou,  Kings College London, United Kingdom \\
\noindent Raffaella  Schneider,  INFN Roma, Italy \\
\noindent Alberto  Sesana,  Universit\`a di Milano-Bicocca, Italy \\
\noindent Lijing  Shao,  Kavli Institute for Astronomy and Astrophysics, Peking University, China \\
\noindent P. Thomas  Sotiriou,  The University of Nottingham, UK \\
\noindent Nicola  Tamanini,  Laboratoire des 2 infinis - Toulouse, France \\
\noindent Thomas  Tauris,  Aarhus University, Greece \\
\noindent Eric  Thrane,  Monash University, Australia \\
\noindent Rosa  Valiante,  INAF Roma, Italy \\
\noindent Maarten  van de Meent,  Max Planck Institute for Gravitational Physics, Potsdam, Germany \\
\noindent Vijay  Varma, Caltech, USA \\
\noindent Justin  Vines,  Max Planck Institute for Gravitational Physics, Potsdam, Germany \\
\noindent Salvatore  Vitale,  MIT, USA \\
\noindent Huan  Yang,  Perimeter Institute and University of Guelph, Canada \\
\noindent Nicolas  Yunes, University of Illinois Urbana-Champaign, USA \\
\noindent Miguel  Zumalacarregui,  Max Planck Institute for Gravitational Physics, Potsdam, Germany and University of California Berkeley, USA \\
} 

%% file: exec.tex
\chapterimage{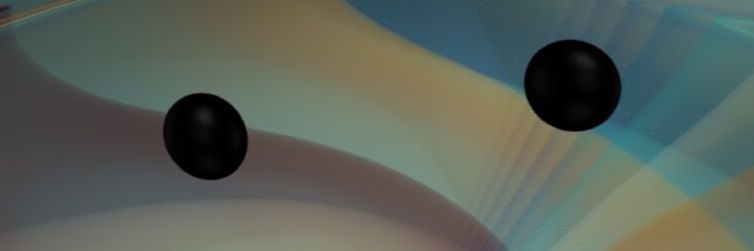}
\chapter*{Executive Summary}
\addcontentsline{toc}{chapter}{\color{ocre} Executive Summary}
\vfill

\noindent
\begin{minipage}{0.49\textwidth}
{\Large\bf W}ithin the last century, we have discovered incredible new truths about the Universe using telescopes and particle detectors. 
We now know it is expanding and composed mostly of dark matter and dark energy that we do not yet understand and neutrinos and ultra high-energy particles, whose origin remains a puzzle. We have learned that it harbors astounding entities such as black holes, regions of spacetime so strongly warped that nothing, even light, that falls inside can ever escape. We have learned that it is gravity that shapes the structure of the Universe, from enormous cosmic fibers of matter to galaxies and solar systems, to planets, stars, and black holes. Gravity drives the most extreme phenomena in the Universe, including incredibly violent collisions of black holes that in nearly an instant release millions of times the energy that our Sun will emit in its entire lifetime.  

\hskip0.5cm 
Gravity encodes information about distant cosmic phenomena in a messenger entirely different from light and particles: {\em gravitational waves,} tiny ripples in the fabric of spacetime emitted by accelerating mass. Unlike light and particles, gravitational waves interact weakly with matter, so they travel vast distances almost entirely unobscured by dust, the Milky Way, or the Earth itself. They offer a crystal clear signature of highly energetic phenomena otherwise hidden from us. 

\hskip0.5cm 
Gravitational waves from a binary black hole merger observed by Advanced LIGO in 2015 gave us our first glimpse of hugely energetic events that are undetectable with light. Detectors on Earth capture gravitational waves emitted by black holes orbiting so quickly that they approach the speed of light before colliding. The first detections also revealed a population of black holes with masses never before observed and charted the existence of black holes formed by collapsed stars farther from Earth than ever before. 
\end{minipage}
\hfill
\begin{minipage}{0.49\textwidth}
\begin{tcolorbox}[standard jigsaw,colframe=gray,colback=black!10!ocre,opacityback=0.1,coltext=black,title=\small\sc  Black Holes: Where the Cosmos meets the Quantum Realm]
\small
A new generation of detectors will allow us to push Einstein's general theory of relativity to the limit and to test alternative theories that aim to resolve the fundamental contradictions between {\em quantum physics}, which describes the Universe at very small sizes, and {\em general relativity}, which describes the Universe at very large sizes. Although the two are incompatible with each other, no experiment yet has discovered new physics outside of either theory. Black holes provide a laboratory that smashes these theories together; squeezing enormous masses into infinitesimally small volumes. By sensing black hole collisions with high fidelity, new detectors would allow us to test completely new regimes of highly warped spacetime that could provide critical insight into this current paradox in physics. The precision of next generation detectors would also allow us to search for a differing gravitational-wave signature of new types of exotic compact objects unlike black holes as we currently understand them.
\vskip5pt
\includegraphics[width=\linewidth]{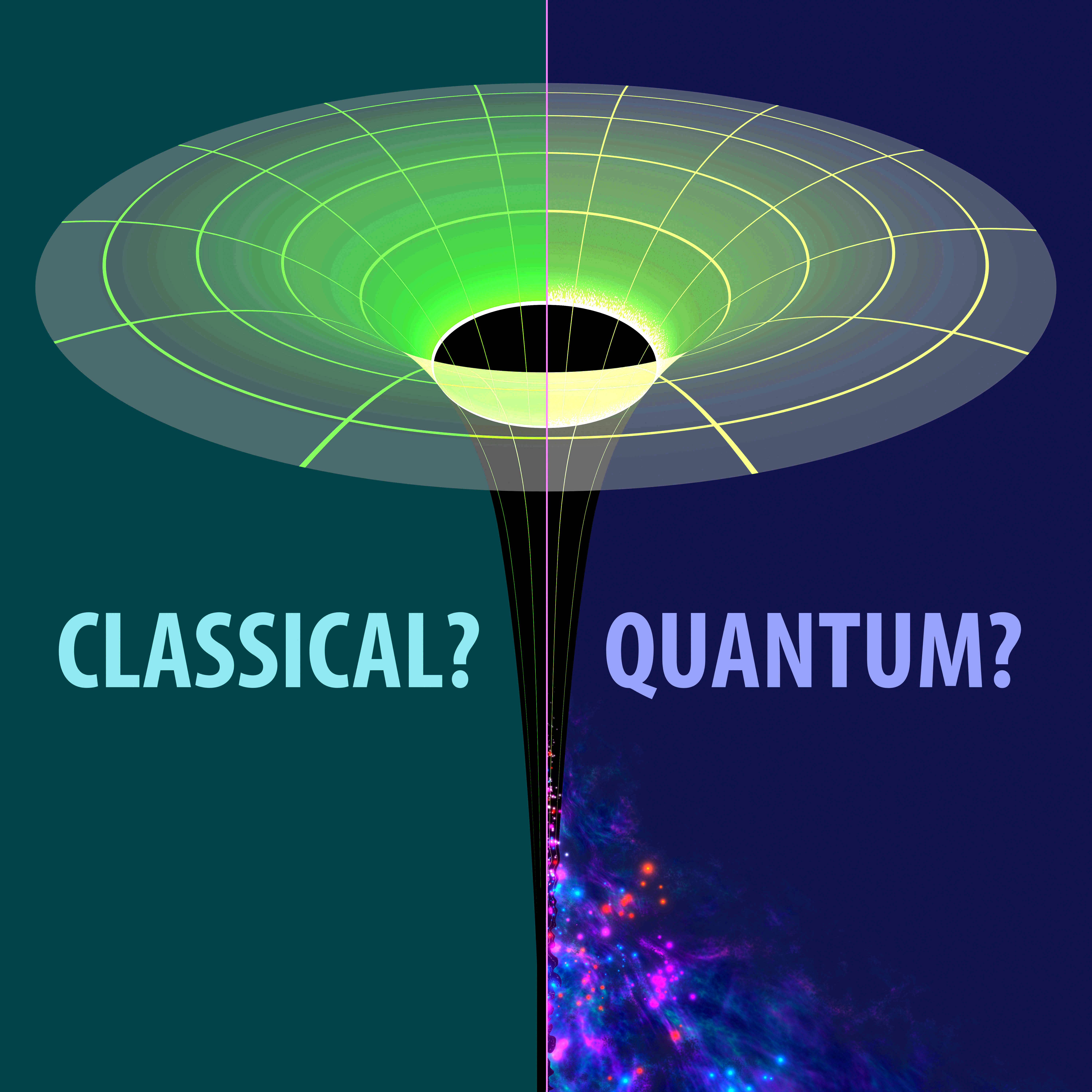}
\end{tcolorbox}
\end{minipage}

A network of next-generation gravitational-wave detectors will survey the extreme Universe with an unprecedented reach into deep space, beyond the cosmic dawn when the first stars began to shine.  

\noindent
\begin{minipage}{0.47\textwidth}
They will  sample  stellar-mass black hole mergers across the visible Universe, observing these systems over the entire history of the cosmos, and chronicle the evolution of black holes from the earliest mergers and inform if they grow from collapsed stars to supermassive black holes at the center of galaxies, billions of times as massive as our Sun. 

\hskip0.5cm 
The 2017 observation of the merger of two neutron stars by LIGO and Virgo and its aftermath with the full spectrum of electromagnetic (EM) radiation  was a spectacular first success of multimessenger astronomy with gravitational waves that reaffirmed theoretical models of the brightest EM events.  Next-generation ground-based gravitational-wave detectors will have the unique capability to observe the violent mergers of neutron stars and their aftermath at far greater distances and with the much higher precision required to address outstanding problems in the physics of dense objects. Neutron stars are only the size of a city yet contain more mass than the sun. They are the densest material objects in the Universe, where intense gravity completely obliterates the atomic structure of matter familiar to us. Even quarks---the tiniest constituents that remain confined inside neutrons and protons in ordinary matter---are likely to be liberated in the neutron star core.  

\hskip0.5cm 
Next generation detectors will allow us to observe neutron star mergers and peek directly into their cores as they tear each other apart by tidal forces before smashing together. These observations are critical to infer new knowledge and understanding about nuclear physics and the states of matter containing quarks.  With next-generation multimessenger astrophysics enabled by new detectors, we will learn how much of the Universe's gold and platinum was produced by neutron star collisions. 

\end{minipage}
\hfill
\begin{minipage}{0.51\textwidth}
\begin{tcolorbox}[standard jigsaw,colframe=gray,colback=black!10!ocre,opacityback=0.1,coltext=black,title=\small\sc Observing the Universe on all Scales]
\small
The diagram shows physical phenomena on different scales explored by gravitational-wave observations. Starting from the scale of the Universe, almost 100 billion light years, represented by the cosmic microwave background at the top, the diagram progressively shows scales that are smaller than the previous ones by the factor shown on the left. On scales about 1000 times smaller are giant galaxy clusters 30 million light years across. Another factor 10 million smaller is the size of a supernova remnant of 10 light years. A trillion times smaller still is the merger environment of binary neutron stars about 100 km across. The core of a neutron star, about 1 km, is a trillion times smaller and contains matter at densities similar to that of atomic nuclei. On scales ten thousand smaller still gravitational waves could probe the nature of dark matter.
\vskip0.25cm 

\includegraphics[width=1.01\linewidth]{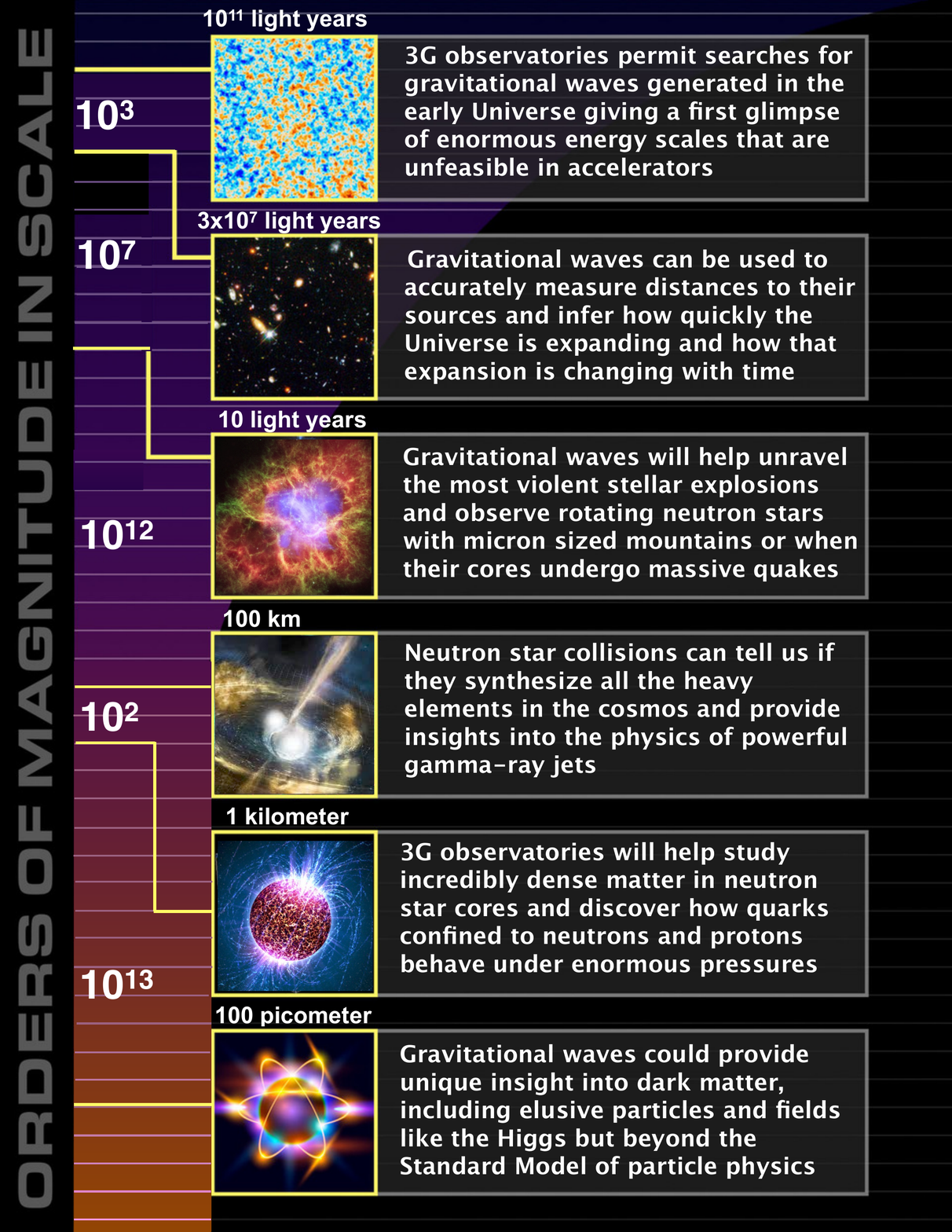}
\end{tcolorbox}
\end{minipage}
\vskip5pt

Furthermore, gravitational waves from binary black holes and neutron stars are {\em standard sirens}---the distance to the source is encoded in the observed gravitational waves. Thus, merging binaries provide a new precision tool for observational cosmology that will help us gain new insight into how the Universe is expanding and evolving and if dark energy is just a cosmological constant or if there is missing new physics associated with the late-time accelerated expansion of the Universe.

We will also achieve unprecedented insight into cosmic explosion mysteries. Multimessenger astronomy with next generation detectors will allow us to better investigate why core-collapse supernovae explode to seed the formation of new stars and whether starquakes cause mysterious bursts of radio emission.  And as with any completely new method of observation, there is also the possibility that next generation detectors will reveal completely new dark phenomena, unseen with light, that we have not yet conceived of. 

Today's gravitational-wave detectors are barely sensitive enough to detect the loudest gravitational waves in the Universe, like a simple radio able to pick up only the loudest signals. Next-generation network detector designs leverage cutting-edge technology to surpass current ground-based detectors, making their ability to measure passing gravitational waves more than ten times better than the current instruments.

\begin{figure*} 
\begin{tcolorbox}[standard jigsaw,colframe=gray,colback=black!10!ocre,opacityback=0.1,coltext=black,title=\small\sc Artists Conception of Einstein Telescope (left) and Cosmic Explorer (right)]
\fbox{
\centering
\includegraphics[width=0.495\textwidth]{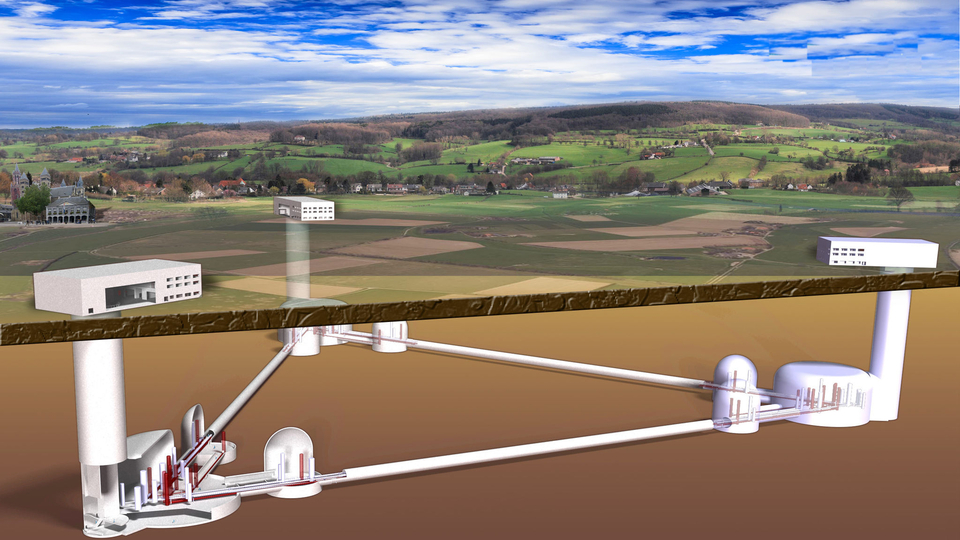}
\hfill
\includegraphics[width=0.495\textwidth]{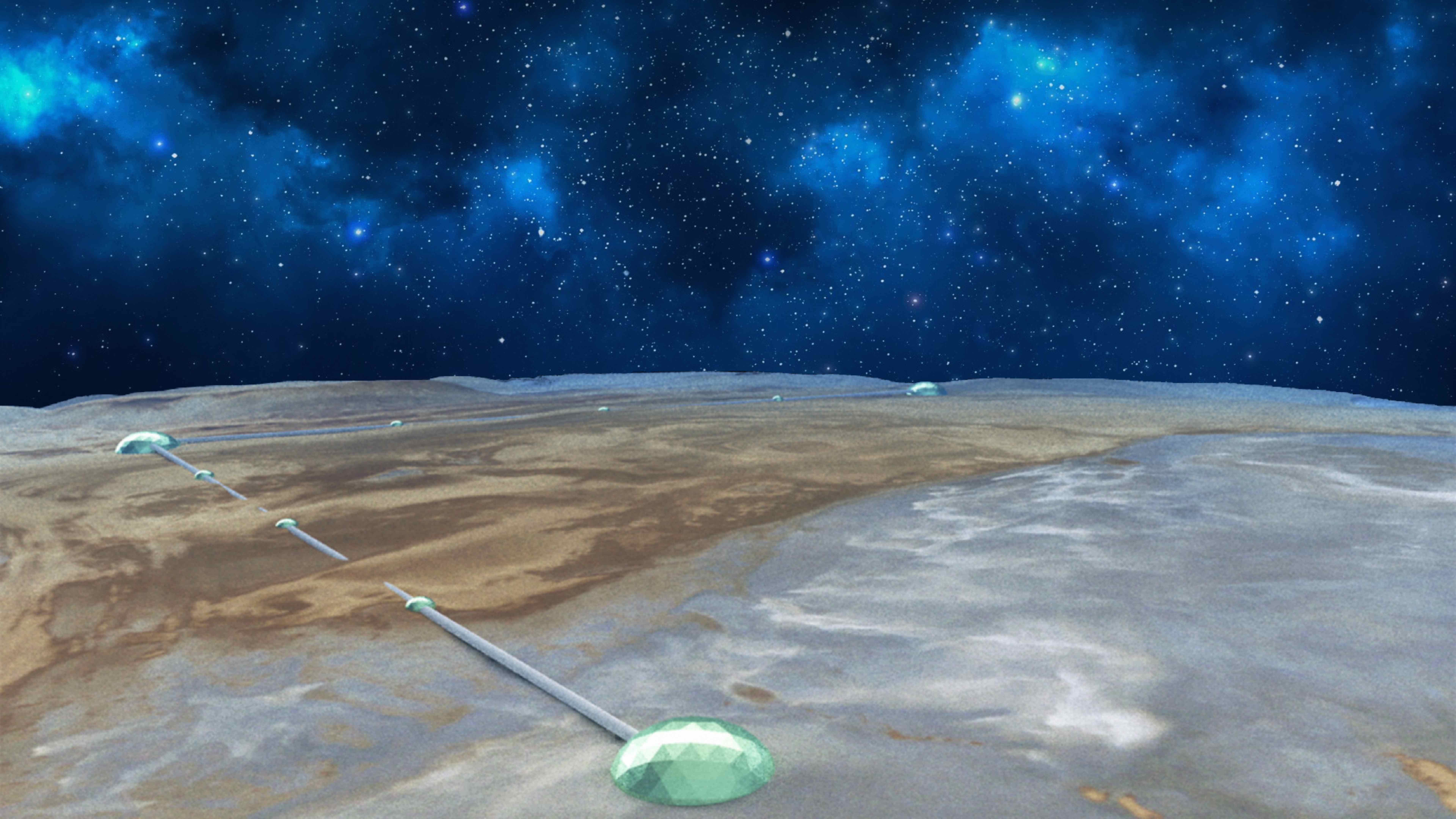}
} 
\caption*{Artists conception of the Einstein Telescope (left panel) and Cosmic Explorer (right panel) observatories.  ET is conceived to be six, V-shaped, underground interferometers, formed out of 10 km sides of an equilateral triangle, while  Cosmic Explorer is conceived to be an L-shaped, overground interferometer, with 40 km arms.}
\end{tcolorbox}
\end{figure*}

More powerful detectors will let us listen to the gravitational-wave universe with unprecedented fidelity, fully revealing the rich physics encoded in the waves but currently hidden by observational uncertainty. Einstein Telescope (ET) is a European design featuring six V-shaped interferometers in a triangular topology with 10 km interferometer arms and Cosmic Explorer (CE) is a U.S. design for one or two interferometers with 40 km L-shaped interferometer arms. ET and CE are expected to detect hundreds of thousands of mergers, as well as tens of thousands of multimessenger sources that would also likely emit EM radiation and particles that telescopes and neutrino and cosmic ray detectors can observe. A network of three detectors distributed around the globe will triangulate the gravitational wave signal's location in the sky, critical information that will guide telescopes on Earth and in space in searches for related EM emission. 

21st century astronomy will be further revolutionized by the launch of the space-based LISA gravitational-wave observatory, expected in 2034. LISA will sense gravitational waves emitted by more massive systems than ground-based detectors, detecting the signature of orbiting black hole systems up to years before ground-based detectors observe them collide. Combining space-based and ground-based observations will allow us to catalog a much broader expanse of the extreme gravitational Universe than ever before. 

Gravitational waves have already given us a first glimpse of the dark, hidden, violent Universe. A global next-generation gravitational wave observatory will propel the field of astrophysics and all foundational science research forward. Observing light, neutrinos and cosmic rays in concert with next-generation gravitational wave detectors will launch enormous advances beyond the current limits of human knowledge, from the quantum realm to the largest cosmological structures in the known Universe. 

%% file: intro.tex
\chapterimage{figures/intro} 
\chapter{Introduction}
\label{ch:introduction}
\vskip-10pt

\section{Prologue}
{\Large\bf O}n 14 Sep 2015, the Laser Interferometer Gravitational-Wave Observatory (LIGO) transformed the way we explore 
the universe. With LIGO, we were able to sense,  
for the first time, the gravitational-wave (GW) sky. The signal came from the merger of two black holes at a distance of 1.3 billion light years. This was also when 
binary black holes were discovered and, for the first time, their merger was observed. The masses of the companions, 29 and 36 solar masses, were unexpectedly 
heavy and the merger converted some 3 solar masses into energy in a mere 200 ms. Since that first discovery, many more black hole mergers have been found by LIGO and the European Virgo, some with masses as small as 2.6 solar masses, heaviest neutron star or lightest black hole, and others as large as 85 solar masses, so large that they could not have formed from the evolution of massive stars. In most cases the component black holes seem to be non-spinning, contrary to what X-ray observations indicate.  

Two years later, on 17 Aug 2017, a new era in multimessenger astronomy began with the observation by LIGO and Virgo of the merger of two neutron stars, followed by the detection, 1.7\,s later, of a gamma-ray burst from the same source by the Fermi gamma-ray space telescope and INTEGRAL. These observations could together localize the event well enough that its host galaxy was quickly found by optical telescopes. The merger produced spectacular fireworks that were captured by telescopes across the entire electromagnetic (EM) spectrum from radio to infrared and optical to X-rays. This treasure trove of data gave us answers to decades old puzzles in fundamental physics and astronomy: verified that GWs travel essentially at the speed of light, confirmed that binary neutron star mergers are progenitors of short gamma ray bursts and prolific sites for the formation of heavy elements, measured the Hubble constant in a completely new way using GWs for the source's distance and EM observations for its redshift and constrained neutron-star radii to be between 9.5 and 13 km by measuring the tidal deformation of neutron stars.

The LIGO and Virgo detectors are yet to achieve their design sensitivities. They will be augmented with new facilities, the Japanese KAGRA and LIGO-India. 
Yet, based on the modest glimpses of the sources discovered to date, we know that the full exploration of the GW sky will require a new generation of detectors of a size that demands new facilities. With the aid of such detectors we will be able to observe sources at the edge of the Universe, unveil the properties of matter at the highest densities in the cosmos, provide a new precision tool for observational cosmology and explore the nature of dynamical spacetimes. A detector network with a leap in sensitivity will resolve signals with far greater precision and fidelity that will pave the way for serendipitous discoveries, observing novel phenomena and unearthing new physics.

\begin{figure*}
\begin{tcolorbox}[standard jigsaw,colframe=gray,colback=gray!10!white,opacityback=0.6,coltext=black, title=\small\sc Sensitivity of ET and CE compared to Advanced LIGO \& the Reach for 3G Observatories]
\vskip-5pt
    \centering
    \includegraphics[width=0.48\textwidth]{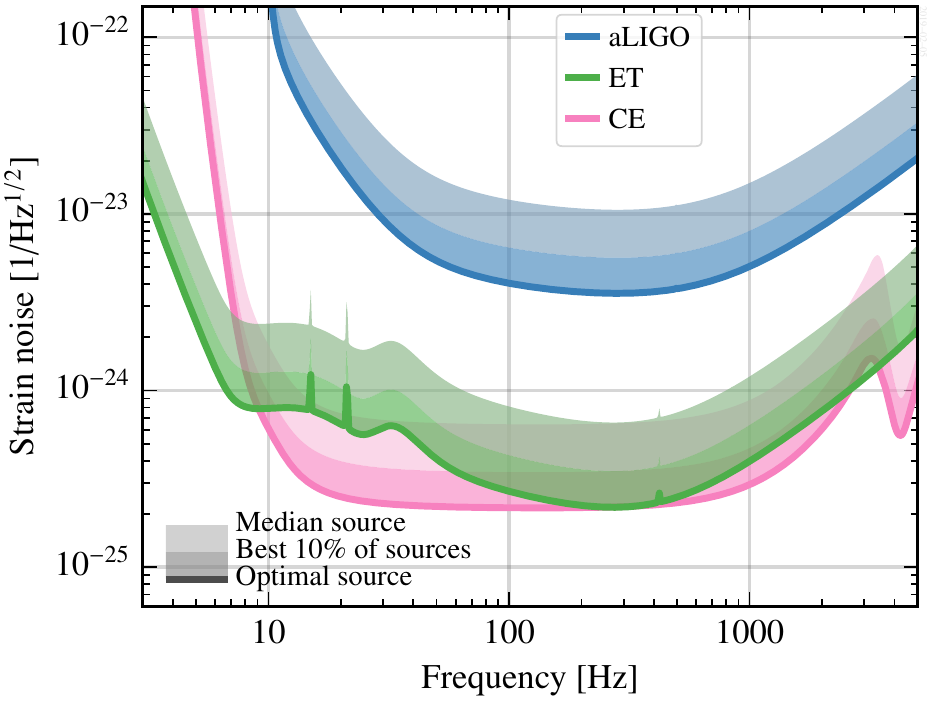}
    \hfill
    \includegraphics[width=0.48\textwidth]{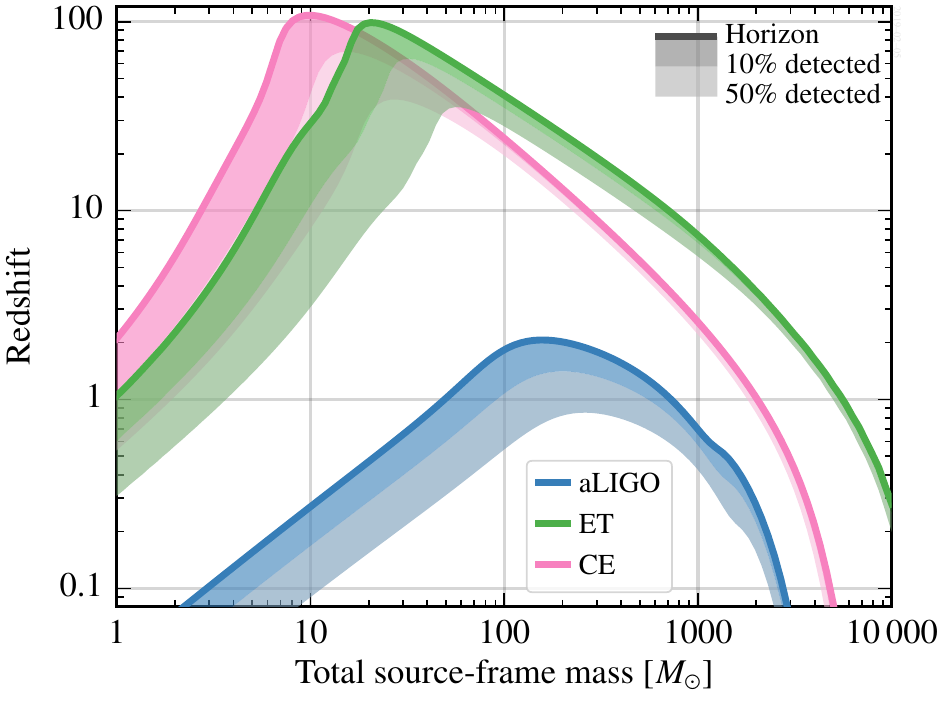}
    \caption{\small GW strain noise for current and future detectors (left) and
    astrophysical reach for equal-mass, nonspinning binaries distributed isotropically in sky 
    and inclination (right).}
    \label{fig:noises_percentiles}
\end{tcolorbox}
\vskip-0.7cm
\end{figure*}

\noindent{\bf Beyond Advanced Detectors:} During 2008-2011 the design study of a third generation (3G) GW observatory in Europe, Einstein Telescope (ET), developed the concept of a triangular interferometer, 10 km on a side, housing six V-shaped interferometers whose combined sensitivity is a factor $\sim 20$ better than Advanced Virgo and pushing the low-frequency sensitivity down from 10 Hz to 3 Hz. A similar effort is currently underway in the US to study the science case for and technical design of a 40 km arm length interferometer called Cosmic Explorer (CE), with sensitivity similar to ET (see Figure \ref{fig:noises_percentiles},  left plot).
For the current study we assume that \emph{the 3G network} consists of one  ET in Europe and one CE each in the US and Australia. A network of at least three sites is required to accurately localize sources in the sky and infer their distances. ET alone could measure the wave's polarization but cannot resolve all the parameter degeneracies to determine the sky position even when the signals last for days. 

The science potential of the 3G network is immediately apparent from the dramatic improvement in strain sensitivity that CE and ET are able to deliver (Figure \ref{fig:noises_percentiles}, right panel). The network makes a leap of 1--2 orders of magnitude in the redshift reach for binary coalescences compared to Advanced LIGO and Virgo.
The network will survey a large redshift range for merging binary black holes and provide a massive catalog of detections to constrain their population and origins. The network will explore a wide parameter space of quantum chromodynamics and study high density matter in a region complementary to heavy ion physics experiments.
The Box below summarizes the science potential of a 3G observatory, elucidated in the next several paragraphs.

\begin{tcolorbox}[standard jigsaw,colframe=ocre,colback=ocre!10!white,opacityback=0.6,
coltext=black,title=\sc Science Targets for the Next Generation of Gravitational Wave Detectors]
\vskip -2 pt
GW astronomy provides a complementary window to EM, neutrino and particle astronomy that could reveal hitherto unseen world.  A new generation of detectors will:
\vskip5pt
\begin{itemize}[leftmargin=*]
\item  {\em determine} the properties of dense matter, {\em discover} phase transitions, and the emergence of quarks
\item  {\em reveal} merging black holes across the cosmos and \emph{search} for seeds of supermassive black holes
\item  {\em investigate} the particle physics of the primeval Universe and {\em probe} its dark sectors
\item  {\em explore} new physics in gravity and in the fundamental properties of compact objects
\item  {\em understand} physical processes that underlie the most powerful astrophysical phenomena
\end{itemize}
\end{tcolorbox}

\section{Extreme Matter, Extreme Environments.}
\label{sec:extreme matter}
Neutron stars are the densest objects in the cosmos and sites of stupendously strong magnetic fields, up to billions of tesla. Six decades after their discovery, we still lack a clear understanding of the equation of state of their deep cores and the origin of their strong magnetic fields.  Neutron stars in binaries are subject to the tidal fields of their companions although the tides raised are extremely small. The extent of tidal deformation depends on the internal structure of neutron stars and the net effect is to accelerate the rate of inspiral allowing to read-off their internal structure from the observed phase evolution of the signal. The merger remnant could be a rapidly rotating, short-lived, hypermassive neutron star that eventually collapses to a black hole. GWs from the merger will lead to tight measurements of NS radii and hence reveal the equation of state of both cold and hot, supranuclear matter and the deconfinement phase transition of quarks and gluons.

The origin of heavy elements in the Universe has been a long-standing problem.  EM observation of GW170817 provided irrefutable evidence that binary  neutron star mergers are prolific sites for the production of lanthanides and other heavy elements.  The 3G network will facilitate EM follow-up of thousands of mergers, a number that is required to confirm if solar  and stellar abundance of heavy elements can be explained by mergers alone or if other production channels, such as supernovae, are necessary.

GW170817 resolved that binary neutron star mergers are progenitors of short gamma-ray bursts. Nevertheless key questions about central engines that produce gamma rays still remain. For example, we do not have a clear picture of the jet properties nor how those properties depend on the progenitor characteristics.  EM follow-up facilitated by the 3G network will allow  a better understanding of the physics of gamma-ray jets, the opening angle of the jet and its distribution; GWs could tell us the nature of the merger remnant and if the central engine is a transient hypermassive neutron star or a promptly collapsed black hole.

3G observatories will detect binary neutron star mergers from epochs far before the peak of star formation activity. Millions of mergers  are expected to be detected by the 3G network. The properties of the detected sources and the  environments in which they occur will provide key data to test astrophysical  models of the formation and evolution of double neutron star and black hole-neutron star binaries, while also informing the history of star formation activity up to redshifts of 5--8.

\section{Observing Stellar-mass Black Holes Throughout the Universe.}
\label{sec:bbh}
\begin{minipage}{0.49\textwidth}
The 3G network will have nearly all-sky sensitivity, detecting stellar-mass black hole binaries of $\sim 10$--100 solar mass from epochs before the first stars formed at redshifts $z\sim30$ (Figures\,\ref{fig:noises_percentiles} and \ref{fig:waterfall-ET-CE}). Consequently, the 3G network could reveal a population of primordial black holes in this mass range formed by quantum processes in the early Universe, in addition to compiling a census of black holes over a range of masses throughout the cosmos. 

\hskip18pt The merger rate of binary black holes observed so far imply that the 3G network will detect hundreds of thousands of mergers each year. This large population will help us study the merger rate 
as a function of redshift up to the beginning of the epoch of reionization. It will also help us explore how these rates are correlated with metallicity and galaxy evolution.
\end{minipage}
\hfill
\begin{minipage}{0.49\textwidth}
\begin{tcolorbox}[standard jigsaw,colframe=gray,colback=gray!10!white,opacityback=0.6,coltext=black, title=\small\sc Visibility of Black Hole Binaries in 3G]
    \vskip-5pt
    {\centering\includegraphics[width=0.90\textwidth]{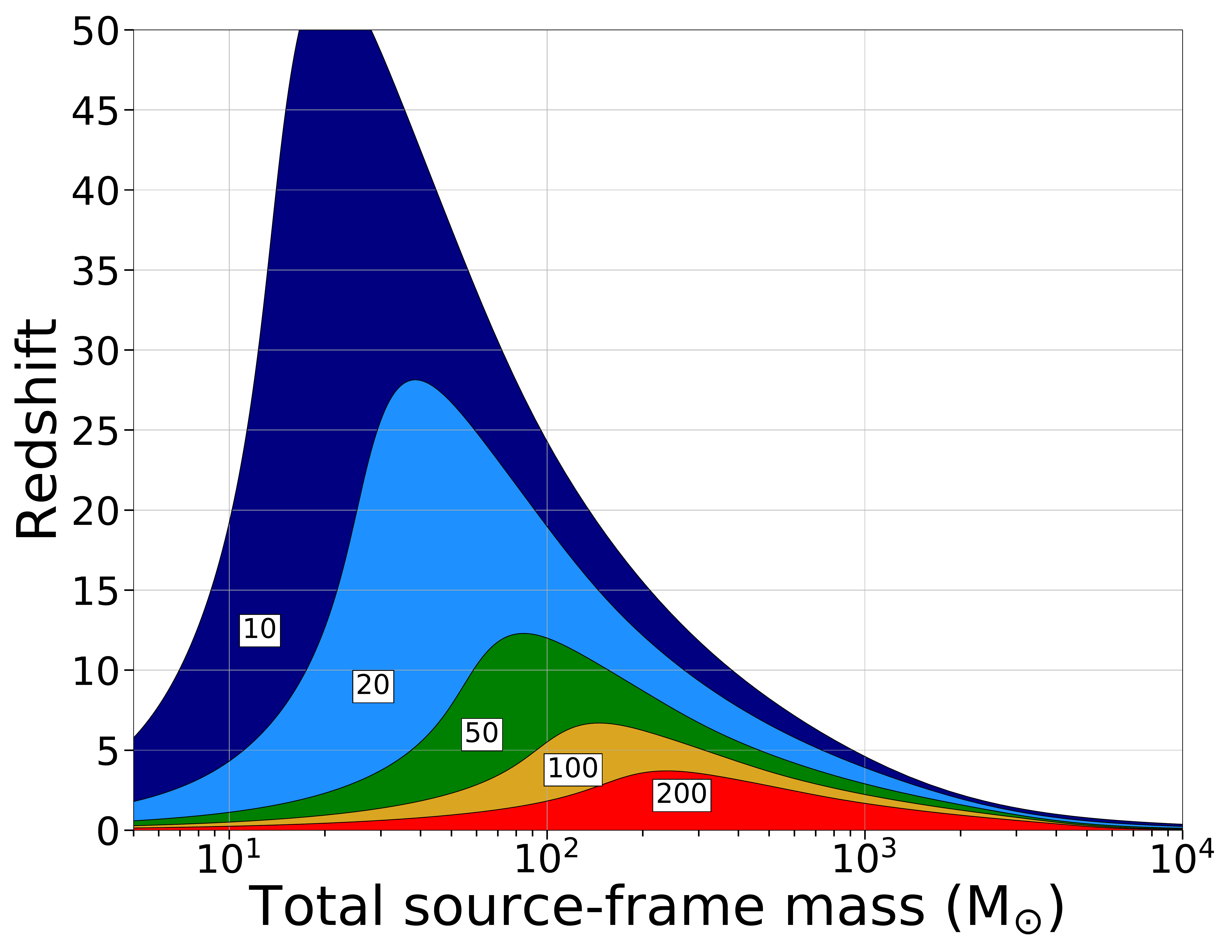}}
    \vskip-5pt
    \captionof{figure}{\small Signal-to-noise ratio contours as a function of binary's total mass and its redshift for equal mass binaries averaged over sky position and orientation in the 3G network.} 
    \vskip-5pt
    \label{fig:waterfall-ET-CE}
\end{tcolorbox}
\end{minipage}
\hfill

Until 2015, it was widely believed that irrespective of how massive a progenitor star was the black hole that resulted from it would be lighter than about 20 solar masses.  LIGO and Virgo have detected many black holes with masses in excess of that number. Indeed, GW190521 revealed a black hole of 85 solar masses that, theory says, cannot form from massive stars.  Understanding how such black holes form and if they grow through repetitive mergers is an outstanding question in astrophysics.  There are a number of competing models. Black holes that form in isolation in globular clusters could sink into the dense cluster cores where they dynamically interact with other holes to form coalescing binaries.  Binaries of massive stars formed in active star formation sites could directly evolve into binary black holes that merge within the Hubble time.  If primordial black holes significantly contribute to dark matter then they could occasionally form merging binaries in galactic halos.  The 3G network will pin down the masses, spins and demographics of black holes, determine principal formation channels and resolve fundamental questions about their origin.

\section{Cosmology and Early History of the Universe.}
\label{sec:cosmology}
GWs from the inspiral and merger of compact binaries can be used to infer the luminosity distance to their sources without the need to calibrate them with standard candles. This is because the orbital dynamics of binary black holes and neutron stars is largely determined by Einstein's theory of gravity. A handful of parameters, e.g. masses and spins of the companion stars, precisely control the pattern of the emitted GWs. The amplitude of that pattern is fixed by the distance to the source, sky position and orientation of the source relative to a detector, which can be inferred with a network of three or more non-collocated detectors.  This contrasts with the dynamics of other astrophysical systems, such as supernovae, that require detailed modelling of their composition and environment, making it extremely hard to predict the emitted GW signal with any precision.  Consequently, with a population of compact binary mergers observed with 3G detectors, and their redshifts obtained by follow-up EM observations, it will be possible to accurately measure cosmological parameters such as the Hubble parameter, dark matter and dark energy densities and the equation-of-state of dark energy, giving a completely independent and complementary measurement of the dynamics of the Universe. 

The cosmological population of point sources create a stochastic GW background. Indeed, with advanced interferometers we could detect the background created by binary black holes and neutron stars throughout the Universe and can do so by cross correlating data from two or more detectors. Such backgrounds would reveal the history of the formation and evolution of these sources and the underlying stellar population. On the contrary, 3G detectors will identify most compact binary mergers in the Universe, giving us a treasure trove of data to study the large-scale distribution of galaxies and their clusters. 

Stochastic GWs could also be produced in the early Universe. As the Universe cools from its primeval hot and dense state it undergoes several phase transitions that are expected to generate GW backgrounds. Detection of such backgrounds would dramatically transform our state of knowledge of the underlying particle theory at energy scales that will never be accessible to terrestrial  accelerators. Defects, such as cosmic strings, associated with symmetry  breaking phase transitions, could also produce stochastic and deterministic signals. The landscape of primordial sources, while uncertain, is a high-risk, high-reward endeavor to pursue in the era of 3G detectors.

\section{Extreme Gravity and Fundamental Physics.}
\label{sec:extreme gravity}
GWs emanate from regions of strong gravity and large curvature,
carrying uncorrupted information from their sources. Imprint in the signal
is the nature of the gravitational field, characteristics of the sources and
the physical environment in which they reside. Their observation in 3G detectors
can put general relativity to the most stringent tests, help explore violations of
the theory in strong fields such as the dynamics of black hole horizons, 
and discover properties of dark matter.

The 3G network offers numerous opportunities to discover failure of general  relativity, e.g. in the form of new particles and fields that violate the strong equivalence principle. It is also possible to detect Lorentz invariance violations or variation in Newton's  constant, both imprint in the propagation of GWs.  One might also see the signature of quantum gravity in the form of parity violation seen in the nature of the GW polarization or in the birefringence of the waves propagating over great distances.  Ultra-light Bosonic fields proposed in certain extensions of the Standard Model could be detected via their effect on the orbital  dynamics of black hole binaries and spin properties of black hole populations observed by the 3G network.

Black holes are the most compelling explanation for the companions in binary coalescences discovered by LIGO and Virgo detectors. The tell-tale  signature of a black hole would be present in the quasi-normal mode spectrum of the merger remnant, whose frequencies and damping times should depend only on the remnant's mass and spin. Signature of additional degrees of freedom would be seen as inconsistency in the  remnant's parameters determined by the different modes. Certain alternatives to black holes could mimic the quasi-normal mode spectra, but they could emit additional signals in the form of echoes of the ingoing radiation reflected from their surface, which could be observable in the 3G network. 

Big Bang cosmology is largely consistent with general relativity but the  accelerated expansion of the Universe in its recent history cannot be explained by the theory, indicating either its failure or the presence of exotic form of matter-energy density, of which we know very little. Observations on galactic  to cosmological scale provide unequivocal indirect evidence for the presence  of weakly interacting dark matter, but none has been directly detected in spite of concerted efforts over the past six decades. The 3G network might detect  various forms of dark matter including axionic and other dark matter fields  around black holes and neutron stars, primordial black holes, etc. 

\section{Sources at the Frontier of Observations}
\label{sec:frontier}
The physics of supernova explosion, glitches in the frequency of pulsars and quakes 
in highly magnetized neutron stars (or magnetars) are open problems in astrophysics. 
Many of these systems will generate
GWs that could be observed with 3G detectors at distances of several
million light years for supernovae and within the galaxy for pulsar glitches
and magnetar flares.  GW observations of these systems
with the 3G network, enhanced by EM and neutrino observatories, will allow us 
to probe extreme astrophysics and address key questions that have hindered progress in our 
understanding of the mechanism behind stellar explosions.

From the observed spectrum of GWs it will be possible to determine the physics of core collapse supernova: the different phases of the collapse, the nature of the explosion that dominates the production of GWs and the asymmetry of the collapse and what triggers that asymmetry. Information about the rotation rate of the progenitor star is also encoded in the observed signal and it should be possible to understand how the initial state of the progenitor star determines the final state of the  collapse, a black hole or a neutron star. Such observations will be greatly aided by all-sky optical and infrared surveys of the stellar population in nearby galaxies, as  well as cosmic rays and neutrinos.

Isolated neutron stars could emit GWs if they are not spherical and don't rotate about their symmetry axis. Indeed, they are persistent sources with the emission lasting for millions of years. Advanced detectors are not likely to detect continuous waves from known pulsars, although all-sky blind searches may reveal hitherto unexpected sources. The 3G network could observe neutron stars whose polar and equatorial radii differ by no more than 10 to 100 microns.  This will provide invaluable information about their crustal strengths and the equation-of-state of high density nucleons in their outer cores. 

Accreting neutron stars, e.g., in low-mass X-ray binaries, could acquire quadrupole deformations from the in-falling matter that could lead to a perpetual source of GWs. Indeed, the 3G network will help resolve if the accretion torque balanced by the GW back-reaction torque is responsible for the observed limiting spin frequencies of neutron stars in low-mass X-ray binaries. The 3G network  will also probe the role of magnetic fields in transient radio emission from magnetars provided the mechanism is caused by crustal quakes that result in the emission of GWs. This could further constrain the equation-of-state of neutron star crusts.

\section{Summary}
\label{sec:intro summary}
LIGO and Virgo discoveries have ushered in a new era in multimessenger physics and astronomy.  GW observations can be used to probe the nature of ultra dense matter, reveal quantum chromodynamic phase transitions,  study the formation and evolution of stellar mass black holes from the epoch of the formation of first stars, measure cosmological parameters, examine phase transitions in the early Universe, test general relativity in dynamical spacetimes, discover the nature of dark matter and other exotic compact objects, and explore the physics of the most violent processes in the cosmos. The mind boggling reach of the 3G network is difficult to fathom but guarantees serendipitous discoveries, with the potential to unearth new physics. Indeed, 3G observatories will operate in a survey mode wherein signals that do not fit our expectations will be flagged off for further study.  The science case for building a new generation of GW detectors that can probe deep into the cosmos and observe a variety of different processes is immensely rich and massively rewarding.  

\clearpage

%% file: extreme-matter.tex
\chapterimage{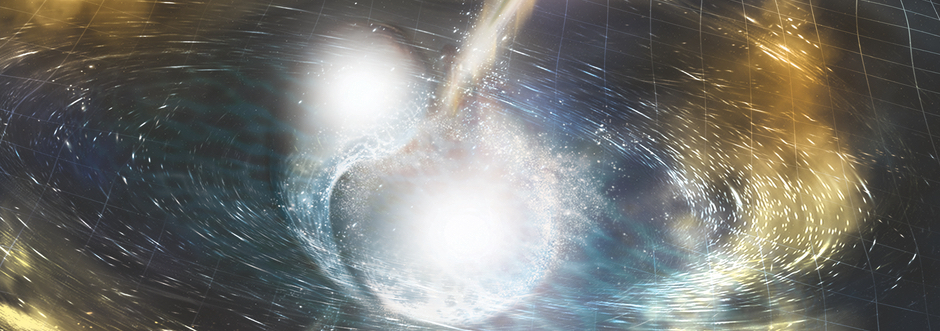} 
\chapter{Extreme Matter, Extreme Environments}
\label{ch:ns}

\vskip10pt
\begin{tcolorbox}[colback=teal!5!white,colframe=yellow!75!black,title=\sc Science Target]
{\em Determine the properties of dense matter, discover phase transitions, and the emergence of quarks.}
\end{tcolorbox}

\noindent The discovery of the binary neutron star (BNS) merger GW170817 \cite{TheLIGOScientific:2017qsa} was a watershed moment in astronomy and astrophysics. Multimessenger observations of the source observations provided incontrovertible evidence that BNS mergers are connected to short-hard gamma-ray bursts (GRBs) \cite{Monitor:2017mdv, Goldstein:2017mmi} and its optical localization \cite{Coulter:2017wya} unveiled that they are prolific sites of heavy element nucleosynthesis~\cite{freiburghaus:1999fr, Abbott:2017wuw, Kasliwal:2017ngb}. It confirmed that GWs and light travel essentially at the same speed, and allowed the first  measurement of the Hubble constant using GW standard sirens  \cite{Abbott:2017xzu, 2018Nature.561..355M} ushering in a new  era in cosmology. Furthermore, the event enabled the first measurement of the neutron star (NS) tidal deformability \cite{Monitor:2017mdv,Abbott:2018exr} and provided the most robust and  stringent constraints on the NS radius and the equation of state of dense matter under extreme conditions inaccessible to experiments and first-principles theoretical calculations \cite{Annala:2017llu, Fattoyev:2017jql, De:2018uhw, Abbott:2018exr, Tews:2018chv,Capano:2019eae}. Circumstantial evidence from the EM counterparts for the formation of a black hole (BH) on a  timescale of tens of milliseconds provided  tighter, albeit model-dependent, constraints on the maximum mass of NSs, and lower bounds on their radii~\cite{Margalit:2017dij, Radice:2017lry, Radice:2018ozg}.

However, a number of important questions were left unanswered (see Sec.~\ref{sec:ns}). Accurate observations of a diverse population BNS mergers with detectors of greater sensitivity and bandwidth will be key to shedding light on the nature of extreme matter in extreme environments produced by the mergers.

\begin{tcolorbox}[standard jigsaw,colframe=ocre,colback=ocre!10!white,opacityback=0.6,coltext=black,title=\sc Key Science Goals]
\small
Multimessenger observations of numerous luminous events in the Universe involving dense matter in extreme environments will uncover several key puzzles in fundamental physics:
\begin{itemize}[leftmargin=*]
\item {\bf Nature of matter at supranuclear densities.}
What are the fundamental properties of the densest matter in the cosmos? How do quarks and gluons manifest in the cores of the most massive neutron stars? 
\item {\bf Production sites of heavy elements.} What elements are produced in NS mergers and how? Are they able to explain the abundances of elements heavier than iron in the solar system and in stars?
\item {\bf Formation and evolution of compact binaries.} How do NS binaries form and evolve? What are their demographics, merger rates, and mass and spin distributions through cosmic time?
\item {\bf Central engines of short-hard GRBs.}  What is the role of the merger remnant and the physics of central engines powering panchromatic EM counterparts to NS mergers? How do they relate to short GRBs? 
\end{itemize}
\end{tcolorbox}

\noindent{\bf Capabilities of Next Generation Detector Networks:}
The 3G network will compile a survey of the Universe  of a large sample of  BNS and NS-BH mergers. Table \ref{table:localize_BNS_100Myr} shows the detection capability of 3G observatories compared to the network of advanced detectors at their design sensitivity. In computing the event rates the local co-moving merger rate was taken to be 1000 Gpc$^{-3}$ yr$^{-1}$ and redshifts at the epoch of merger were sampled  assuming the Madau-Dickinson star formation rate, with an exponential time delay (an e-fold time of 100 Myr) between formation and merger~\cite{Vitale:2018yhm}.  Tens of thousands of well-localized events detected by the 3G network will provide ample opportunities  for EM follow-up of these mergers as opposed to a handful of them by the current network. 
\vskip-10pt

\begin{table}[h!]
\parbox{0.51\linewidth}{
In addition to discovering a great number and diversity of mergers through cosmic time, the wide band sensitivity of 3G detectors will enable tracking the full inspiral, merger and post-merger GW signals. 
These unique capabilities will help address key science questions on the properties of dense matter.  The 3G network will accurately measure the masses and spins NSs and determine their long-sought equation of state, probe the merger dynamics, state of the merger remnant and BH formation, and explore, for the first time, properties of matter at even greater density and temperature if the remnant does not promptly collapse to a BH.}
\hfill
\parbox{.47\linewidth}{
\small
\begin{tcolorbox}[standard jigsaw,colframe=gray,colback=gray!10!white,opacityback=0.6,coltext=black,title=\small \sc BNS Event Rates in 2G \& 3G Networks]
\vskip-5pt
\caption
{
\small Expected number $N$ of BNS detections per year,  the number of events localized to within $1$, $10$ and $100$ deg$^2$ ($N_1,$ $N_{10}$ and $N_{100}$,  respectively) and the median localization error $M$ in square degrees, in a network consisting of LIGO and Virgo (HLV), LIGO, Virgo, KAGRA and LIGO-India (HLVKI) and the 3G network.}
\label{table:localize_BNS_100Myr}
\centering
\vskip-5pt
\begin{tabular}{lccccc}
\hline
 Network      &   $N$    &   $N_1$ & $N_{10}$ & $N_{100}$ & $M$ \\
\hline
 HLV          &       48 &       0 &      16  &      48   & 19 \\
 HLVKI        &       48 &       0 &      48  &      48   &  7 \\
 3G           &    990k  &     14k &    410k  &    970k   & 12 \\
\end{tabular}
\end{tcolorbox}
}
\vskip-10pt
\end{table}

The sub-arcsecond localization of the panchromatic EM counterpart will provide information about environment and geometry of the event, its host galaxy, the physical state and evolution of the ejected material and the nucleosynthesis of heavy elements.  Thus, the combination of information derived independently from the GW and EM signals will be immensely powerful to build a complete, self-consistent astrophysical picture, enable cosmological applications, and constrain formation scenarios.

\vskip-10pt
\begin{table}[h!]
\parbox{0.51\linewidth}{
\section{Nature of Matter at Highest Densities} 
\label{sec:ns}
Neutron stars are precious laboratories for the subatomic physics of matter under unique
conditions. 
The multitude of phenomena connected with multimessenger emissions from BNS mergers 
is of broad interest to nuclear and particle astrophysics. 
Our current understanding of the NS interior 
is captured in Fig.~\ref{fig:neutronstar_profile}. 
The theoretical understanding of  
matter up to densities present in terrestrial nuclei 
($\rho_0 \simeq 2.5 \times 10^{14}~\mathrm{g\,cm}^{-3}$) 
are fairly advanced and relevant to the NS crust. However, nuclei dissolve at higher densities $\rho\gtrsim \rho_0/2$ into a uniform liquid of neutrons, with a small admixture of other particles including protons, electrons, and muons~\cite{Lattimer:2004pg}. In NSs with large masses, densities in the core may be sufficiently high for exotic states of matter to appear. Furthermore, at densities $\gtrsim 2\rho_0$--$3 \rho_0$ the distance between nucleons becomes comparable to their size, and their quark sub-structure is expected to manifest and phase transitions to new states of matter containing deconfined quarks may occur \cite{Baym:2017whm}.} 
\hfill\parbox{0.47\linewidth}{
\small
\begin{tcolorbox}[standard jigsaw,colframe=gray,colback=gray!10!white,opacityback=0.6,coltext=black,title=\small\sc Internal Structure of a NS ]
\vskip-5pt
\captionof{figure}{\small Composition of matter in the interior of a NS predicted by theory. Quark degrees of freedom become important at the densities
encountered in the inner core. The nature of the transition to matter containing de-confined quarks is unknown. }\vskip-5pt
\label{fig:neutronstar_profile}
{\hskip-8pt\includegraphics[width=1.10\textwidth]{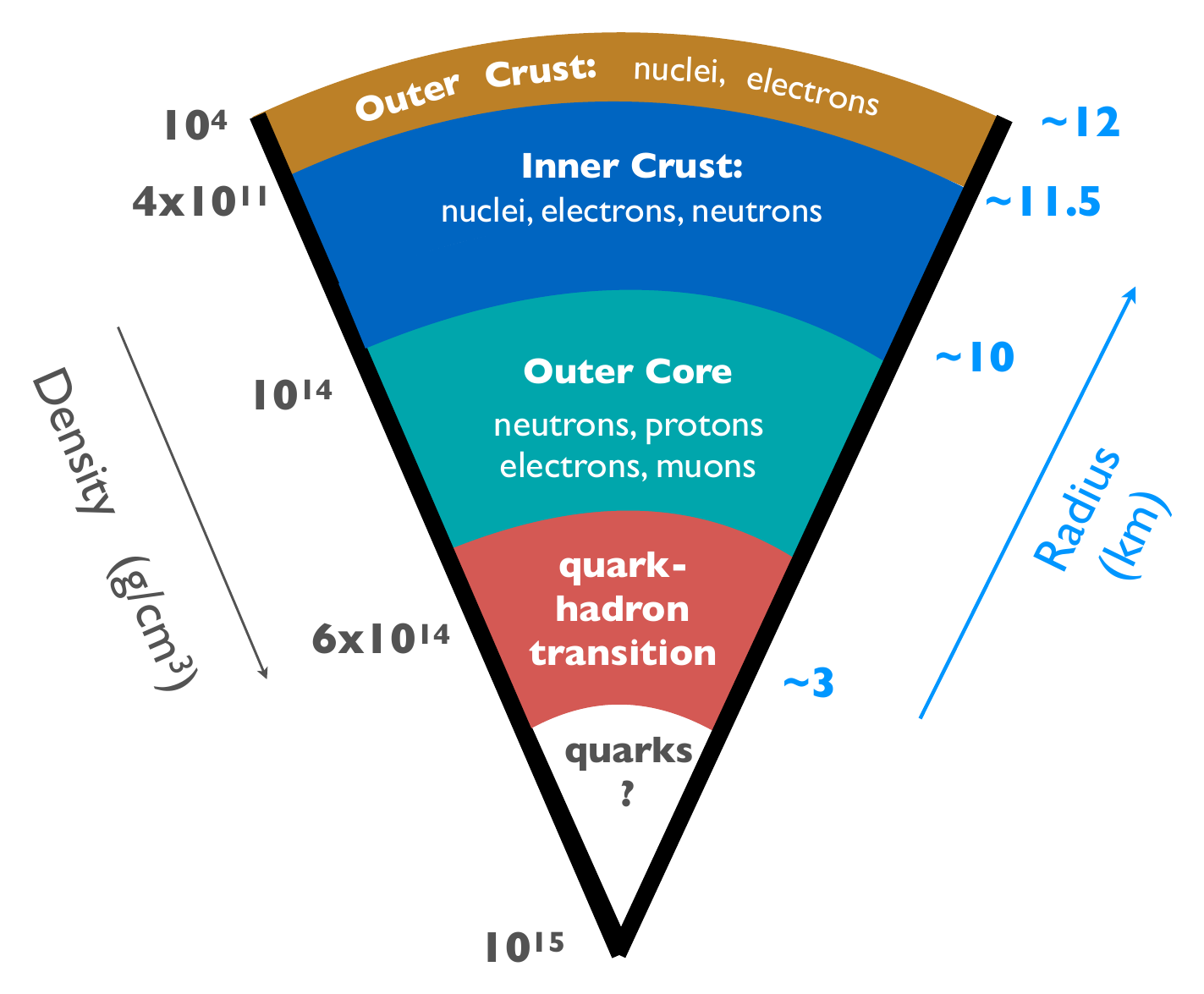}}
\end{tcolorbox}
}
\vskip-10pt
\end{table}

The nature and location of the transition from hadronic to quark matter remains unknown, but is a fundamental question with broad implications.
The properties of matter inside NSs directly affect their global characteristics, masses and radii, motivating significant observational
\cite{Watts:2016uzu,Ozel:2016oaf} and theoretical
\cite{Gandolfi:2011xu,Hebeler:2013nza,Lattimer:2012nd,Oertel:2016bki,Tews:2018kmu} efforts to constrain the properties of NS matter and measure NS masses and radii. Radio observations of pulsars have
yielded accurate mass measurements of a handful of NSs \cite{Ozel:2016oaf}. 
The discovery of a massive NSs with $M\simeq 2$ M$_\odot$
\cite{Demorest:2010bx,Antoniadis:2013pzd,Cromartie:2019kug} has had far-reaching implications for the equation of state of
dense matter~\cite{Lattimer:2010uk}. However, accurate measurements of the NS
radius from X-ray observations have been more challenging since
they rely on poorly tested models of EM emission from or near the NS surface. 
Efforts to model and interpret X-ray data from accreting NSs during
bursts, and in quiescence, suggest that NS radii are in the
range $9$--$13$ km \cite{Ozel:2016oaf, Nattila:2017wtj}, albeit
with untested model assumptions. For a few pulsars, NASA's NICER mission is anticipated to provide reliable radius measurements, using a different method \cite{Watts:2016uzu}. Results from the first NICER observations are promising, although the errors associated with the extracted NS radius remains large \cite{Raaijmakers:2019qny,Bogdanov:2019qjb}. 

\begin{table}[h!]
\parbox{0.49\linewidth}{
\hskip8pt The 3G will probe the  properties of dense matter in a \emph{diverse population} of NSs by measuring a variety of matter-dependent GW signatures during the inspiral phase of BNS systems. These arise from tidal effects, including the tidal excitation of a NS's internal oscillation modes~\cite{Kokkotas:1995xe, Lai:1993di, Shibata:1993qc, Flanagan:2006sb} that can provide direct information about phase transitions,   rotational deformations~\cite{Poisson:1997ha}, spin-tidal couplings  \cite{Landry:2018bil, Abdelsalhin:2018reg}, and the tidal disruption of the NS by a BH companion \cite{Lattimer:1974slx, Shibata:2011jka}. The 3G network will measure the radii of several NSs over a wide mass range including both light and heavier NSs to within 0.5 to 1 km, and discern subdominant matter effects on GWs. Both these aspects are critical for measuring the properties of dense matter and discovering phase transitions. Fig.~\ref{fig:mr} shows a projection of the precision with which 3G detectors will allow us to measure the NS mass and radius.

\hskip8pt
The 3G network will further open an exceptional window onto fundamental properties of \emph{matter in a completely unexplored regime}, at higher temperatures and yet greater densities than encountered in NSs, which is accessible only during the merger and post-merger epochs in BNS collisions. 
}
\hfill
\parbox{0.49\linewidth}{
\begin{tcolorbox}[standard jigsaw,colframe=gray,colback=gray!10!white,opacityback=0.6,coltext=black,title=\small\sc Measuring NS Radius]
\includegraphics[width=1.00\textwidth]{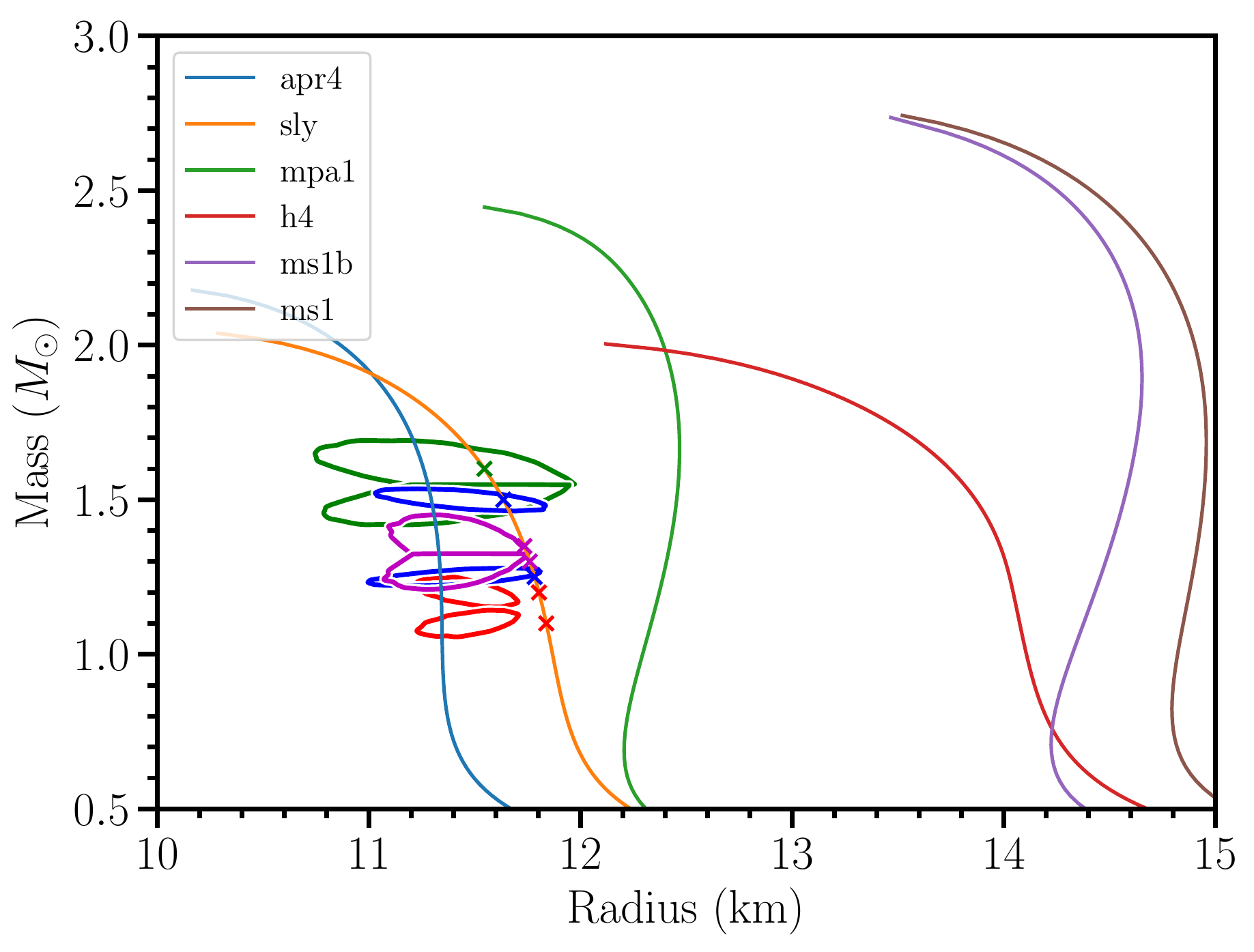}
\captionof{figure}{\small The plot shows $90\%$  credible region for eight NSs  with masses in the range $1.1\mbox{--} 1.6 M_\odot$ in four binaries, optimally oriented at a distance of $400~{\rm Mpc}$, with signal-to-noise  ratio of $100\mbox{--}140$ in a 3G detector. Crosses (curves) represent  the masses and radii (six equations of states) chosen for the simulation. In this example, the NS radius is measured to within $0.5\mbox{--}1~{\rm km}$. }
\label{fig:mr}
\end{tcolorbox}
}
\vskip-10pt
\end{table}
The merger  outcome depends on the companion masses and the equation of state of NS matter \cite{Duez:2009yz, Faber:2012rw, Paschalidis:2016agf, Baiotti:2016qnr}.  Above a critical total mass of the binary the merger will result in prompt collapse to a BH, while for very low-mass progenitors a stable remnant may form. 
For a wide range of parameters, the merger outcome is a short-lived hypermassive NS that involves complex microphysics and generates a significant amount of  GWs~\cite{Xing:1994ak, Shibata:1999wm, Oechslin:2001km, Shibata:2002jb, Bauswein:2011tp, Bauswein:2012ya, Takami:2014tva, Bernuzzi:2015rla, Lehner:2016lxy, Maione:2017aux} ultimately collapsing to a BH. The rich GW signals from the merger and post-merger regimes have frequencies in the range 1--5 kHz and are thus difficult to measure with advanced detectors; detailed studies of the complex physics driving the dynamics of NS binary mergers and beyond are a major unique capability of the 3G detector network. 

\begin{figure*}
\begin{tcolorbox}[standard jigsaw,colframe=gray,colback=gray!10!white,opacityback=0.6,coltext=black, title=\small\sc Phase Diagram of Dense Matter ]
\begin{minipage}{0.55\textwidth}
\center{\includegraphics[width=0.90\textwidth]{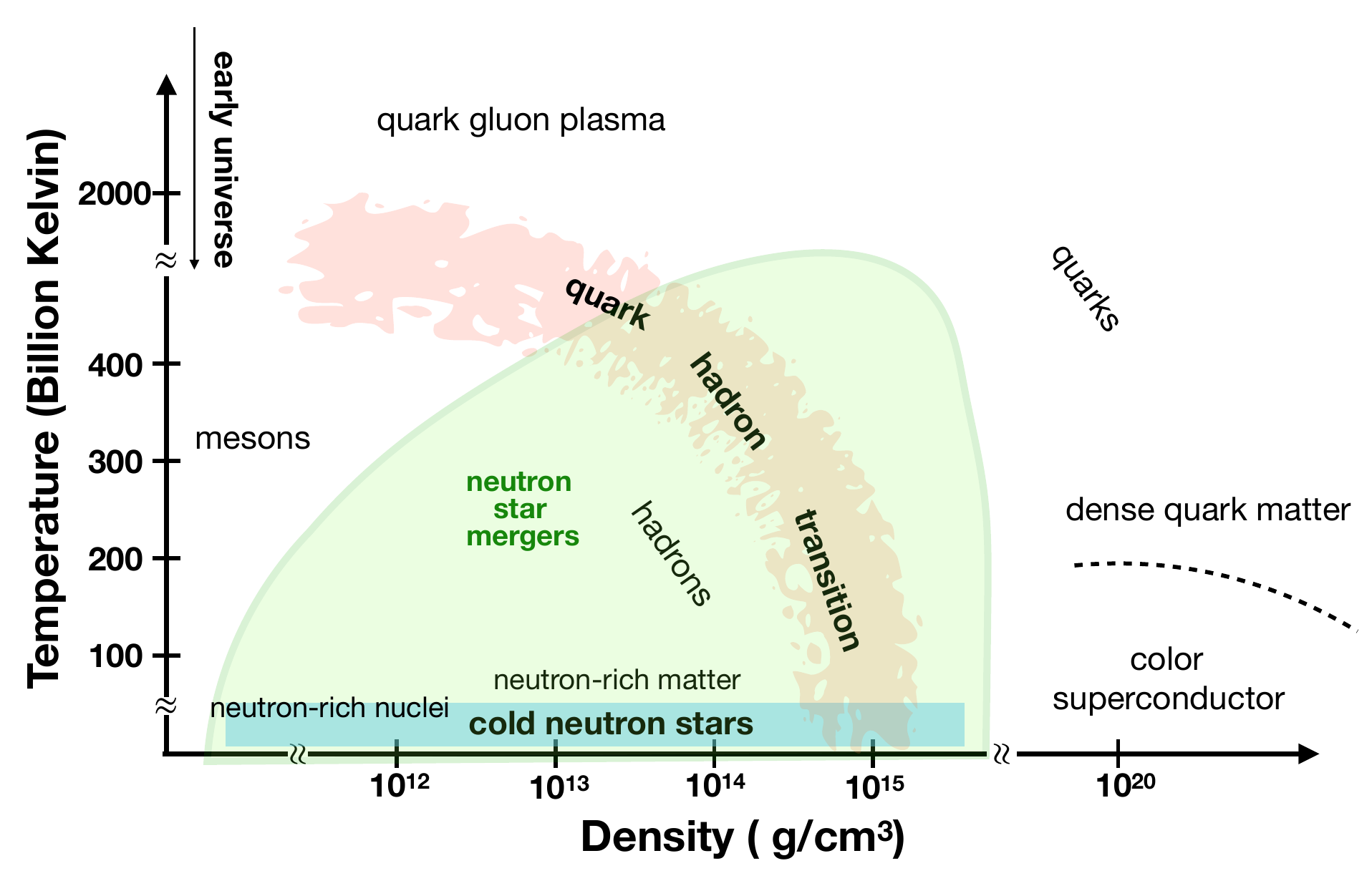}}
\end{minipage}
\hfill
\begin{minipage}{0.40\textwidth}
\captionof{figure}{\small The phase diagram of dense matter conjectured by theory and focusing on the regime relevant for NSs, complementary to the parameter space probed by heavy ion collisions. The figure shows approximate  boundaries between different phases of matter and the most relevant degrees of freedom. The regions occupied by cold NSs (blue shaded region) and the those encountered during the merger (green shaded region) are also shown. }
\label{fig:phase_diagram}
\end{minipage}
\end{tcolorbox}
\vskip-15pt
\end{figure*}

Various phases of strongly interacting matter that will be uniquely accessible with 3G detectors are depicted in Fig.~\ref{fig:phase_diagram}, which focuses on the regimes relevant to NS binaries that are complementary to those explored by heavy ion collisions. Matter encountered during BNS mergers, shown as light-green shaded region, explores a large swath of the phase diagram of dense matter. The 3G network will enable unprecedented measurements of the new physics encountered during the coalescence and post-merger epochs, with EM and neutrino counterparts providing complementary information to obtain a deeper and more complete understanding of extreme states of matter.  

In summary, observations by the 3G network will shed light on 
many critical questions about the nature of NSs and the fundamental subatomic physics of matter:
Are NSs composed solely of similar constituents as nuclei on earth or do they contain condensates of exotic particles or quark matter phases? Do NS mergers produce novel phases of matter not realized inside nuclei and heavy-ion collisions? What is the nature of the transition from nuclear to quark matter?  How do nuclear reactions and neutrinos shape NS merger dynamics? 

\section{Nucleosynthesis in Neutron Star Binary Mergers}
A long-standing puzzle in astrophysics is how the elements
heavier than iron came into being. About half of these
elements are believed to have been created by a nuclear process of rapid
neutron capture (the r-process) but it is unclear which
astrophysical sites are the main contributors.  
GW170817 and its associated thermal EM counterpart provided the first
direct identification of a NS merger as a prolific site of r-process
nucleosynthesis \cite{GBM:2017lvd}.
However, determining the degree to which
NS mergers contribute to cosmic chemical abundance and
evolution will require a more extensive sample of the
rates, locations, timescales, and nucleosynthetic yields of
the various types of merger events. 

Heavy elements can be synthesized in BNS or NS-BH tidal disruptions
when clouds of neutron-rich material are expelled, either
dynamically during the merger or through winds blown off the
remnant accretion disk. The subsequent
radioactive decay of the freshly synthesized elements powers
a thermal ultraviolet/optical/infrared EM transient called a kilonova.
The brightness and color of the kilonovae are diagnostic of
both the total mass of r-process elements and the relative
abundance of lighter to heavier elements \cite{Kasen:2017sxr}. 

Whereas historical studies of chemical evolution have relied
on observing fossil traces of r-process elements mixed into
old stars, multimessenger observations provide the
unique opportunity to study heavy element formation
at its production site and to determine how the initial
conditions of an astrophysical system map to the final
nucleosynthetic outcome.  Answering the basic question of
the extent to which BNS and NS-BH mergers are the site of
r-process production will require multimessenger observations of a large sample of events. GW measurements with the 3G network will pin down the rate of mergers and the binary properties, such as the binary type (BNS or NS-BH), companion masses, spin--orbit alignment, and the merged remnant lifetime, while optical/infrared photometry of the associated kilonovae will determine the average r-process yields and probe the relative abundance distribution of heavy elements.
These observations will also illuminate the key physics driving the r-process and kilonova, such as the equation of state of dense matter, the fundamental interactions of neutrinos and the  magneto-hydrodynamics of accretion.

Statistical studies of multimessenger observations will reveal how r-process production depends on host galaxy type, location and redshift, allowing us to piece together the history of when and where the heavy elements were formed over cosmic time.  Such studies can determine the distribution of delay times between star formation and mergers, thereby addressing whether some of these mergers occurred promptly enough to explain the enrichment of the oldest metal poor stars and the extent to which compact binaries receive strong kicks that may expel them from their host galaxies, a factor that is important for understanding whether mergers can explain the unusually high r-process enhancement seen in some dwarf galaxies \citep{Ji:2015wzg}.

In the era of 3G detectors, optical kilonovae will be detectable by the Vera Rubin Observatory out to 3 Gpc and infrared characterization photometrically by WFIRST/Euclid and spectroscopically by JWST/GMT/TMT/E-ELT would be out to $1$ Gpc ($z<0.2$). The multimessenger information about these event will enable answering questions such as: How nuclear reactions and neutrinos shape NS merger dynamics and nucleosynthesis? How do the properties of nuclei that are far from stability impact the EM emission from material ejected during NS mergers? How do the progenitor properties impact the nucleosynthesis and kilonovae?

\section{Formation, Demographics and Merger Sites of Compact Binary Mergers}
Observations of NS binary merger systems systems are essential to advance our understanding of how nature assembles these systems--either via standard (isolated) binary star evolution or via dynamical encounters in dense stellar environments. Precise measurements of NS masses and spins in these systems across cosmic time will provide key evidence for their origin in different types of supernova explosions and fossil records of close binary progenitor star interactions and accretion history after birth. Any signs of NS masses being different in NS-BH mergers compared to BNS mergers will yield crucial information on their formation process and the evolution of massive stars.

A key question about compact binary mergers is their demographics, as this could reveal their formation mechanism. Localization of merger events to less than galactic scales ($\sim 30$ kpc) is essential to unambiguously infer associations of mergers with their host galaxies.  Without an EM counterpart the vast majority of GW events will have error boxes that greatly exceed the typical radii of potential host galaxies. The census of the binaries, their locations, and environments will provide deep insights into the formation and evolution of NS binaries and their connection with the progenitor stars~\citep{Chaurasia:2005aq,Bloom:1998zi,Abbott:2017ntl}.

EM follow-up of NS binary mergers will be critical in pinning down
host galaxies. For binaries involving a NS and a BH with a mass ratio that is not too large, depending on the BH's spin, the NS may get tidally disrupted. The debris may result in accretion disk around the BH and lead to EM counterparts that might rival the absolute visual magnitude of the GW170817 kilonova, and, unless they occur in globular cluster cores, will be detectable out to $z=0.5$ in the reddest filters. 

Based on our current understanding, galaxies are assembled by the merger of smaller proto-galaxies and star formation peaks near $z\sim 2$ \citep{Madau:2014bja}. Identification of kilonovae beyond $z\sim 0.5$ requires hour-long integrations on 8m class facilities such as LSST or Subaru, rendering the identification of the host galaxies of binary NS mergers near the peak of star formation more challenging in the absence of a gamma-ray burst jet pointing towards the Earth, even with ELTs.  Nevertheless, at redshifts $z<0.5$ 3G detectors will work in concert with astronomy facilities
to enable thousands of host galaxy identifications from NS
binary mergers where the kilonova counterpart is observable. At larger distances, the identification will be possible only through the detection of an associated gamma-ray burst afterglow,  which can be much more luminous than a kilonova if the jet is directed towards the Earth.

\section{Jet Physics in Neutron Star Binary Mergers}
Relativistic explosions and compact-object mergers can generate collimated, energetic jets of fast-moving material and radiation. Prior to GW170817, our understanding of jet physics came from studies of gamma-ray bursts, active galactic nuclei and X-ray binaries. Multimessenger observations provide an entirely new
perspective on this topic. For instance, the panchromatic study of GW170817 revealed that there was both a narrow ultra-relativistic jet \citep{2018Nature.561..355M, 2018arXiv180800469G, Margutti:2018xqd, Lamb:2018qfn} and a wide-angle mildly relativistic cocoon from surrounding material ejected during the merger \citep{Nakar:2018cbe, Kasliwal:2017ngb, Mooley:2017enz}. 
This event opened up many questions for future observations to answer. Specifically, what is the connection to the class of cosmological short hard gamma-ray bursts? Does a wide-angle mildly relativistic cocoon always accompany a binary NS merger? Does the jet always successfully escape the cocoon or is it sometimes choked? How do the observed jet properties vary as a function of viewing angle, mass ratio, hypermassive NS lifetime, remnant spin, and ejecta mass? Do mergers produce prompt EM signals or even precursors? What is the distribution of the time delays between the EM and GW signal arrival times? What are the characteristics of a jet from a NS-BH merger? A census of NS binary mergers, and full GW and EM coverage of the signals, joint multimessenger parameter inference will be key in understanding the physical origin of jets, ubiquitous around relativistic sources. For the first time, a direct measurement of the BH spin in a source emitting a collimated jet, will enable to establish the close correlations between the jet power, the spin and the inflow rate from the debris disk, which determines the conditions for launching the jet. The sensitivity of gamma-ray, X-ray and radio telescopes will enable studying jet physics out to 500\,Mpc, thus requiring a sample of the order of a thousand events localized to better than few square degrees to map the full parameter space provided by the 3G network.

\begin{table}
\parbox{1.0\linewidth}{
\begin{tcolorbox}[standard jigsaw,colframe=ocre,colback=ocre!10!white,opacityback=0.6,coltext=black,title=\small\sc Facilities for Observing EM Counterparts to GWs]
\small
\caption{Present ($P$) and future ($F$) EM facilities that are able to observe faint/distant counterparts to GWs. Detection Limit ({\bf DL}, 1 hr exposure time) for UV, optical, and near-IR facilities are expressed in AB magnitudes, for X-rays in $10^{-16}\,\rm erg\,s^{-1}\,cm^{2},$ and for radio in $\mu$Jy.  Distance reach ({\bf D} in Mpc) of facilities for GW170817-like events are also shown.}
\vskip-5pt
\label{tab:facilities}
\parbox{.50\linewidth}{
	\begin{tabular}{cllr} 
		\hline
		\hline
	    & {\bf Facility} & {\bf DL} & {\bf D} \\
        \hline 
         Gamma-rays 
         & {\it Fermi} $P$ & S/N 5 & 80 \\ 
         & AMEGO $F$ & S/N 5 & 130 \\ 
        \hline
         &  {\it Swift} $P$ & S/N 5 & $\sim$80 \\ 
         & {\it Chandra} $P$ & 30 & 150 \\
         X-rays
         & {\it ATHENA}  $F$ & 3 & 480 \\
         & {\it Lynx} $F$ & 6  & 450 \\%
         & STROBE-X $F$ & S/N 5 & 120 \\
        \hline UV 
        &  {\it HST} (im) $P$ & 26 & 2000 \\
        &  {\it HST} (spec) $P$ & 23 & 400 \\
        \hline
        Optical 
        & Subaru $P$ & 27 & 3200 \\
        Imaging 
        & LSST $F$ & 27 & 3200 \\ 
        \hline
        IR 
        & {\it WFIRST} $F$ & 27.5 & 4800  \\
        Imaging
        & {\it Euclid} $F$ & 25.2 & 1700 \\
        \hline
        \end{tabular}
        }
        \hskip20pt
\parbox{.50\linewidth}{
	\begin{tabular}{cllr} 
        \hline
	    & {\bf Facility} & {\bf DL} & {\bf D} \\
        \hline
        &   Keck/VLT & 23 & 500 \\
        & Gemini Obs. & 23 & 500 \\
        Optical
        & GMT $F$ & 25 & 1265 \\
        Spec.
        & TMT $F$ & 25.5 & 1592 \\
        & E-ELT $F$ & 26 & 2005 \\
        \hline
        & Keck/VLT & 21.5 & 481  \\
        Infrared 
        & GMT $F$ & 23.5 & 762 \\
        Spec.
        & TMT $F$ & 24 & 960 \\
        & E-ELT $F$ & 24.5 & 1208 \\
        \hline
        \multirow{4}{*}{Radio} 
        & VLA (S) $P$ & 5 & 91 \\
        & ATCA (CX) $P$ & 42 & 51 \\
        & ngVLA (S) $F$ & 1.5 & 353 \\
        & SKA-mid (L) $F$ & 0.72 & 634 \\
		\hline
        \hline
	\end{tabular}
        }
        \end{tcolorbox}
        }
        \vskip-5pt
\end{table}     

\section{Outlook for Extreme Matter and Extreme Environments}
 Observations of BNS and NS-BH mergers with a network of 3G detectors will transform our understanding of the fundamental properties of matter in unexplored regimes of density and temperature and, in conjunction with EM facilities, will address longstanding questions about the formation of heavy elements in the universe, the central engines of highly energetic EM transients, and the formation and evolution of NS binary systems. 
 \vfill

\begin{tcolorbox}[colback=teal!5!white,colframe=red!75!black,title=\sc Science Requirements]
{
The unique capabilities of 3G detectors required to accomplish these science goals are:
\begin{itemize}
    \item  {\em an order-of-magnitude greater sensitivity} than 2G detectors enabling observations of the NS binary population in the cosmos and measurements of loud-source properties with exquisite accuracy, 
    \item {\em a wider frequency range} than 2G detectors that will allow tracking the entire GW signals from the inspiral through the merger, tidal disruption and beyond, and
    \item synergies with panchromatic EM facilities, that will be critical to fully capitalize on the rich multimessenger science potential of these sources for cosmology, fundamental physics, and astrophysics.
\end{itemize}
}
\end{tcolorbox}

%% file: bbh.tex
\chapterimage{MassPlot_GW190521_errorBars.png} 

\chapter{Observing Stellar-mass Black Holes Throughout the Universe}

\begin{tcolorbox}[colback=teal!5!white,colframe=yellow!75!black,title=\sc Science Target]
{\em Reveal merging black holes across the cosmos and \emph{search} for seeds of supermassive black holes.}
\end{tcolorbox}

\textbf{Merging binary black holes are sources unique to GW astronomy}---they are the most frequently observed sources to date. We now know that binary black holes (BBHs) form ubiquitously in galaxies and so far appear to be completely dark. These mergers are unrivalled laboratories for testing extreme gravity, and exquisite astronomical sources for gaining insight into the origin and evolution of massive stars in the Universe. 
With a leap in sensitivity and increased frequency bandwidth, the 3G network will observe them back to the early Universe, chart how the population evolves with time, discover a broader range of masses, and connect stellar-mass black holes (BHs) with the supermassive BHs (SMBHs) found in the centres of galaxies, obtain precision measurements of BH properties, and finally resolve the mysteries of their formation.  
GW astronomy is perfectly suited to studying BHs, and with 3G detectors we would achieve a complete picture of the family of stellar-mass BBHs.
\vfill

\begin{tcolorbox}[standard jigsaw,colframe=ocre,colback=ocre!10!white,opacityback=0.6,coltext=black,title=\sc Key Science Goals]
The 3G network will uncover BBHs throughout the cosmos, back to the beginning of star formation and detect new sources, if they exist, beyond stellar-mass binaries, such as intermediate-mass BBHs. 

\begin{itemize}[leftmargin=*]
\item {\bf Discover BBHs throughout the Universe.} 
What is the merger rate as a function of cosmic time and how does it relate to the star formation rate, metallicity and galaxy formation and evolution? 

\item {\bf Reveal the fundamental properties of BHs.} 
What are the mass and spin demographics of BHs throughout the Universe? Are they correlated and do they evolve with redshift? What do they reveal about the formation and evolutionary origin of BHs? 

\item {\bf Uncover the seeds of SMBHs.} 
GW observations have proven that intermediate-mass BHs can form at least from BBH mergers. The 3G network promises to explore their population, reveal if intermediate-mass BH mergers occur in nature and serve as the long-sought seeds of SMBHs? 
\end{itemize}
\end{tcolorbox}

The first three observing runs of Advanced LIGO and Advanced Virgo  have yielded the discovery of \emph{more than 50} BBH systems. Already these detections have revolutionized the astrophysics of stellar-mass BHs \cite{TheLIGOScientific:2016htt,LIGOScientific:2018jsj,LIGOScientific:2020stg,Abbott:2020mjq,Abbott:2020khf} and provided first new tests of general relativity \cite{TheLIGOScientific:2016src,Yunes:2016jcc,Abbott:2017vtc,Abbott:2017oio,LIGOScientific:2019fpa}. Through the end of the 2020s, the current advanced detector network will continue to be enhanced as sensitivities reach design goals and a new detector in Japan, KAGRA, comes online \cite{Akutsu:2018axf,Aasi:2013wya}. In the BBH domain, we will be able to detect a pair of $10 M_\odot$ BHs out to a cosmological redshift of $z \simeq 1$ when the Universe was $6~\mathrm{Gyr}$ old \cite{Aasi:2013wya}. The annual BBH detection rates are forecast to be several hundreds of mergers and science benefits will compound through accumulated observing time and growing detected samples \cite{Mandel:2009nx,Stevenson:2017dlk,Talbot:2017yur,Barrett:2017fcw,Zevin:2017evb,Fishbach:2018edt}.

Beyond this horizon, step-wise sensitivity improvements with the next generation of ground-based GW observatories will be required if we are to pursue major science questions that cannot be answered by the current and near-term GW facilities \cite[e.g.,][]{Sathyaprakash:2012jk,Evans:2016mbw}. Current-generation GW detectors are able to provide constraints on the merger-rate densities in the local Universe and broad constraints on component masses \cite{LIGOScientific:2018jsj,Roulet:2020wyq}. However, precise measurements of, for example, spin magnitudes and tilts are of paramount importance to understand the origin and the evolutionary physics of binary systems \cite{Mandel:2009nx,Rodriguez:2016vmx,Vitale:2015tea,Stevenson:2017dlk,Talbot:2017yur,Kimball:2019mfs,Bavera:2019fkg,Sedda:2020vwo}. 
This information is essential to obtain insights on the formation channels of compact binaries. So far we have been surprised by the properties of individual exceptional sources, but the population constraints are too weak to distinguish among formation path possibilities. We highlight here how 3G GW ground-based detectors will enable us to survey deeper, to observe a wider range of frequencies, and to make more precise physical measurements; how observations can be synergistically combined between 3G and space-based GW observatories, and how these results will be transformational in the study of BBH astrophysics. 

\section{A survey of BHs throughout cosmic time}

With a 3G detector network, for the first time, we will detect BBH mergers at redshifts beyond $z \sim 1$ and we will measure the evolution of the BBH merger rate out to redshifts of $z \gtrsim 10$ when the Universe was $< 500~\mathrm{Myr}$ old \cite{Aasi:2013wya,Fishbach:2018edt,Vitale:2018yhm}. GW astronomy would thereby gain a synoptic view of the evolution of BHs across cosmic time, beyond the peak of the star-formation rate, which took place at $z \sim 2$ when the Universe was $3~\mathrm{Gyr}$ old \cite{Madau:2014bja}, back to the cosmic dawn around $z \sim 20$ when the Universe was only $200~\mathrm{Myr}$ old and the first stars were forming in pristine dark matter halos.

Measurements of the merger rate as a function of redshift combined with high fidelity measurements of the BH physical parameters will enable conclusive constraints on the BBH formation channels. Stellar-origin BH formation tracks cosmic star formation \cite{Dominik:2013tma,Mapelli:2017hqk,Chruslinska:2018hrb,Mapelli:2018wys,Neijssel:2019irh,Santoliquido:2020axb}, while the density of primordial BHs is expected to be independent of the star formation density \cite{Sasaki:2018dmp,Scelfo:2018sny}; different binary formation channels are predicted to lead to different distributions of delay times between formation and merger \cite{Dominik:2012kk,Kinugawa:2015nla,Mandel:2015qlu,Marchant:2016wow,Rodriguez:2016avt,Barrett:2017fcw,Fragione:2018vty,Qin:2018nuz,Rodriguez:2018rmd,Choksi:2018jnq,Safarzadeh:2020qrc}. Therefore, determining the merger rate as a function of redshift provides a unique insight into the lives of binary BHs. \emph{Only next-generation GW detectors can survey the complete redshift range of merging BBHs and provide a sufficiently large catalog of detections to constrain the full BBH population and their origins}. 

\begin{figure*}
\begin{tcolorbox}[standard jigsaw,colframe=gray,colback=gray!10!white,opacityback=0.6,coltext=black, title=\small\sc Sensitivity Required to Observe Cosmic Binary Black Holes]
\centering
  \vskip-5pt
  \includegraphics[width=0.45\textwidth]{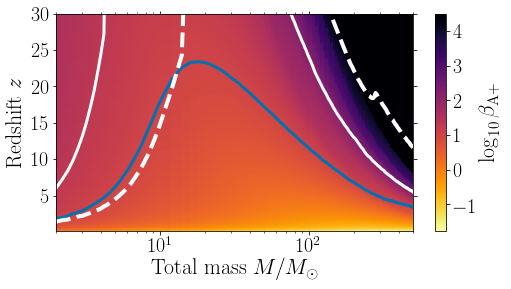}
  \includegraphics[width=0.45\textwidth]{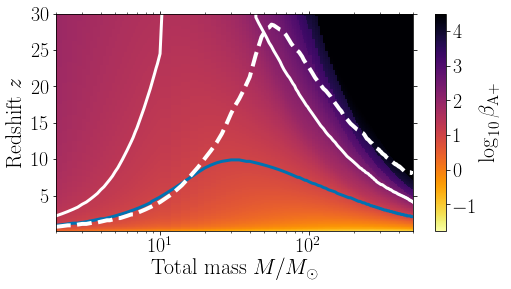}
  \vskip-5pt
  \captionof{figure}{\small Colour maps show the boost factor relative to the LIGO A+ design $\beta_\mathrm{A+}$ required to see a binary with a given total source mass $M = m_1 + m_2$ out to given redshift. 
The colour bar saturates at $\log_{10} \beta_\mathrm{A+} = 4.5$; some high-mass systems at high redshift are not detectable for any boost factor as there is no signal above $5~\mathrm{Hz}$. 
Panels are for mass ratios $q = m_2/m_1 = 1$ (\textbf{left}) and $q = 0.1$ (\textbf{right}). 
The blue curve highlights the reach at a boost factor of $\beta_\mathrm{A+} = 10$. 
The solid and dashed white lines indicate the maximum reach of Cosmic Explorer \cite{Evans:2016mbw} and the Einstein Telescope \cite{Sathyaprakash:2012jk}, respectively; sources below these curves would be detectable.}
\label{BCO-detect}
\end{tcolorbox}
\vskip-20pt
\end{figure*}

To capture BBH mergers across the stellar mass spectrum (up to total masses of $M = m_1 + m_2 \simeq 200 M_\odot$) all the way back to the end of the cosmological dark ages ($z \simeq 20$), a major advance in GW detector sensitivity is required.
This cannot be delivered by the maximal sensitivity planned for the current ground-based detector facilities. We quantify this sensitivity step by the boost factor $\beta_\mathrm{A+}$ relative to the LIGO A+ design \cite{Barsotti:2018} between $5~\mathrm{Hz}$ and $5~\mathrm{kHz}$ (and no sensitivity outside this range). In Figure~\ref{BCO-detect}, we show this boost factor, required to detect an optimally-oriented, overhead binary at a signal-to-noise ratio (SNR) of $8$, as a function of the binary's total mass and redshift. The boost factors $\beta_\mathrm{A+}$ needed to acquire a complete census of BBH mergers throughout the Universe are well within the design aspirations the 3G network; for these specific sensitivity assumptions, BBH mergers of total mass $M \sim 10$--$40 M_\odot$ can be detected out to $z \sim 10^2$.

Observations of the cosmological distribution of coalescing binaries would complement planned EM surveys designed to study stars and stellar remnants back to cosmic dawn \cite{Whalen:2012yk,Koopmans:2015sua,Cassano:2018zwm,Kalirai:2018qfg,Katz:2019akl}, as well as millihertz GW observations made by the \textit{Laser Interferometer Space Antenna} (\textit{LISA}) \cite{Audley:2017drz}, which can observe systems ranging from local stellar-mass binaries (days to years before they enter the frequency range of terrestrial detectors) \cite{Sesana:2016ljz,AmaroSeoane:2009ui} to SMBH systems in the centres of galaxies \cite{Klein:2015hvg,Babak:2017tow}. In X-rays SMBHs are detectable due to gas accretion in galactic nuclei. X-ray observatories like \textit{Athena} \cite{Barret:2013bna} and the mission concept \textit{Lynx} \cite{LynxTeam:2018usc} would detect SMBHs back to high redshift ($z \gtrsim 7$); \textit{Lynx} would observe $10^3 M_\odot$ BHs to $z \sim 5$ and $10^2 M_\odot$ BHs to $z \sim 2$, while \textit{Athena} will survey these in the nearby Universe. Next-generation GW detectors have the unique potential to observe stellar-mass BH systems all the way back to the early Universe.

\section{Expanding the BH mass spectrum}

EM astronomy has benefited enormously from advancing observing facilities to cover an expanded range of frequencies. These enable new probes of previously known sources, and allow for the discovery of new types of previously unobserved sources. \emph{3G GW detectors have the unique capability to push the frequency range down to $\simeq 1~\mathrm{Hz}$ and up to $\simeq 5~\mathrm{kHz}$}, while improving performance across the band in between. 

The merger frequency for a coalescing binary scales inversely with the mass of the binary, hence observing at lower frequency opens up the potential of detecting more massive BHs. The first intermediate-mass BH (mass in excess of $100 M_\odot$) has already been observed as the result of a BBH merger \cite{Abbott:2020tfl}, but there may be multiple formation paths for these BHs.  Reaching down to frequencies of $\simeq 1~\mathrm{Hz}$ is the most robust means to chart the population of intermediate-mass BHs, and discover any mergers of intermediate-mass BHs---which may be the process through which SMBHs form  \cite{Volonteri:2012tp,Latif:2016qau,Bernal:2017nec,Woods:2018lty}. 
SMBHs are observed up to redshift $z=7.54$ \cite{Banados:2017unc} as quasars, at lower redshifts as active galactic nuclei \cite{Merloni:2015dda}, and today in massive galaxies in their quiescent state \cite{Kormendy:2013dxa}, and cover a mass range from $\sim 10^4M_\odot$ \cite{2017ApJ...836..237N,2015ApJ...809L..14B,Graham:2018bre,Graham:2018byx} up to $> 10^{10}M_\odot$ \cite{McConnell:2011mu,2015Natur.518..512W,2018MNRAS.474.1342M}. 
SMBHs may have light seeds ($\sim 10^{2}$--$10^{3} M_\odot$), formed from massive stars in low metallicity halos which evolve into BHs beyond the pair instability gap \cite{Hirano:2015wxa}, or heavy seeds, formed from supermassive (proto)-stars  of $\sim 10^{4}$--$10^{6} M_\odot$ growing through continued and fast accretion within their birth clouds, which eventually collapse down to BHs \cite{Latif:2013pyq,2014MNRAS.442.2036D,Umeda:2016smj,Habouzit:2016nyf, 2016agnt.confE...4V,Regan:2017vre}. 
In particular, the observation of high-redshift BHs with mass $\gtrsim 100 M_\odot$, beyond the (pulsational) pair-instability supernova mass gap (where supernova explosions are hypothesised to completely disrupt the  star leaving behind no remnant) \cite{Belczynski:2016jno,Spera:2017fyx,Woosley:2016hmi,Giacobbo:2017qhh,Marchant:2018kun,Farmer:2019jed}, would be key to understand not only the properties of very massive ($\gtrsim 250 M_\odot$) metal-poor stars \cite{Hartwig:2016nde,Ezquiaga:2020tns}, but also the assembly of the first massive BHs in the Universe \cite{Ferrarese:2000se,Peng:2007mv,Volonteri:2009vh,2016MNRAS.457.3356V,Silk:2017yai}.

\begin{figure*}
\begin{tcolorbox}[standard jigsaw,colframe=gray,colback=gray!10!white,opacityback=0.6,coltext=black, title=\small\sc Waveform from a Binary Black Hole Coalescence]
\centering
    \includegraphics[width=0.451\textwidth]{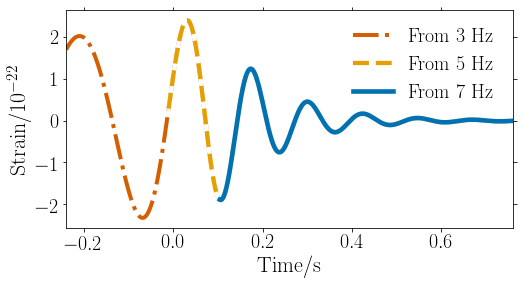}
    \includegraphics[width=0.453\textwidth]{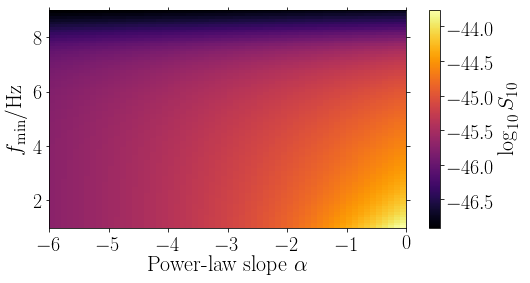}
    \captionof{figure}{
\textbf{Left:}  The waveform from the final stages of inspiral, merger and ringdown of a $100 M_\odot + 100 M_{\odot}$ BBH at a redshift of $z=10$. 
Highlighted is the time evolution of the waveform from $3$, $5$ and $7~\mathrm{Hz}$. 
\textbf{Right:} Requirements on the low-frequency noise power spectrum $S_n(f)$ necessary to detect an overhead, face-on $100 M_\odot + 100 M_\odot$ BBH merging at $z=10$. 
We assume a power-law form $S_n(f) \propto f^\alpha$ extending down to a minimum frequency $f_\mathrm{min}$ with the specified normalization $S_{10}$ at $f=10~\mathrm{Hz}$.}
\label{fig:low-f}
\end{tcolorbox}
\vskip-20pt
\end{figure*}

In Figure~\ref{fig:low-f}, we illustrate the importance of sensitivity in the $1$--$10~\mathrm{Hz}$ regime.  Even with detectors sensitive to $3~\mathrm{Hz}$, we see only one cycle of a $100 M_\odot + 100 M_\odot$ circular binary with non-spinning components at $z =10$ before merger. This system is not observable above $10~\mathrm{Hz}$.  Therefore, the objective to observe the most massive BBHs of stellar origin and the potential seeds of SMBHs in the Universe's early history requires new detectors sensitive to currently inaccessible frequencies below $\sim 10~\mathrm{Hz}$.

The detectability of intermediate-mass BHs places requirements on low-frequency sensitivity. We can model the low-frequency noise power spectral density of the detector as a power-law $S_n(f) = S_{10}(f/10~\mathrm{Hz})^\alpha$ and assume that the power law extends to some minimal frequency $f_\mathrm{min}$, below which the detectors have no sensitivity.  In Figure~\ref{fig:low-f}, we show the combination of power law $\alpha$, minimum frequency $f_\mathrm{min}$ and the normalisation $S_{10}$ necessary to detect an optimally located and oriented merger of two $100 M_\odot$ intermediate-mass BHs at $z=10$. There is a trade-off between the power-law slope, minimal frequency and overall normalization, such that a range of specifications can fulfill the science requirements.  


\section{High-precision measurements of binary properties}

A leap in sensitivity combined with the increased frequency bandwidth of the 3G detectors will enable high-precision measurements of the properties of individual binaries \cite{Vitale:2016icu,Vitale:2018nif,Hall:2019xmm}. Parameter uncertainties are inversely proportional to the SNR \cite{Cutler:1994ys}. The increase in SNR made possible by the greater sensitivity will lead to exquisite measurements of the loudest events. Increased bandwidth enables the coalescence to be tracked for a longer time, improving estimates of quantities like the spins. Masses, spins, merger redshifts, orbital eccentricities and (where possible) associations with host galaxies all give complementary insights into binary physics. High-precision measurements of individual systems allow us to make detailed studies of their origins and fundamental physics \cite{Mishra:2010tp, Gossan:2011ha, Bhagwat:2016ntk, Berti:2018vdi, Isi:2019aib, Carullo:2019flw}. Combining many events together lets us study the properties of the population.  \emph{The unique and critical advantage of BBH observations with the 3G network is the combination of high-precision measurements for a very large number of detected sources}, something that cannot be delivered by the current detectors.

As an example, consider a highly precise reconstruction of the BH mass spectrum. At high masses, a gap is predicted to exist between $\sim 50 M_{\odot}$ and $\sim 130 M_{\odot}$ due to (pulsational) pair-instability supernovae \cite{Woosley:2007qp,Woosley:2016hmi,Marchant:2018kun,Farmer:2019jed}, although nature has some ways to form BHs in it \cite{Abbott:2020tfl}. At lower masses, there is potentially a gap between the maximum neutron star mass and the minimum stellar BH mass \cite{Ozel:2010su,Farr:2010tu,Kreidberg:2012ud,Wyrzykowski:2019jyg,Abbott:2020khf}. Determining the precise bounds for these gaps would provide insight into the mechanics of supernova explosions \cite{Belczynski:2011bn,Fryer:2011cx,Fryer:2017iry,Zevin:2020gma}, insights into the neutron star equation of state \cite{Kiziltan:2013oja,Annala:2017llu,Margalit:2017dij,Alsing:2017bbc,Abbott:2018exr,Most:2020bba,Fattoyev:2020cws,Tsokaros:2020hli}, and even details of nuclear reaction rates \cite{Farmer:2019jed,Farmer:2020xne}. It can be shown that: (i) for the high-mass gap, if the desired accuracy on the mass gap boundary measurement is $\sigma_g \sim 1 M_\odot$, with a conservative individual mass uncertainty for near-threshold detections of order $\sigma_m \sim 10 M_\odot$, $N \gtrsim 500$ detections are required; (ii) for the low-mass gap, $\sigma_g \sim 0.3 M_\odot$ and $\sigma_m \sim 3 M_\odot$, would require $N \gtrsim 1500$ BBH detections. To provide robust answers to questions regarding massive star evolution and BBH formation, we need to trace the dependence of the boundaries of the mass gaps on metallicity and hence redshift. Therefore, it is desirable to observe $\sim 1000$ sources in each redshift bin of width $\Delta z = 0.1$, since we may expect knowledge of the star formation rate and metallicity distribution to be available at this resolution on the timescale of the 3G network \cite{Madau:2014bja}. Such observations would provide $\sim 3\%$ fractional accuracy on the merger rate per redshift bin, sufficient to determine the redshift evolution of the rate, and constrain details of the binary evolution at that redshift \cite{Zevin:2017evb,Barrett:2017fcw}.

\begin{figure*}
\begin{tcolorbox}[standard jigsaw,colframe=gray,colback=gray!10!white,opacityback=0.6,coltext=black, title=\small\sc Expected BBH Merger Rate]
\begin{minipage}{0.60\textwidth}
\vskip-5pt
{\hskip-10pt \includegraphics[width=1.05\textwidth]{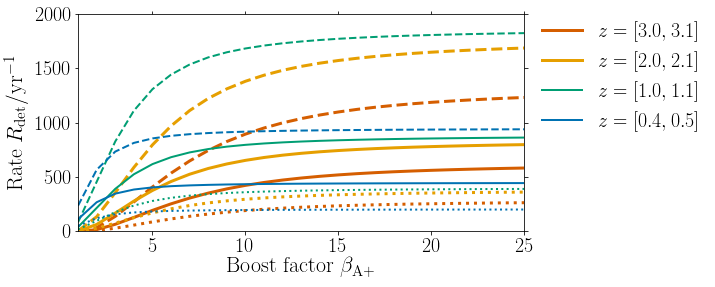}}
\end{minipage}
\hfill
\begin{minipage}{0.37\textwidth}
  \captionof{figure}{\small Expected rate of BBH detections $R_\mathrm{det}$ per redshift bin as a function of $A+$ boost factor $\beta_\mathrm{A+}$, for $z=[0.4,0.5]$, $z=[1,1.1]$, $z=[2,2.1]$, $z=[3,3.1]$.  Constant BBH merger rate densities of $53$ ($112$, $24$) $\mathrm{Gpc}^{-3}\,\mathrm{yr}^{-1}$ are shown with solid (dashed, dotted) curves, assuming equal component masses distributed according to $p(m) \propto m^{-1.6}$ \cite{LIGOScientific:2018jsj}.} 
  \vskip-10pt
  \label{fig:redshiftdistance}
\end{minipage}
\end{tcolorbox}
\vskip-15pt
\end{figure*}

With this in mind, we plot the number of expected BBH detections for a next-generation detector as a function of its boost factor relative to A+ in Figure~\ref{fig:redshiftdistance}. This assumes a BBH merger rate that does not evolve in redshift and is roughly consistent with current GW observations \cite{LIGOScientific:2018jsj}. From this, the target of $\sim 1000$ detections per redshift bin is achievable with boost factors of $\beta_\mathrm{A+} \sim 10$ after only $2$ years of observing time.
These factors are possible only with next-generation GW detectors. 

We note that observing across a broader range of frequencies gives a more complete picture of BBH properties. The precession of component spins misaligned with the orbital angular momentum occurs over many orbits \cite{Apostolatos:1994mx,Blanchet:2013haa}. Its imprint is easier to discern over longer inspirals, and hence becomes more apparent with low-frequency data. Orbital eccentricity is rapidly damped through GW emission \cite{Peters:1964zz}.This means that it is near immeasurably small for current GW detectors \cite{Lower:2018seu,Romero-Shaw:2019itr,Lenon:2020oza}, but reachable for 3G detectors. Both the spins and the orbital eccentricity are indicative of the formation channel; enabling their measurement for large samples will have a transformative effect on our ability to answer questions about BBH origins. 

\section{Multiband gravitational-wave observations}

Joint observations of GW events by \textit{LISA} at millihertz frequencies and 3G detectors at higher frequencies maximises their science potential. If \textit{LISA} had been observing in 2010, it would have detected GW150914 years before it was observed by LIGO \cite{Sesana:2016ljz}.  \textit{LISA} will potentially see up to hundreds of stellar-mass BBH mergers  of $M>20$--$30 M_\odot$, up to $z\approx0.3$ \cite{Sesana:2016ljz,2016MNRAS.462.2177K}. 
A small fraction of these will sweep across the detector band within few years that will eventually be detected by ground-based detectors. Multiband observations enables unrivalled measurements of BBH properties.

\textit{LISA} would provide a precise measurement of the system's eccentricity to a precision of $\Delta{e}<0.001$ \cite{2016PhRvD..94f4020N}, sky localization to $0.1~\mathrm{deg}^2$, and time to coalescence within few seconds, several weeks prior to coalescence \cite{Sesana:2016ljz}. This enables EM telescopes to be pointed in the right direction before the merger, permitting a much deeper coverage from radio to gamma-ray than what is possible without any early warning. 
Alternatively, one can use the information extracted by the 3G network to dig out sub-threshold \textit{LISA} events \cite{Wong:2018uwb}. From an astrophysical standpoint, the eccentricity information from \textit{LISA} can be combined with the spin measurement from 3G detectors to better constrain different formation channels \cite{2016ApJ...830L..18B, 2017MNRAS.465.4375N, Samsing:2018ykz}. Multiband observations will also facilitate tests of general relativity \cite{2016PhRvL.117e1102V, Carson:2019rda, Gnocchi:2019jzp, Liu:2020nwz} by enhancing the sensitivity to specific deviations arising in the long inspiral as predicted, for example, from dipole radiation not predicted in general relativity \cite{2016PhRvL.116x1104B}. 

\begin{figure*}
\begin{tcolorbox}[standard jigsaw,colframe=gray,colback=gray!10!white,opacityback=0.6,coltext=black, title=\small\sc BBH Mergers in 3G and LISA]
\begin{minipage}{0.45\textwidth}
\includegraphics[width=1.00\textwidth]{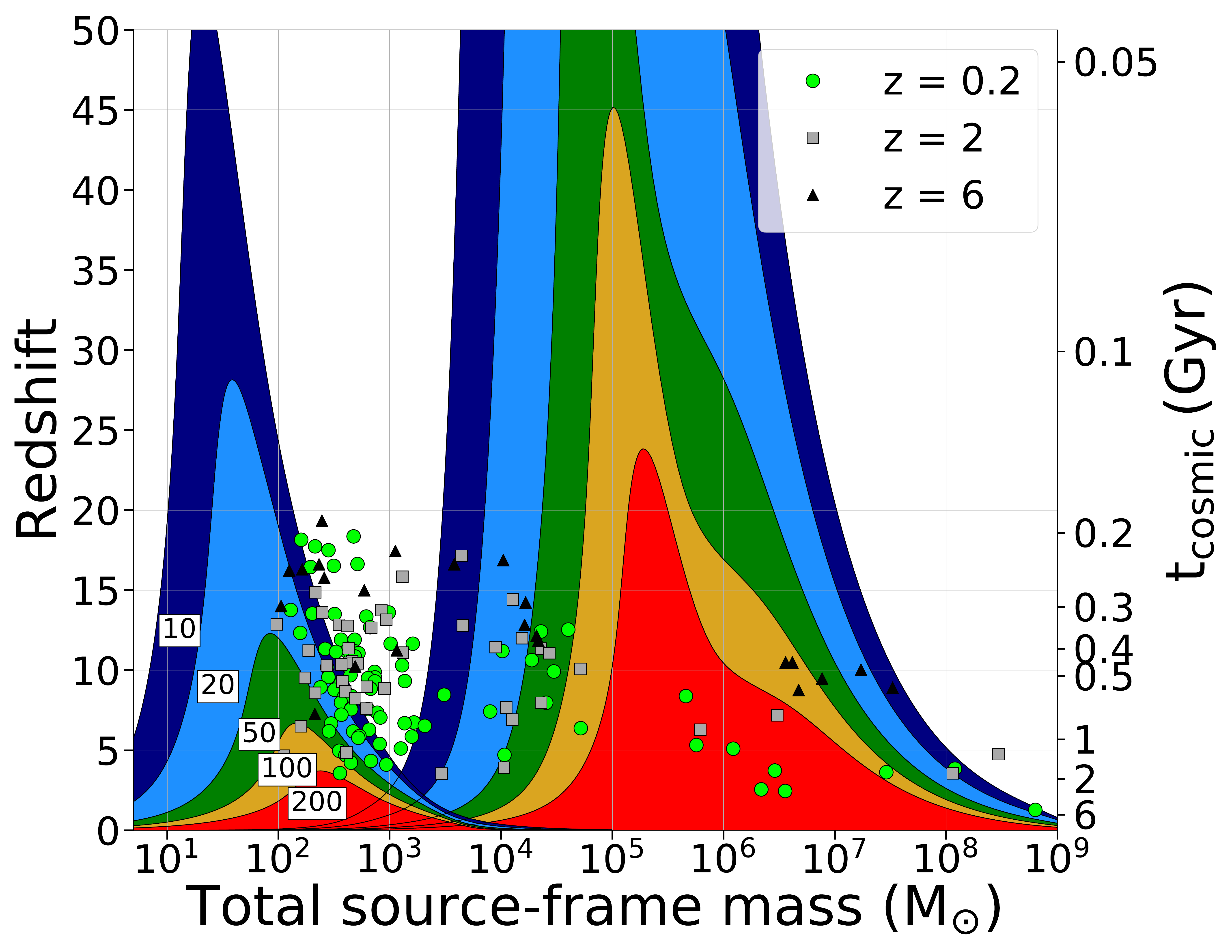}
\end{minipage}
\hfill
\begin{minipage}{0.45\textwidth}
\captionof{figure}{\small 
Contours of constant SNR in the space of redshift (or cosmic time) and total-mass assuming equal-mass binary companions. The plot also shows the cosmological growth of light seeds that formed when the Universe was less than 0.3 Gyr old ($z>13$), and the growth of heavy seeds that form at $z\sim 12.$ Seeds growing through merger transit into the supermassive domain and form quasars at $z=6$ (triangles), $z=2$ (squares) and $z=0.2$ (circles) \cite{Valiante:2020zhj}. Mergers above $10^5\,M_\odot$ host
at least one heavy seed. Mergers lighter than $10^5\,M_\odot$ come from only the light seed population.
}
\label{fig:ET+LISA} 
\end{minipage}
\end{tcolorbox}
\vskip-20pt
\end{figure*}

Figure\,\ref{fig:ET+LISA} shows the cosmological growth of the earliest light and heavy seeds that transit into the supermassive domain, inferred using a semi-analytical model for the formation of quasars at $z=6,$ $z=2$ and $z=0.2$ \cite{Valiante:2020zhj}. Binaries of light seeds ($\sim 10^2\,M_\odot$) are accessible to the 3G network with an SNR of 10--20 at $6 < z < 15.$ They then enter the {\em LISA} domain with larger SNRs as they grow to a few $10^4\,M_\odot.$ Mergers above $10^5\,M_\odot$ come from at least one heavy growing seed in the binary, lighter mergers arise from the light-seed population.
Combining the observations in the two different frequency domains will provide the first ever census of coalescing BBHs forming in the Universe. The comparison between the detection rate between light and heavy seed BHs in the two GW bands will be instrumental in determining, at statistical level, the relative contribution of light and heavy seeds in building up the  population of SMBHs, and the role of mergers versus accretion in determining their growth. Figure~\ref{fig:ET+LISA} highlights that intermediate-mass BHs are a prime multiband GW astronomy target  \cite{2010ApJ...722.1197A,Amaro-Seoane:2018gbb}.  As well as both 3G and space-based detectors being able to see the same intermediate-mass BH binary signal at different phases in the inspiral, two GW bands allow the intermediate-mass BH population to be explored in different regimes. 
While \textit{LISA} will be sensitive to mergers of $M \ge 10^3 M_\odot$ binaries out a redshift of $z>20$, 3G detectors will be able to access $M \sim 100 M_\odot$ populations at comparable redshifts. 
\emph{Multiband GW observations will quantify the continuity between the stellar-mass, intermediate-mass, and SMBH populations}. 

\section{Outlook for black hole gravitational-wave astronomy}

The 3G network will enable the measurement of the cosmological evolution of the mass and spin distributions of BBHs and probing their dependence on star formation history and metallicity evolution with redshift. 
They will make robust discovery of intermediate-mass black holes, if they exist, and reveal the boundaries of any mass gaps.  The precise measurements of physical properties for large numbers of BH systems, back to the cosmic dawn, would lead to constraints on the physics of massive star evolution in single and binary systems,  
and to place bounds on the different formation channels of merging BBHs.  Additionally, 3G observations could solve the long-standing mystery of the nature of SMBHs' seeds. 3G data would complement those from future EM and space-based GW observatories, enabling the maximum scientific return from these facilities.  We have an unparalleled opportunity to advance the frontiers of stellar astrophysics, the fundamental physics of compact objects, and the formation mechanisms for the entire spectrum of BHs. 
\vskip5mm

\begin{tcolorbox}[colback=teal!5!white,colframe=red!75!black,title=\sc Science Requirements]{
The 3G network will transform our BH studies by enabling cosmological probes of their formation across cosmic time: 
\begin{itemize}
    \item an improvement of a factor of 10-20 in strain sensitivity will allow us to probe the complete BH population to the edge of the universe, 
    \item reaching down to frequencies of about 1\,Hz is needed to detect a population of intermediate-mass BHs, and quantitatively probe mergers of such BHs and uncover their link to LISA sources and SMBH growth, and 
    \item both the leap in sensitivity and expansion to low frequencies are needed for precise BBH properties, masses, spins, and possibly eccentricities, to firmly uncover their formation origins. 
\end{itemize}


}
\end{tcolorbox}

%% file: cosmology.tex
\chapterimage{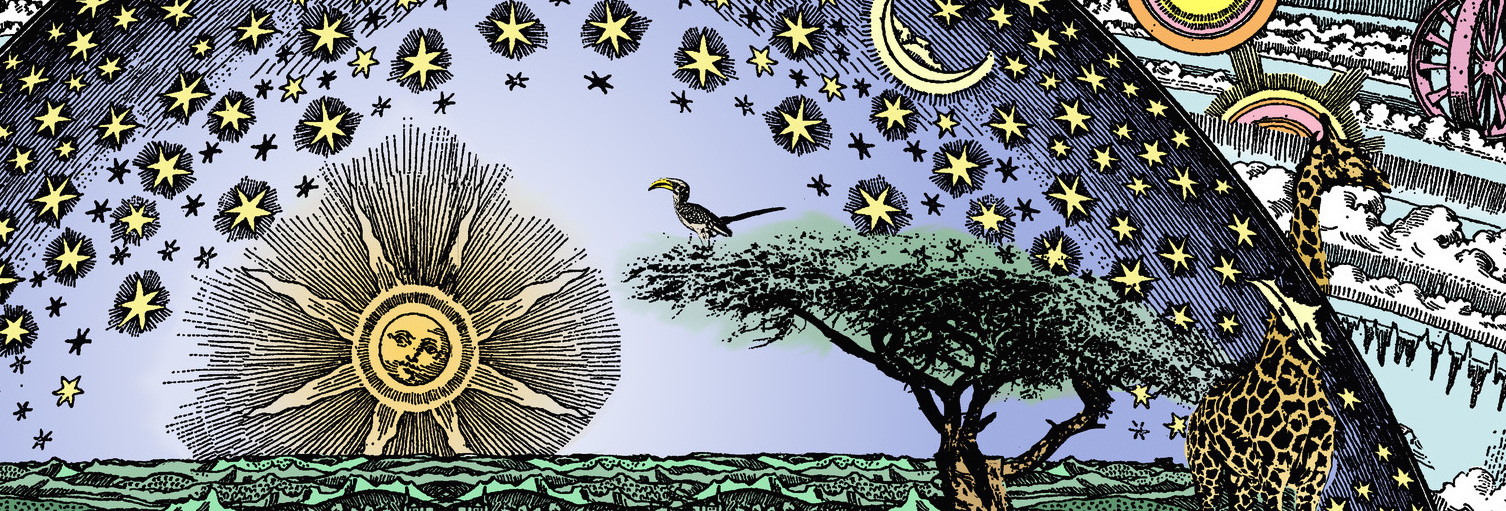} 
\chapter{Cosmology and the Early History of the Universe}
\begin{tcolorbox}[colback=teal!5!white,colframe=yellow!75!black,title=\sc Science Target]
{\em Investigate the particle physics of the primeval Universe and probe its dark sectors.}
\end{tcolorbox}

\textbf{Gravitational waves (GWs) offer unique new probes of the early universe and its composition}---the weakness of gravity relative to other known forces implies that GWs decoupled very early from the primordial plasma in the Universe, a fraction of a second after the Big Bang. Detection of such waves would provide unique opportunities to study the evolution of the very early Universe and of the physical laws that apply at very high energy scales, inaccessible in traditional laboratories. Particle physics processes that drove inflation and phase transitions in the early Universe would leave imprints in the form of spectral, spatial, and polarization properties of the stochastic GW background. In the more recent history, merging neutron star (NS) and black hole (BH) binaries generate signals that precisely determine the luminosity distances of the binaries, independently of the cosmic distance ladder. Such coalescing binaries can be used as \emph{standard sirens} to measure the expansion and acceleration rates of the Universe as a function of redshift, thereby inferring fundamental properties of the dark sectors and of gravity itself.

\begin{tcolorbox}[standard jigsaw,colframe=ocre,colback=ocre!10!white,opacityback=0.6,coltext=black,title=\sc Key Science Goals]
Future GW observations will enable exploration of {\em particle physics}, \emph{early Universe}, and {\em cosmology}:
\begin{itemize}[leftmargin=*]
\item {\bf Standard Siren Cosmology.} What is the precise value of the Hubble constant? Is dark energy fully described by a cosmological constant, or does its equation-of-state vary with redshift?
\item {\bf Early Universe.} What particle physics laws and energy scales drove the Universe's early evolution? How did it transition from one evolutionary phase into another and to the present Universe?
\item {\bf Modified Theories of Gravity.} Do GWs propagate from their sources in the same way as EM waves do? How do modified theories of gravity affect the propagation of GWs from their sources?
\end{itemize}
\end{tcolorbox}

Gravity assembles structures in the Universe from the smallest scale of planets to the largest scale of galaxy clusters and the Universe itself. GW observations can, therefore, elucidate the Universe's evolution and its constituents, complementary to the EM, neutrino, and particle observations.  Indeed, GW observations forever changed the role of gravity in our exploration of the Universe. In particular, the multi-messenger nature of GW170817 \cite{Monitor:2017mdv,GBM:2017lvd} generated a treasure trove of data ushering in a new era in cosmology \cite{Abbott:2017xzu}. 

General relativity (GR) completely determines the time evolution of the amplitude and frequency of GWs generated by binaries of NSs and BHs. Matched filtering the GW data with the predicted GR waveform readily infers the luminosity distance to the binary's host galaxy~\cite{Schutz:1986gp}. Thus, coalescing binaries have been hailed as \emph{standard sirens}. Just like the traditional standard candles (e.g. type Ia supernovae) they provide a tool for measuring the dynamics of the Universe, but are not susceptible to the systematic biases of the cosmic distance ladder. Standard sirens can be used to make completely independent measurement of the Universe's expansion rate as a function of redshift and to infer cosmological parameters describing the dark sector.

The measured compact binary merger rates \cite{LIGOScientific:2018jsj, TheLIGOScientific:2017qsa} imply that future detectors will observe a stochastic background formed from the astrophysical population of binaries at cosmological distances \cite{TheLIGOScientific:2016htt, Abbott:2017xzg}. Such observations will reveal the history of star formation from a time when the Universe was still assembling its first stars and galaxies. Buried under this background could be stochastic signals of primordial origin  (see, e.g., \cite{Grishchuk:1974ny, Starobinsky:1979ty, Rubakov:1982df, Giovannini:1998bp,  Boyle:2007zx, Figueroa:2016dsc, Caprini:2018mtu, Kosowsky:1991ua, Kamionkowski:1993fg, Hindmarsh:2017gnf, Caprini:2009yp})
that provide insights into the physics and energy scales of the earliest evolutionary phases in the history of the Universe---scales not accessible to current or planned particle physics experiments.  The 3G network will probe those energy scales with its excellent sensitivity to stochastic backgrounds.

\section{Standard Siren Cosmology}
{\bf Hubble Constant:} GWs from a compact binary merger support a direct measurement of the source's luminosity distance. Combined with the redshift obtained from the EM counterpart, or with galaxy catalogs, this gives an estimate the present value of the Hubble parameter $H_0$~\cite{Schutz:1986gp, Abbott:2017xzu}. It is estimated that about ten mergers with EM counterparts would be required to reach an accuracy of 5\% and 200 to reach 1\% \cite{2010ApJ...725..496N, Chen:2017rfc, 2019PhRvL.122f1105F}.  While binary NS events are promising based on GW170817, NS-BH and BH-BH mergers due to precession of the orbital plane induced by spin-orbit coupling \cite{Vitale:2018wlg} or the presence of higher multipole modes in the observed waveform \cite{Graff:2015bba, Borhanian:2020vyr}, can break the degeneracy between the orbital inclination and luminosity distance, and provide more accurate distance measurements. EM observations could also break this degeneracy \cite{Hotokezaka:2018dfi}.   There is significant potential in statistical methods as well, where binary mergers without EM counterparts are combined with galaxy catalogs to make inferences \cite{DelPozzo:2011yh}. For example, 3G detectors could localize some sources within a volume where on average only one galaxy is present \cite{Vitale:2018nif, 2017ApJ...848L..16S, Borhanian:2020vyr}, although the method is limited by the peculiar velocity at the redshift of interest and the distance uncertainty $\sim 1\%$ from GW observations.

\begin{figure*}
\begin{tcolorbox}[standard jigsaw, colframe=gray, colback=gray!10!white, opacityback=0.6, coltext=black, title=\small\sc Measuring Cosmological Parameters and Testing Modified Gravity Theories]
\centering
\includegraphics[width=0.40\textwidth]{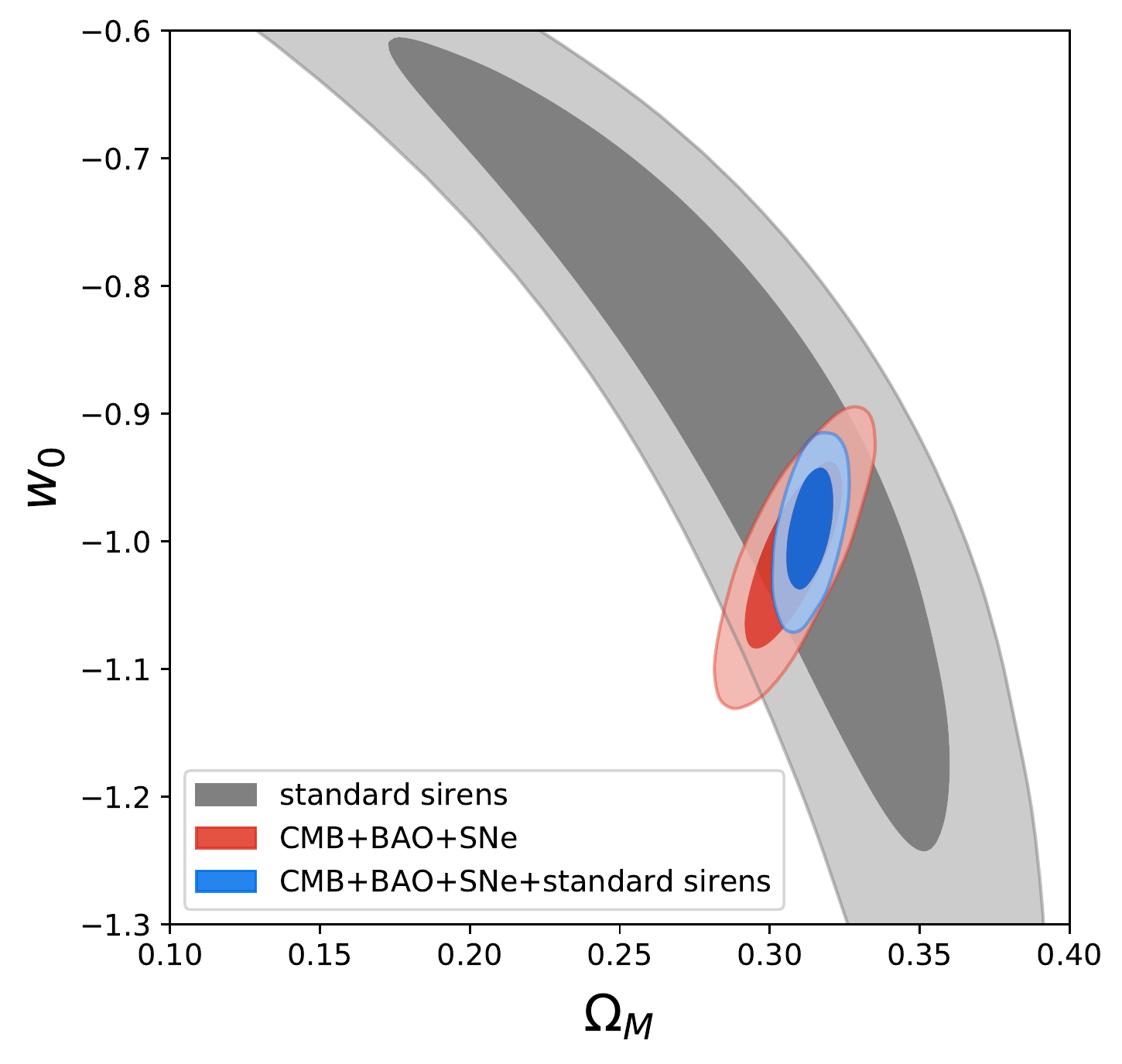}\hspace{10mm}
\includegraphics[width=0.40\textwidth]{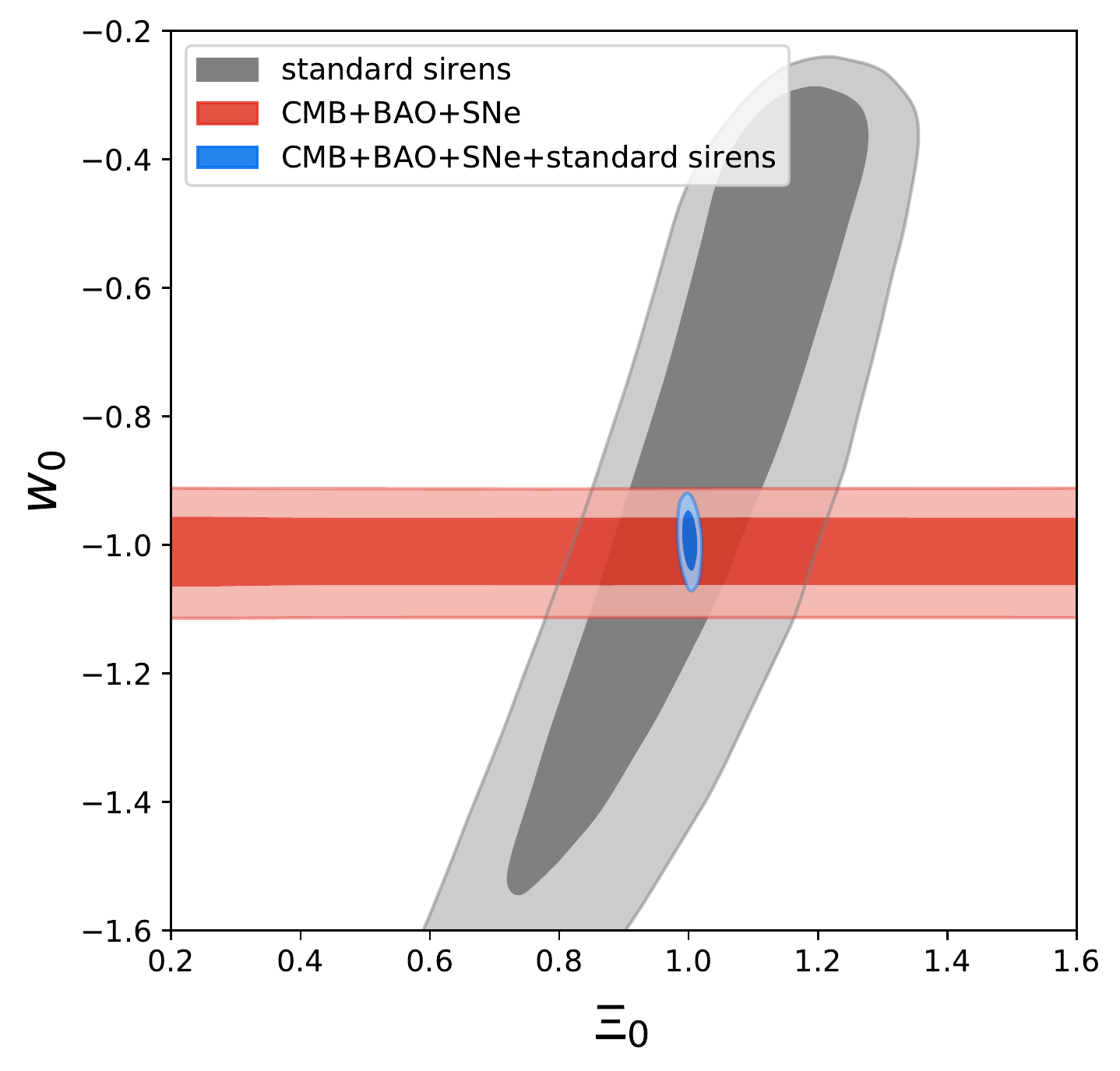}
\caption{\small Left:  $1\sigma$ and $2\sigma$
contours of the two-dimensional likelihood in the plane of the matter energy density $\Omega_\mathrm{M}$ and the dark energy equation of state parameter $w_0$, with the combined  contribution from {\em Planck}
CMB data + baryon acoustic oscillations (BAO) + supernovae (red), the contribution from  $10^3$ standard sirens at Einstein Telescope (gray), and the total combined result (blue). Right: two-dimensional likelihood in the $(\Xi_0,w_0)$ plane is shown, where $\Xi_0$ captures the deviation between the GW and EM luminosity distances.  The mock catalog of standard sirens is generated  assuming  $\Lambda$CDM as the fiducial model ($w_0=-1$ and $\Xi_0=1$) and the plot shows the accuracy in the reconstruction of $w_0$ and $\Xi_0$ (plot from \cite{Belgacem:2018lbp}).}
\label{fig:w0wa}
\end{tcolorbox}
    \vskip-0.5cm
\end{figure*}

\noindent {\bf Precision Cosmology:} GWs offer a new approach to probing the dark energy properties either by using redshift measurements from EM counterparts, or by using statistical methods \cite{Dalal:2006qt, Nissanke:2009kt, Sathyaprakash:2009xt, Zhao:2010sz, Taylor:2012db, Camera:2013xfa, Cai:2016sby, Belgacem:2017ihm, Belgacem:2018lbp}. The 3G network will measure standard sirens out to large redshifts, $z\, \sim 10$, significantly farther than what is possible with the standard candles. Being susceptible to a completely different set of systematic errors from those due to type Ia supernovae this approach will provide a complementary probe precision cosmology and of comparable sensitivity to those due to the cosmic microwave background, baryon acoustic oscillation, and supernovae. An example is shown in Figure \ref{fig:w0wa} (left) \cite{Belgacem:2017ihm, Belgacem:2018lbp}, where addition of standard sirens improves the accuracy of the measured dark energy equation of state parameter $w_0$. 

\noindent{\bf Modified Gravity Theories:} Standard sirens are sensitive to another powerful signature of the dark energy sector that is not accessible to EM observations. A generic modified gravity theory induces modifications, with respect to the standard model of cosmology, in the cosmological background evolution and perturbations. Indeed, theories with extra dimensions \cite{Deffayet:2007kf}, some scalar-tensor theories \cite{Saltas:2014dha, Lombriser:2015sxa, Arai:2017hxj, Amendola:2017ovw, Linder:2018jil}, as well as a nonlocal modification of gravity \cite{Maggiore:2013mea, Maggiore:2014sia, Belgacem:2017cqo, Belgacem:2017ihm, Belgacem:2018lbp}, are characterized by GWs propagating at the speed of light but with their amplitude decreasing differently with the scale factor than in GR. Consequently, the standard sirens would measure a different luminosity distance compared to their EM counterparts. The 3G network will be sufficiently sensitive to search for this deviation, and probe multiple classes of modified theories of gravity in the context of their dark energy content (Figure \ref{fig:w0wa}, right) \cite{Belgacem:2017ihm, Belgacem:2018lbp}.

\section{Early History of the Universe}
\begin{figure}[t]
\begin{tcolorbox}[standard jigsaw, colframe=gray, colback=gray!10!white, opacityback=0.6, coltext=black, title=\small\sc Stochastic Background Sources and Detector Sensitivities]
\centering
\includegraphics[width=1.00\textwidth]{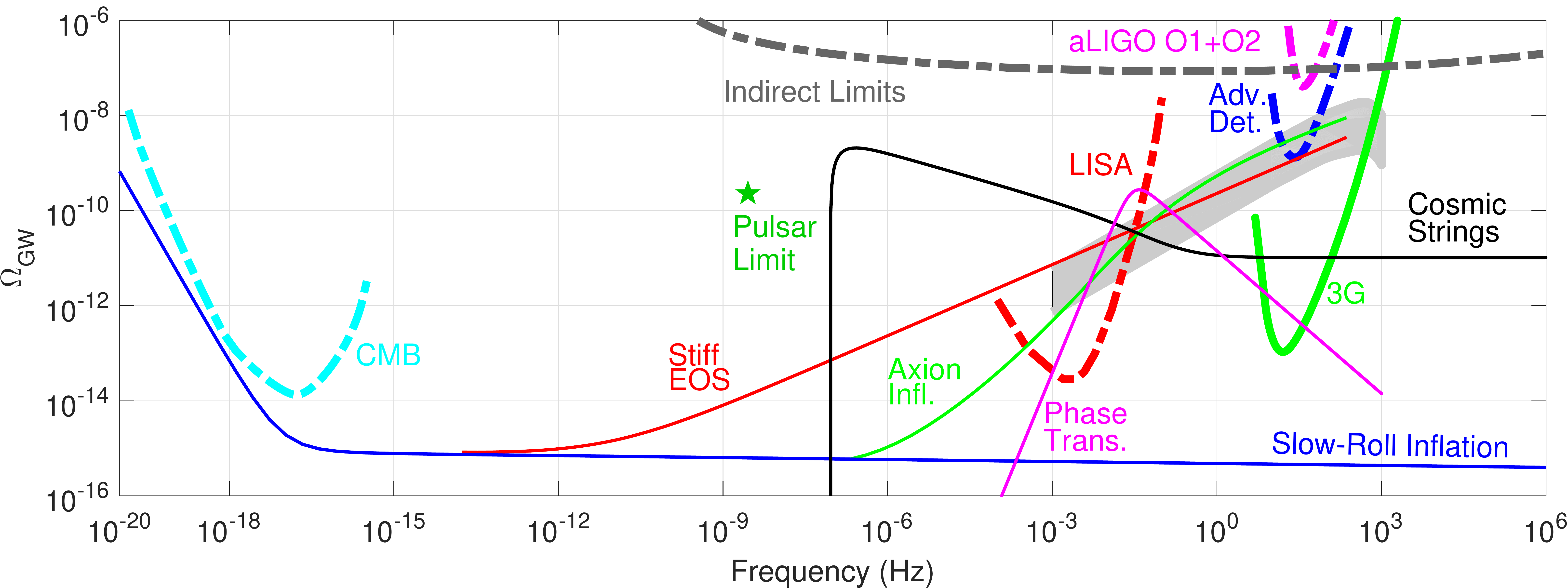}
\caption{\small Stochastic GW background for several proposed model spectra in comparison with past measurements (Advanced LIGO upper limit \cite{TheLIGOScientific:2016dpb}, constraints based on the big bang nucleosynthesis and cosmic microwave background (CMB) observations, low-$l$ CMB observations, and pulsar timing \cite{Lasky:2015lej}),  and future expected sensitivities \cite{Thrane:2013oya}, (the final sensitivity of Advanced LIGO \cite{TheLIGOScientific:2016wyq}, Cosmic Explorer \cite{Evans:2016mbw}, and LISA, all assuming 1 year of exposure \cite{LisaProposal:2017,Axen:2018zvb}). The gray band denotes the expected amplitude of the background due to the cosmic population of compact binary mergers, based on the observed coalescieng binary systems \cite{2018PhRvL.120i1101A}.} \label{fig:landscape}
 \end{tcolorbox}
\vskip-20pt
\end{figure}

Stochastic GW backgrounds could either be astrophysical in origin, generated by a myriad of individual sources, or it could be generated by quantum processes associated with inflation and spontaneous symmetry phase transitions breaking in the early Universe. Figure \ref{fig:landscape} shows examples of GW energy density spectra for some of the cosmological background models in comparison with the best current upper limits and future expected detector sensitivities. Cosmological background is arguably the most fundamentally impactful observation that GW observatories could make. The astrophysical background, however, may mask the primordial background over much of the accessible spectrum, while still carrying important information about the evolution of structure in the Universe. Techniques are being developed to identify and estimate these various contributions to the stochastic GW background \cite{Sachdev:2020bkk}.

\noindent {\bf Irreducible GW background from inflation:} Inflation represents the leading framework to explain the properties and initial conditions of the observed Universe. During inflation massless fields experience quantum fluctuations, and due to the accelerated expansion small fluctuations with wavelengths initially smaller than the Hubble radius are amplified and stretched to super-Hubble scales. This applies, in particular, to tensor perturbations \cite{Grishchuk:1974ny, Starobinsky:1979ty, Rubakov:1982df} 
that re-enter the Hubble radius after inflation and turn into a stochastic GW background. In the standard slow-roll inflationary model (shown in Figure \ref{fig:landscape}), the energy density spectrum of this background 
is likely below the proposed sensitivity of the 3G network, although upgrades to 3G detectors could be sufficiently sensitive to observe this background.

\noindent{\bf Beyond the irreducible background from inflation:} Additional processes during or immediately after inflation could lead to significant amplification of the stochastic background in the frequency band of 3G detectors. For example, coupling of axions to the scalar field of inflation could extend GW production with significant enhancement at higher frequencies \cite{Barnaby:2010vf, Cook:2011hg} as shown in Fig.\ \ref{fig:landscape}. Furthermore, the resulting background is distinctly chiral (i.e.\, only one polarization is excited) and non-Gaussian  \cite{Cook:2013xea}, unique predictions which  can be used to differentiate it unambiguously from other GW backgrounds.  Similar trends in the background spectrum are also possible if inflation is followed by a phase with a stiff equation of state ($w > 1/3$), that could be detectable with 3G detectors~\cite{Giovannini:1999bh, Giovannini:1998bp, Boyle:2007zx,  Figueroa:2016dsc}.


\noindent {\bf First order phase transitions:} Following the end of inflation, the Universe has undergone quantum chromodynamic, electroweak, and, possibly, other phase transitions.  Currently, an experimentally verified physical model lacks energy scales higher than the electroweak scale. Several proposed extensions of the Standard Model predict the occurrence of phase transitions. Any experimental confirmation that such phase transitions took place in the early Universe would constitute a step change in our understanding of particle theory at energy scales inaccessible to terrestrial experiments.

To be an efficient direct source of GWs, a phase transition must be of first order. First-order phase transitions proceed through the nucleation of bubbles of the, energetically more favourable, true vacuum in the space-filling false vacuum. The dynamics of the bubble expansion and collision is phenomenologically rich, and the sources of GWs are the tensor anisotropic stresses generated by these multiple phenomena: the bubble wall’s expansion~\cite{Kosowsky:1991ua,  Weir:2016tov}, the sound waves in the plasma~\cite{Hindmarsh:2017gnf}, and the subsequent magnetohydrodynamic turbulence~\cite{Kahniashvili:2008pe, Caprini:2009yp}. The nature of the phase transition and its energy scale determine the amplitude and the spectral shape of the GW background. An example of such a background is shown in Figure \ref{fig:landscape} which is potentially within reach of the 3G network \cite{Axen:2018zvb}.

\noindent {\bf Cosmic Strings:} Topological defects such as cosmic strings may arise in the aftermath of a phase transition \cite{Jeannerot:2003qv}. Often, the string tension is the only free parameter and it defines the energy scale of the phase transition and the accompanying spontaneous symmetry breaking scale that leads to the formation of cosmic strings. It is also possible to form a network of fundamental cosmic (super)strings. Cosmic strings predominantly decay by the formation of loops and the subsequent GW emission by cosmic string cusps and kinks \cite{Vachaspati:1984gt, Sakellariadou:1990ne}. Searches for individual bursts of GWs from cosmic strings and for the stochastic background from a string network have placed a strong constraint on the string tension for the three well-known models \cite{Blanco-Pillado:2017oxo, Ringeval:2017eww, Abbott:2017mem, Jenkins:2018nty}. The 3G network will either detect cosmic strings or improve on these bounds by eight orders of magnitude, depending on the model (see Fig.~\ref{fig:landscape}).

\noindent {\bf Dark Photons:} A dark photon is proposed to be a light but massive gauge boson in an extension of the Standard Model. If sufficiently light, the local occupation number of the dark photon could be much larger than one, so it can then be treated as a coherently oscillating background field that imposes an oscillating force on objects that carry dark charge. The oscillation frequency is determined by the mass of the dark photon. Such effects could result in a stochastic background that could be measured by 3G detectors, potentially exploring large parts of the parameter space of such models~\cite{Pierce:2018xmy}. 

\section{Astrophysical Binary Foregrounds and Large Scale Structure}

The cosmological population of compact binary mergers will give rise to a stochastic foreground of GWs \cite{Regimbau:2016ike, Mandic:2016lcn, Dvorkin:2016okx, Nakazato:2016nkj, Dvorkin:2016wac, Evangelista:2014oba}.  The amplitude of this foreground is estimated to be $\Omega_{\text{GW}} \sim 10^{-9}$ at 25 Hz, and is likely to be detected by Advanced LIGO and Advanced Virgo \cite{Abbott:2017xzg}. The 3G network, thanks to better low-frequency sensitivity,  will probe $\Omega_{\text{GW}}\sim10^{-13}$--$10^{-12}$~\cite{ETBook, Cutler:2005qq, Regimbau:2016ike}. The 3G network could, therefore, detect the predicted spatial anisotropy \cite{Cusin:2018rsq, Jenkins:2018uac} and non-gaussianity~\cite{Jenkins:2018uac} in the energy density of this foreground. In particular, measurements of higher order correlation functions would be extremely useful in understanding the large scale structure of the Universe.

Figure \ref{fig:landscape} shows that the amplitude of this foreground is several orders of magnitude stronger than most cosmological GW backgrounds, hence masking the cosmological signals discussed above. However, a large fraction of the binary merger signals will be individually detected by the 3G network, allowing for the possibility to subtract this foreground to probe a cosmic background of $\Omega_{\rm GW}\sim 10^{-13}$ \citep{Cutler:2005qq, Regimbau:2016ike, Sachdev:2020bkk}. Small errors in subtraction could lead to a substantial residual foreground---novel methods are being explored to enable more effective subtraction \cite{Smith:2017vfk}.

Astrophysical sources other than compact binary mergers could also contribute to the astrophysical GW foreground, including isolated NSs \cite{Surace:2015ppq, Talukder:2014eba, Lasky:2013jfa}, core collapse supernovae \cite{Crocker:2017agi, Crocker:2015taa} and population III binaries~\cite{Kowalska:2012ba}.  Distinguishing these unresolved foreground sources from the cosmological background will require using spatial correlations (both GW-GW and GW-EM correlations) and careful measurements of the foreground spectra. Studying the properties of the astrophysical foreground will provide unique new information on galactic and stellar physics and allow for new types of constraints on astrophysical models. 

\section{Outlook for gravitational-wave cosmology and the early Universe}

3G observatories will revolutionize our understanding of the evolution of the Universe, from its earliest moments to the recent past. Measurements of the frequency spectrum, spatial anisotropy, and polarization content of the primordial stochastic GW background are likely to reveal imprints of the physical laws and processes that drove the earliest phases of the Universe's evolution and take place at energy scales inaccessible to current or planned particle physics experiments.  Observation of the astrophysical, compact-binary stochastic GW background and its properties would provide information on the formation and evolution of matter in the Universe, going back to the time when the first stars and galaxies were formed. More recent compact binary mergers will be used as standard sirens to provide novel and independent measurements of the expansion and acceleration of the late Universe, constraining the fundamental nature of gravity and cosmological parameters. While the irreducible inflationary stochastic GW background is likely not within reach of the 3G network, it may be within reach of the follow-up upgrades, which should be factored into the site selection and facility design of the 3G detectors.
\vskip8mm

\begin{tcolorbox}[colback=teal!5!white,colframe=red!75!black,title=\sc Science Requirements]{
With 3G observatories we can investigate the fundamental physics of the primeval Universe and probe its dark sectors:
\begin{itemize}
    \item a factor of 10-20 improvement beyond the sensitivity of advanced detectors is critical to infer cosmological parameters at a precision that is competitive with current measurements,
    \item lowering the frequency response down to 5 Hz will allow the observation of astrophysical foregrounds and their subtraction via better estimation of source parameters, and
    \item a factor of 10 improvement in strain sensitivity (100 in energy density) over advanced detectors is required to observe GW background from early Universe phase transitions.
\end{itemize}

}
\end{tcolorbox}

%% file: extreme-gravity.tex
\chapterimage{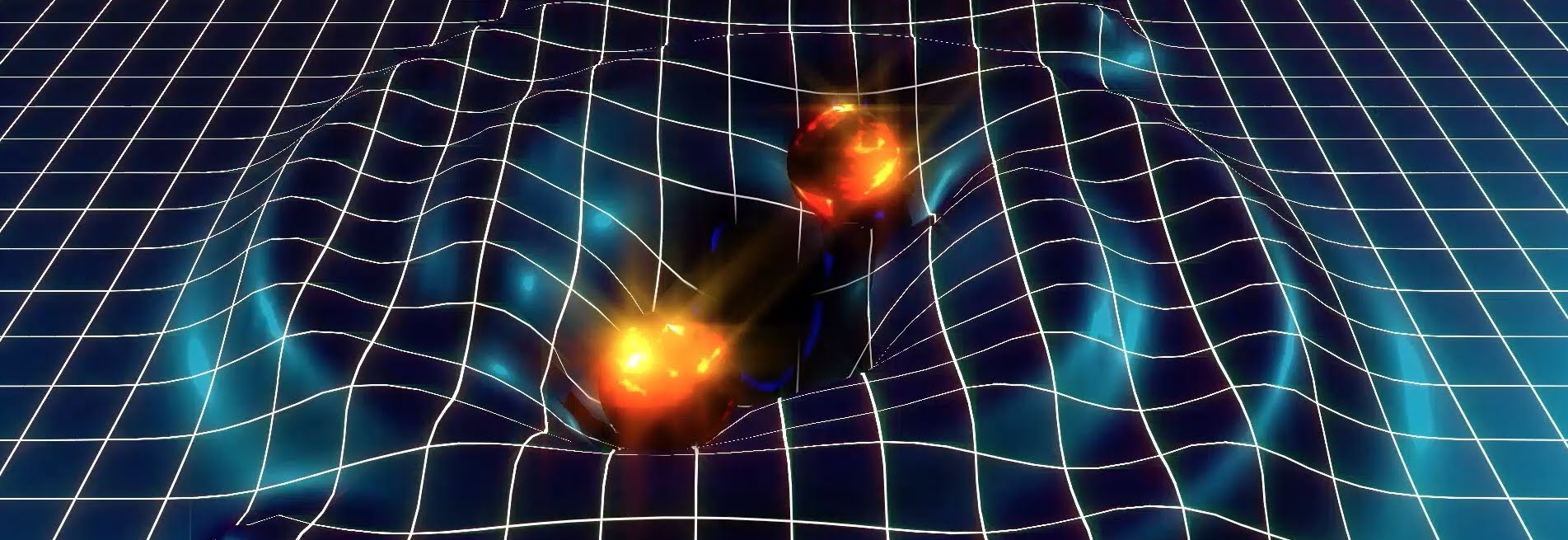} 
\chapter{Extreme Gravity and Fundamental Physics}

\begin{tcolorbox}[colback=teal!5!white,colframe=yellow!75!black,title=\sc Science Target]
{\em  Explore new physics in gravity and in the fundamental properties of compact objects}
\end{tcolorbox}

General Relativity (GR) is a mathematically elegant and physically appealing theory that has been tremendously successful in explaining all relevant astronomical observations and laboratory experiments\cite{Will:2005va, Psaltis:2008bb, Yunes:2013dva}. Nonetheless, it is widely believed that GR is at best incomplete \cite{Yunes:2013dva, Rham:2015mxa}, representing an approximation to a more complete theory that cures some or all of its deep conceptual problems. Black hole (BH) information loss, spacetime singularities,  cosmological constant and the lack of a viable formulation of quantum gravity, have all added to the suspicion that GR violations will eventually show up in observations \cite{Chapline:2000en, Baker:2014zba, Rham:2015mxa, Ishak:2018his}. Over the past decade new insights into the relationship between entanglement entropy and the architecture of spacetime \cite{Bianchi:2012ev} on the one hand and the connection between asymptotic symmetries, the BH entropy and infrared behaviour of quantum gravity \cite{Hawking:2016msc} on the other are all hinting towards a modified theory of gravity. At the same time, the discovery of gravitational waves (GWs) has provided a powerful new tool to test GR in a realm of the theory that is inaccessible to other experiments and observations. 

GW150914 was not only the first direct detection of GWs but also the first ever observation of a binary black hole (BBH) \cite{Abbott:2016blz}.  Since then tens of BBH mergers have been observed \cite{LIGOScientific:2018mvr}, GW190521 being the most massive system discovered so far \cite{Abbott:2020tfl} that converted 9 solar masses to pure energy in a mere 100 ms. BBH mergers are arguably the most powerful phenomena in nature, save for the Big Bang, the signals coming from a region where both the spacetime curvature and gravitational field are extremely large.  They have helped test Einstein's gravity in regimes where the theory has never been tested before \cite{LIGOScientific:2019fpa}. 3G observatories will make a step change in studying gravity and the nature of ultra-compact objects.   
\begin{tcolorbox}[standard jigsaw,colframe=ocre,colback=ocre!10!white,opacityback=0.6,coltext=black,title=\sc Key Science Goals]
The 3G GW observatories will enable unprecedented and unique science in \emph{extreme gravity} and \emph{fundamental physics}:
\vskip5pt
\begin{itemize}[leftmargin=*]
\item {\bf The nature of gravity:} Are the building-block principles and symmetries in nature, e.g. Lorentz invariance and equivalence principle, invoked in the description of gravity valid at all scales?
\item {\bf The nature of compact objects:} Are black holes and neutron stars the only ultra-compact objects in the Universe? If other compact objects exist what are their signatures in gravitational waves?
\item {\bf The nature of dark matter:} Is dark matter composed of particles or dark objects or is it a manifestation of failure of general relativity? 
\end{itemize}
\end{tcolorbox}

GWs are copiously produced in regions of strong gravity and relativistic motion. Yet the waves carry pristine information about their sources because they interact feebly with matter and remain unscathed as they propagate over billions of light years to Earth.  This makes them ideal for testing GR in new ways that could reveal subtle departures from the theory \cite{thorne.k:1987, Sathyaprakash:2009xs, Yunes:2016jcc, Berti:2018vdi}.  We can now directly probe the two-body dynamics (specifically, the complex orbital motion predicted in GR such as precessional effects), the spacetime structure near BHs and the dynamics of their horizons, and if the Universe depicted by electromagnetic (EM) and particle messengers is the same as the one revealed by GWs. In addition, imprint in the observed GWs are the signatures of compact objects that can be deformed in a way that BHs cannot be. This should help us detect a new class of ultra-compact compact objects consisting of boson condensates or other exotic particles and fields should they exist. Finally, the presence of dark matter around compact binaries or their accumulation in neutron star (NS) cores, will also modify the observed signals. Hence, GWs could be used to measure the properties of dark matter particles---their masses and interaction cross sections with hadronic matter.
\vskip10pt

\noindent
\begin{minipage}{0.48\textwidth}
\section{Nature of Gravity.}
Probing the nature of gravity and its possible implications on fundamental physics is a high-reward, even if uncertain, prospect of GW observations.  
To our knowledge, astrophysical BHs and relativistic stars exhibit the largest curvature of spacetime accessible to us. They are, therefore, ideal systems to observe the behaviour of spacetimes under the most extreme gravitational conditions.   New physics indicative of departures from the basic tenants of GR could reveal itself in high fidelity waveforms expected to be observed in the 3G network.  Such signals would provide a unique access to extremely warped spacetimes and gain invaluable insights into GR or what might replace it as the theory of gravity. 

\hskip0.5cm Figure \ref{fig:gravitytests}
provides a perspective of the reach of different missions/facilities
and their target regime with respect to characteristic spacetime
curvature ($R$) and gravitational potential $\Phi$ (which for binary
systems can be equated with $v^2/c^2$, where $v$ is the 
characteristic velocity of companion stars and $c$ the speed of light).
\end{minipage}
\hfill
\begin{minipage}{0.50\textwidth}
\begin{tcolorbox}[standard jigsaw,colframe=gray,colback=gray!10!white,opacityback=0.6,coltext=black, title=\small\sc Probing Gravity at all Scales]
\hskip-10pt\includegraphics[width=1.10\textwidth]{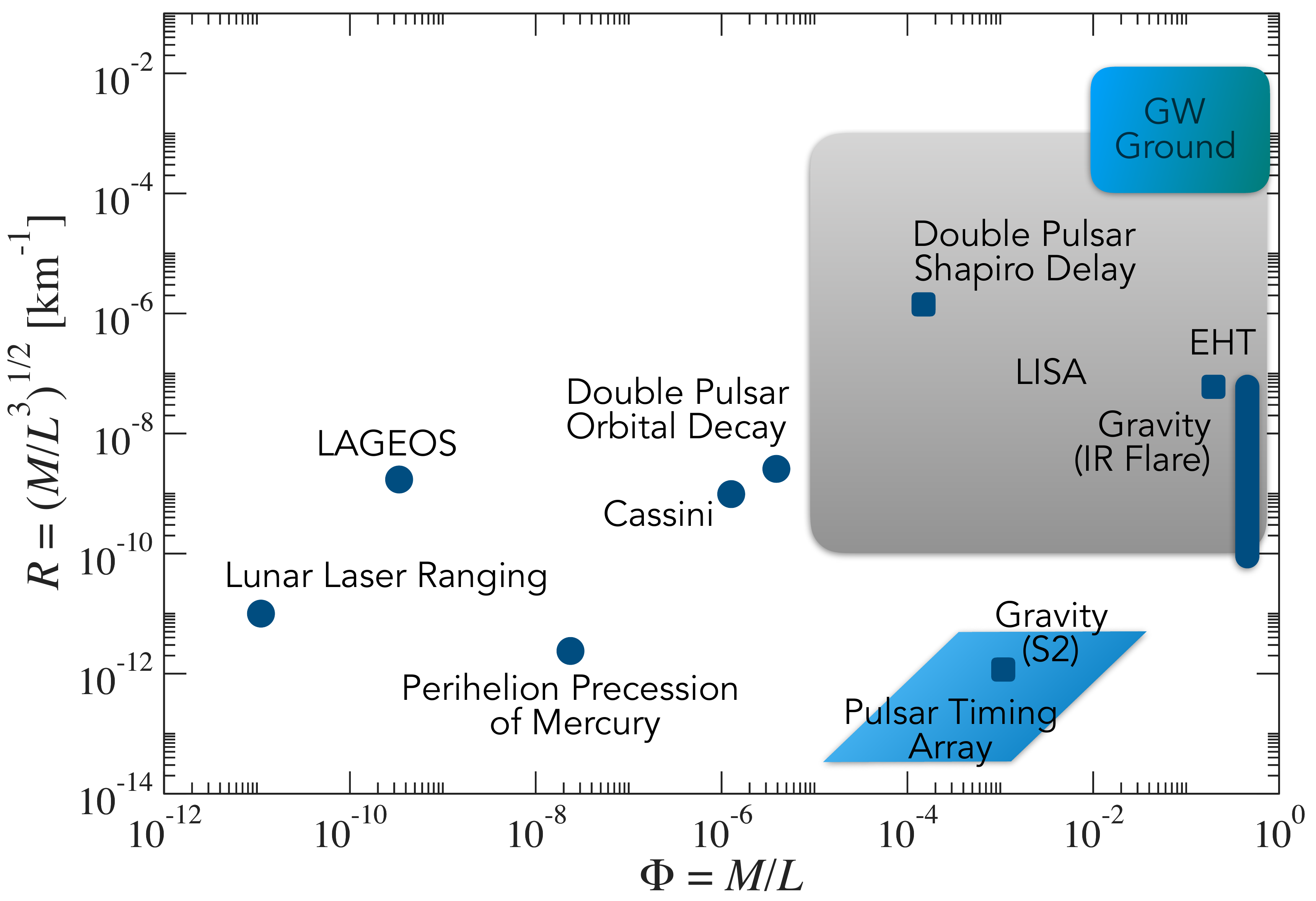}
\captionof{figure}{\small Illustration of the reach in curvature scales vs potential
scales targeted by different, representative, past/current/future missions. In this figure, $M$ and $L$ are the
characteristic mass and length involved in the observable associated to each mission. For instance, in
observables associated to binary systems $M$ is the total mass and $L$ the binary's separation, in
this case $M/L$ is related to  $v^2/c^2$ through the virial theorem.}
\label{fig:gravitytests}
\end{tcolorbox}
\end{minipage}
\vskip5pt

To this end, beyond having access to the sensitivity and frequency windows of 3G detectors, guidance
from theory, together with further refinements in data analysis, will be of utmost importance to harness
this potentially revolutionary opportunity.
On the theoretical front a major challenge in extracting new physics with GWs is that, in principle, one needs to model the characteristics of the emitted signal for the desired physical scenarios beyond the framework of GR and then confront it with the data~\cite{Will:2014kxa,Berti:2015itd}. The powerful perspective of effective field theory \cite{Donoghue:1995cz, Burgess:2007pt, Weinberg:2008hq, Goldberger:2004jt}) allows one to build extensions to GR with higher-order corrections and search for new physics,
even before a new fundamental theory and its low-energy phenomenology is fully developed (see, e.g., Refs.~\cite{Sennett:2017lcx, Endlich:2017tqa, Cardoso:2018ptl}). Among possible departures under scrutiny are the following:

\noindent{\bf New fields, particles and polarizations:} 
GR has not only passed every experimental and observational test that it has been subject to but it is also a very robust theory. In fact,  any physically meaningful departures from GR necessarily require the presence of extra degrees of freedom, e.g. scalar and vector fields, in addition to the metric tensor \cite{Lovelock:1971yv}. Such additional degrees of freedom also generically arise in the low-energy limit of quantum gravity theories, which often lead to violations of the strong equivalence principle. Among possible theories, those with an additional scalar field, e.g. the Brans-Dicke theory, are relatively simple~\cite{Brans:1961sx, Fujii:2003pa}.  Yet they could give rise to exciting new phenomenology in strong gravitational fields indicative of failure of GR \cite{Palenzuela:2013hsa, Shibata:2013pra}. Such theories, therefore, serve as excellent proxies of the type of new physics we can hope to discover. Additionally, if a binary’s companions can become dressed with a scalar configuration \cite{Damour:1993hw, Kanti:1995vq, Mignemi:1992nt, Antoniou:2017acq}, we can expect the emission of scalar GWs, in addition to tensorial ones, with the dominant component being dipolar emission \cite{Will:2014kxa}. Such emissions lead to additional polarizations, beyond the two in GR, that could be detected directly with a network of detectors \cite{TheLIGOScientific:2016src, Abbott:2018lct} or indirectly inferred from their effects on the source’s dynamics and consequent impact on the observed GWs~\cite{Will:2014kxa}.

\noindent{\bf Graviton mass and speed of GWs:} Recently, the possibility that gravitons could have mass has resurfaced in theoretical physics within extensions of GR~\cite{deRham:2010kj,Hassan:2011hr}. In a massive graviton theory GWs would be dispersed as they propagate from their sources to Earth, causing a change in the phase evolution of the observed signal relative to GR.  The current best bound on the graviton mass $m_g$ comes from modifying the GR dispersion relation for GWs and this sets the bound $m_g < 5.0 \times 10^{-23}\,{\rm eV}/c^2$ ~\cite{LIGOScientific:2019fpa}. The 3G network could improve this bound by two orders-of-magnitude by detecting sources from as far as $z \sim 50$ and further constrain massive graviton theories. 


\noindent{\bf Lorentz invariance and parity violations:} 
Lorentz invariance is regarded as a fundamental property of the Standard Model of particle physics, tested to a spectacular accuracy in particle experiments~\cite{Mattingly:2005re}. In the gravitational sector, constraints are far less refined. Theories with Lorentz invariance violation (e.g., Ho\v{r}ava--Lifschitz~\cite{Horava:2009uw} and Einstein-\ae{}ther~\cite{Jacobson:2000xp}) give rise to significant departures from GR on the properties of BHs \cite{Eling:2006ec, Barausse:2011pu}, existence of additional polarizations \cite{Sotiriou:2017obf}, and the propagation of GWs through dispersion and birefringence~\cite{Kostelecky:2016kfm,Abbott:2017vtc}. Furthermore, parity violations in gravity arise naturally within some flavors of string theory~\cite{Green:1987mn}, loop quantum gravity \cite{Ashtekar:1988sw} and inflationary models~\cite{Weinberg:2008hq}. The associated phenomenologies are, to some degree, understood from effective theories~\cite{Jackiw:2003pm}. For instance, they give rise to BHs with nontrivial pseudo-scalar configurations that violate spatial parity~\cite{Yunes:2009hc}. The resulting scalar dipole leads to a correction to the GWs produced in a binary inspiral and merger signal~\cite{Sopuerta:2009iy, Yagi:2012vf, Okounkova:2017yby}. Additionally, parity violating theories can exhibit birefringence, thus impacting the characteristics of GWs tied to their handedness~\cite{Yagi:2017zhb}.   All of these effects will be greatly constrained by the 3G network as it will observe sources at  redshifts of $z\sim 50$ and higher \cite{Yagi:2017zhb} and will have the ability to measure additional polarizations.


\noindent{\bf Ultra-light Bosonic Clouds:} 
Ultralight bosons have been proposed in various extensions of the Standard Model~\cite{Essig:2013lka}. When the Compton wavelength of ultralight bosons (masses in the range $10^{-21}$ eV--$10^{-11}$\,eV) is comparable to the horizon size of a spinning stellar-mass or supermassive BH, superradiance can cause BH spin to decay, populating bound Bohr orbits around the BH, with an exponentially large number of particles \cite{Arvanitaki:2009fg, Pani:2012vp, Brito:2013wya}. Such bound states, in effect gravitational atoms, have bosonic clouds with masses up to $\sim$ 10\% of the BH mass~\cite{Arvanitaki:2010sy, Brito:2014wla, East:2017ovw}. Once formed, the clouds annihilate over a longer timescale through the emission of coherent, nearly-monochromatic, GWs~\cite{Arvanitaki:2010sy, Arvanitaki:2014wva}.

Presence of such clouds could be detected via blind searches in the Milky Way \cite{Arvanitaki:2014wva, Arvanitaki:2016qwi, Baryakhtar:2017ngi, Brito:2017zvb, Brito:2017wnc} or directed searches aimed at a candidate BH, such as that formed in a merger event~\cite{Arvanitaki:2014wva, Arvanitaki:2016qwi, Yoshino:2013ofa, Baryakhtar:2017ngi, East:2017mrj},  or observations of a stochastic background from an unresolved population~\cite{Brito:2017zvb, Brito:2017wnc}. Annihilation of such clouds can also impact a binary's dynamics and thus the GWs produced during the inspiral~\cite{Baumann:2018vus}. Such a modified GW signal is a promising target for 3G detectors with a few to hundreds of events per year expected for bosons in the $\sim10^{-13}$--$10^{-12}$ eV range \cite{Arvanitaki:2016qwi}. Measuring the spin and mass distribution of merging BBHs can provide evidence for characteristic BH spindown from superradiance \cite{Arvanitaki:2016qwi, Baryakhtar:2017ngi}, which would allow exploration of a new parameter space for ultralight bosons \cite{Vitale:2016icu}.  In addition, the presence of such clouds can be probed through the imprint of finite-size effects on the compact objects in a binary system~\cite{Baumann:2018vus}. Some dark-matter candidates alternative to weakly interacting massive particles (e.g., fuzzy dark matter~\cite{Hui:2016ltb}, axion-like particles, and other ultralight bosons \cite{Essig:2013lka}) predict exotic compact objects (or ECOs) either in the form of boson stars or in the form of condensates that form spontaneously due to BH superradiant instabilities.

\noindent{\bf Large, non-local, quantum effects:}  Semi-classical arguments have been put forward to support the possibility of exotic states of matter or dressed compact objects with further structure stemming from quantum gravitational origin.  Examples of electromagnetically dark but horizonless compact objects include fuzzballs~\cite{Mathur:2005zp,Mathur:2008nj}, gravastars \cite{Mazur:2004fk}, dark stars \cite{Barcelo:2007yk, Carballo-Rubio:2017tlh}, and others~\cite{Danielsson:2017riq, Berthiere:2017tms}.  Additionally, new non-local physics at the horizon scale has been suggested by firewall arguments~\cite{Almheiri:2012rt} as well as other quantum effects~\cite{Giddings:2013kcj, Giddings:2014nla, Giddings:2017mym, Cardoso:2017cqb}. These scenarios can generically give rise to signatures that can potentially be detected by the 3G network. Thus, GWs facilitate a unique window to these arguably speculative ideas, with far reaching consequences if observed. 

\noindent
\begin{minipage}{0.45\textwidth}
\section{Nature of Compact Objects.}
Observational evidence so far suggests that compact massive objects in the Universe exist in the form of BHs and NSs. Binary systems composed of such objects provide ideal scenarios to unravel both astrophysical and fundamental physics puzzles such as elucidating the connections of strong gravity with the most energetic phenomena in our Universe, exploring the final state conjecture \cite{1969NCimR...1..252P} (namely, the end point of gravitational collapse is a Kerr BH), and probing the existence of horizons.

\noindent{\bf Nature of black holes:}  Kerr BHs are characterized by just two parameters, their mass and spin angular momentum. This  remarkable property implies that the oscillations of a perturbed BH are rather unique.  Indeed, a perturbed BH returns to its quiescent state by losing the energy in its deformation into GWs. The emitted waves consist of a spectrum of damped sinusoids called {\em quasi-normal modes} whose frequencies and decay times are determined by the BH's mass and spin. 
\end{minipage}
\hfill
\begin{minipage}{0.52\textwidth}
\begin{tcolorbox}[standard jigsaw,colframe=gray,colback=gray!10!white,opacityback=0.6,coltext=black, title=\small\sc Test of Black Hole Nature]
\vskip-5pt
{\hskip-10pt \includegraphics[width=1.10\textwidth]{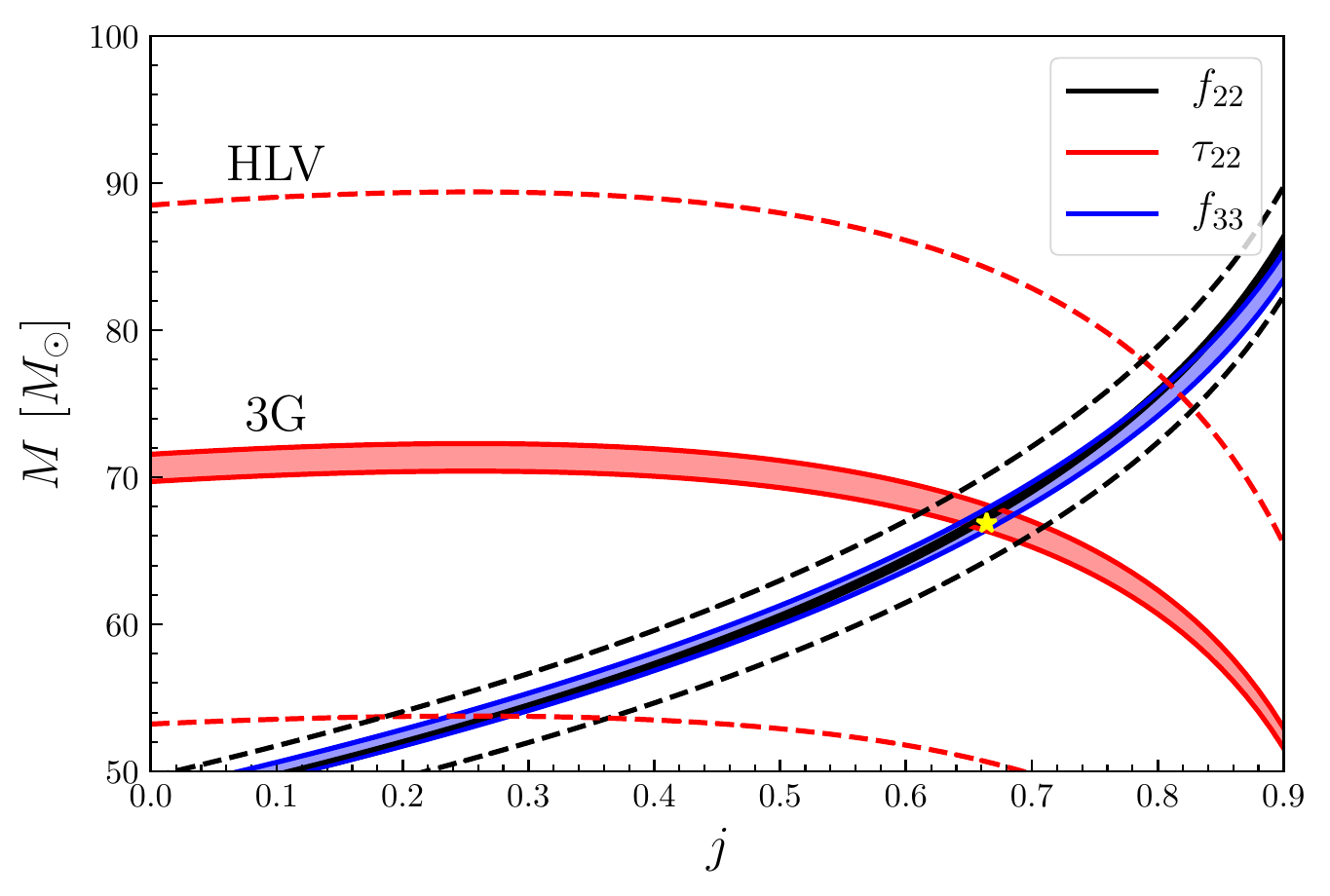}}
\vskip-5pt
\captionof{figure}{\small Projections of the $95\%$ confidence intervals for the fundamental quasi-normal mode frequency $f_{22},$ the corresponding damping time $\tau_{22}$ and a sub-dominant mode $f_{33})$, as function of the mass ($M$) and dimensionless spin ($j\equiv J/M^2$) of the final compact object for a \emph{single} GW150914-like event with advanced detectors at design sensitivity (HLV, dashed lines) and the 3G network (solid lines).
The yellow star corresponds to the true value of the BH's mass and spin used in the simulation. 
}
\label{fig:BH-nature}
\end{tcolorbox}
\end{minipage}

These parameters of the BH can be inferred from the measurement of a {\em single} mode frequency and its decay time.  Detection of several modes would facilitate multiple null tests of the Kerr nature of BHs~\cite{Dreyer:2003bv, Gossan:2011ha, Meidam:2014jpa, Berti:2016lat, Bhagwat:2017tkm, Yang:2017zxs, Brito:2018rfr, Carullo:2018sfu}. Such tests look for consistency in the masses and spins inferred from the different modes (see Fig.~\ref{fig:BH-nature}). The sensitivity of a 3G network is necessary for the precision with which such consistency tests can be carried out. Multiple loud events expected to be detected by the 3G network will provide exquisite tests of the Kerr nature of compact objects.

\noindent{\bf Beyond black holes:} 
From a phenomenological standpoint, BHs and NSs are just two species of a larger family of compact objects. More exotic species are theoretically predicted in extensions to GR, but also in particular scenarios within GR~\cite{Cardoso:2017cqb, Barack:2018yly}.  For instance, extremely compact objects (ECOs) arise from beyond-standard model fundamental fields minimally coupled to gravity (e.g., boson stars ~\cite{Liebling:2012fv}), in Grand Unified Theories in the early Universe (e.g., cosmic strings~\cite{Jeannerot:2003qv}), from exotic states of matter, as dressed compact objects with further structure stemming from quantum gravitational origin \cite{Giddings:2013kcj, Giddings:2017mym} or new physics at the horizon scale (e.g., firewalls~\cite{Almheiri:2012rt}), or as horizonless compact objects in a variety of scenarios, for example, fuzzballs, gravastars, and dark stars~\cite{Mathur:2005zp, Mazur:2004fk, Barcelo:2007yk, Carballo-Rubio:2017tlh, Danielsson:2017riq, Berthiere:2017tms}.
GW observations provide a unique discovery opportunity in this context, since exotic matter might not interact electromagnetically or any EM signal from the surface of an ECO might be highly redshifted~\cite{Cardoso:2017cqb}. Example GW signatures from the inspiral epoch include dipole radiation as well as a variety of matter effects as in the case of NSs \cite{Barack:2018yly}.


An ECO could be parameterized by the  gravitational redshift $z_g$
near its surface. This parameter can change by several orders of magnitude depending on the model. BHs have $z_g \rightarrow \infty$ while NSs and the most compact theoretically constructed  boson stars have $z_g \sim {\cal O}(1)$. For sufficiently large values of $z_g$ compact objects could behave like BHs with increasing precision. Studies of the geodesic motion and quasi-normal modes indicate that ECOs with $z_g \lesssim 1.4$ display internal structure effects that can be discerned in future GW observations. For larger values of $z_g,$ ECOs mimic BHs \cite{Abramowicz:2002vt, Cardoso:2017cqb, Cardoso:2019rvt} as departures are redshifted to ever smaller values. Interestingly, models of near-horizon quantum structures---motivated by various scenarios~\cite{Mathur:2005zp, Mathur:2008nj, Mazur:2004fk, Almheiri:2012rt}---can reach redshifts as high as $z_g \sim {\cal O}(10^{20})$ for ECOs in the frequency band of ground-based detectors. GWs could be our only hope to detect or rule them out.

Additionally, while the ringdown signal that follows from the merger of a compact binary can be qualitatively similar to that of a BH, quasi-normal modes of, e.g., gravastars, axion stars and boson stars, are different from Kerr BHs~\cite{Berti:2018vdi}.  In addition to gravitational modes, matter modes might be excited in the ringdown of an ECO, akin to the fluid modes excited in a remnant NS \cite{Barack:2018yly}. In the case of certain BH mimickers the prompt ringdown signal is identical to that of a BH. However, these objects generically support quasi-bound trapped modes which produce a modulated train of pulses at late time. These modes appear after a delay time whose characteristics are key to test Planckian corrections at the horizon scale. The 3G network will have unprecedented ability extract all these modes and explore the existence of ECOs~\cite{Cardoso:2017cqb}.

\section{Nature of Dark Matter.}
\label{sec:NatureDarkMatter}

The exquisite ability of 3G detectors to probe the population and dynamics of electromagnetically dark objects  throughout the Universe and harness deep insights on gravity can help reveal the nature of dark matter and answer key questions about its origin.

\begin{figure*}
\begin{tcolorbox}[standard jigsaw,colframe=gray,colback=gray!10!white,opacityback=0.6,coltext=black, title=\small\sc Primordial Black Holes as Dark Matter]
\begin{minipage}{0.48\textwidth}
\centering
\includegraphics[width=1.0\textwidth]{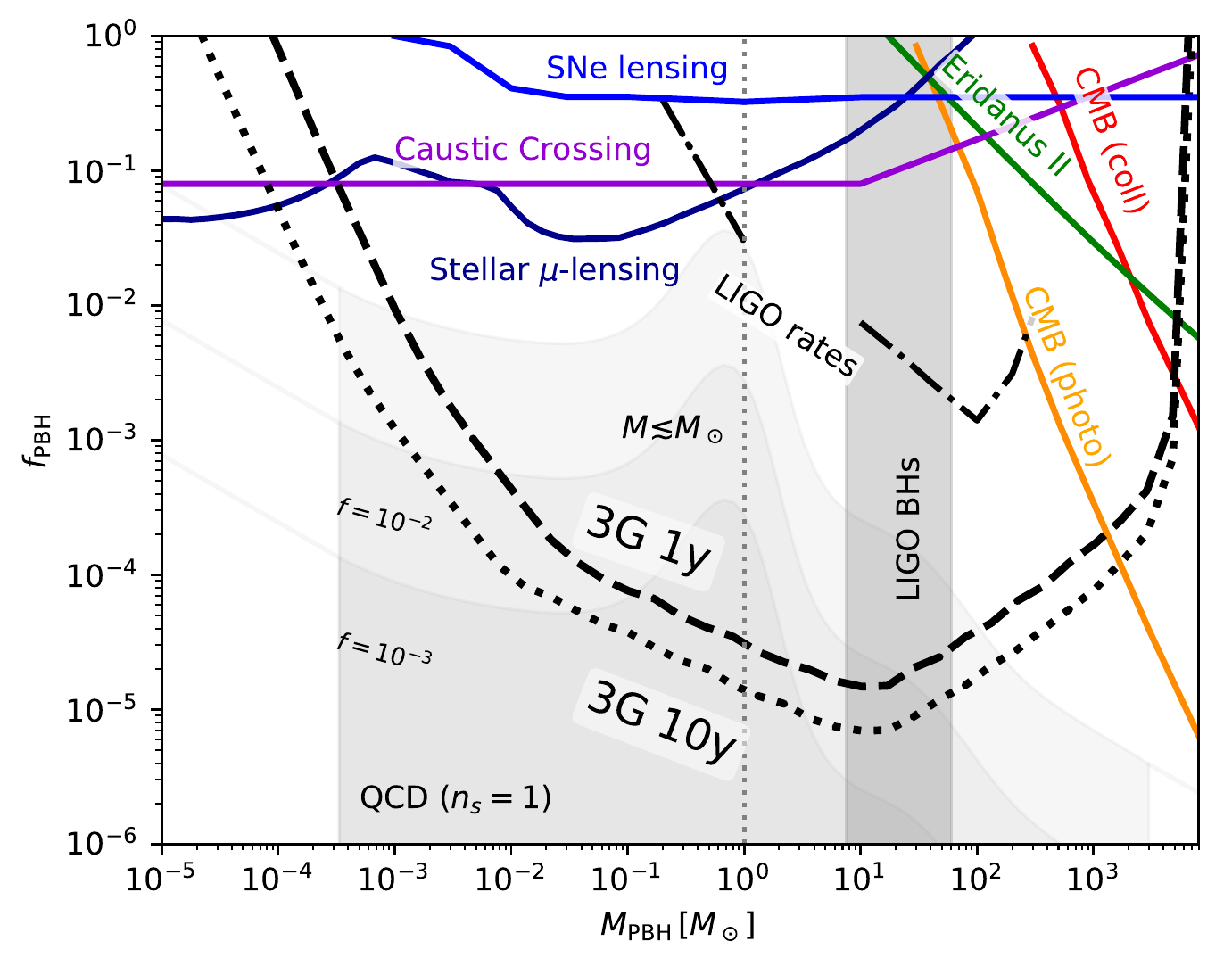}
\end{minipage}
\hfill
\begin{minipage}{0.48\textwidth}
\captionof{figure}{\small{\bf PBHs as dark matter:} The plot shows upper limits based on the observation of microlensing of stars \cite{Tisserand:2006zx, 2011MNRAS.413..493W}, microlensing of supernovae (SNe) \cite{Zumalacarregui:2017qqd}, caustic disruption \cite{Diego:2017drh,Venumadhav:2017pps,Oguri:2017ock}, distortions in the cosmic microwave background by Planck \cite{Ali-Haimoud:2016mbv,Bernal:2017vvn} and the stability of Eridanus II dwarf galaxy \cite{Brandt:2016aco,Li:2016utv}. GW constraints correspond to LIGO searches for sub-solar compact objects \cite{Abbott:2018oah} and stellar-mass mergers \cite{Ali-Haimoud:2017rtz}. The shaded regions correspond to the mass function imprinted by the QCD phase transition for different PBH abundances and a primordial scale invariant spectrum \cite{Byrnes:2018clq}. 3G detectors can improve the current bounds by many orders of magnitude.
}
\label{fig:DM}
\end{minipage}
\end{tcolorbox}
\vskip-10pt
\end{figure*}

\noindent{\bf Black holes as dark matter candidates:} 
LIGO and Virgo discoveries have revived interest in the possibility that dark matter could be composed, in part, of primordial black holes (PBHs) of masses $\sim 0.1$–$100\,M_\odot$ ~\cite{Clesse:2016vqa, Bird:2016dcv, Sasaki:2016jop}. Such BHs might have formed from the collapse of large primordial density fluctuations in the very early Universe or during inflation \cite{Carr:1974nx, Sasaki:2018dmp}. The exact distribution of their masses and spins depends on the model of inflation, which might be further affected by processes in the early Universe such as the quantum chromodynamics phase transition (QCD)~\cite{Byrnes:2018clq, Bianchi:2012ev}. Detection of binaries composed of objects much lighter than stellar-mass BHs or mass and spin distributions showing an excess in a certain range, could point towards the existence of PBHs \cite{Abbott:2018oah}. Identifying mergers at redshifts $z>30,$ when first stars were yet to form, would be another hint towards this formation channel \cite{Koushiappas:2017kqm}.  With the sensitivity to observe mergers at redshifts as large as $z\sim 50,$ the 3G network will be uniquely positioned to determine the mass, spin and redshift distributions of BHs, which will be crucial to test the hypothesis that dark matter consists of PBHs \cite{Kovetz:2016kpi}. Figure\,\ref{fig:DM} shows the current bounds and discovery potential of 3G detectors.

\noindent{\bf Detection of dark matter with compact objects:} Apart from probing whether dark matter can be partially made up of PBHs, GWs can also scrutinize models where dark matter consists of particles beyond the Standard Model, e.g., weakly interacting massive particles~\cite{Steigman:1984ac}, fuzzy dark matter \cite{Hui:2016ltb} or axion-like particles \cite{Essig:2013lka}. Indeed, BBHs evolving in a dark-matter rich environment will not only accrete the surrounding material, but also exert a gravitational drag on the dark matter medium, which affects their orbital dynamics \cite{Eda:2013gg, Macedo:2013qea, Barausse:2014tra}. Even though their magnitude is small, drag and accretion could have a cumulative effect over a large number of orbits that could be detected by a combination of observatories in space and 3G detectors~\cite{Barack:2018yly}.

Additionally, dark matter that interacts with standard model particles can scatter, lose energy, and be captured in astrophysical objects \cite{Press:1985ug, Gould:1989gw, Goldman:1989nd, Bertone:2007ae}. The dark matter eventually thermalizes with the star, and accumulates inside a finite-size core. The presence of dark matter would change the core's equation of state imprinting its signature into GWs emitted during the inspiral and merger of such objects in a binary system~\cite{Ellis:2017jgp}. In certain models, asymmetric dark matter can accumulate and collapse to a BH in the dense interiors of NSs. The core can grow by accumulating the remaining NS material, in effect turning NSs into light BHs in regions of high dark-matter density such as galactic centers \cite{Bramante:2017ulk, Kouvaris:2018wnh}. This provides a mechanism for creating light BHs that could be observed by 3G detectors. However, BHs that result from implosion of dark matter accreting NSs will always be heavier than $\sim 1\,M_\odot,$ any BH candidates of mass $\lesssim 1\,M_\odot$ could only be primordial in origin.

\section{Outlook on Exploring Extreme Gravity and Fundamental Physics}
Einstein's description of gravity led to a revolution in our thinking of the very nature of spacetime itself. Gravity is the manifestation of the curvature of spacetime caused by matter and energy density. GR has so far passed every test to which it has been subject to, yet some of its predictions are deeply troubling. The physical singularity at the Big Bang, loss of information when matter and energy fall into a BH are but examples of predicaments faced by the theory for which no satisfactory resolutions exist.  Moreover, observations are hinting that our knowledge of the constituents of the Universe is underwhelmingly poor and breakthroughs in the detection of new particles and fields is keenly awaited. 
\vskip8mm

\begin{tcolorbox}[colback=teal!5!white,colframe=red!75!black,title=\sc Science Requirements]{
GW observations by the 3G network could provide answers to some of the most fundamental questions about spacetime and matter. The capabilities required to answer them are:
\begin{itemize}
    \item high fidelity observations (SNR>1000) of a large number of sources to discover rare events that carry unique signatures of new particles and fields, dark matter and violations of GR,
    \item a network of at least one ET and two CEs to constrain or detect additional polarizations,
    \item access to frequencies below the sensitivity of current detectors to track GWs over a much longer period to test GR predictions and to detect dipole radiation, and 
    \item ability to identify black holes at large redshifts $z\ge 50$ to detect birefringence, improve bounds on graviton mass and discover primordial black holes.
\end{itemize}

}
\end{tcolorbox}

%% file: explosions.tex
\chapterimage{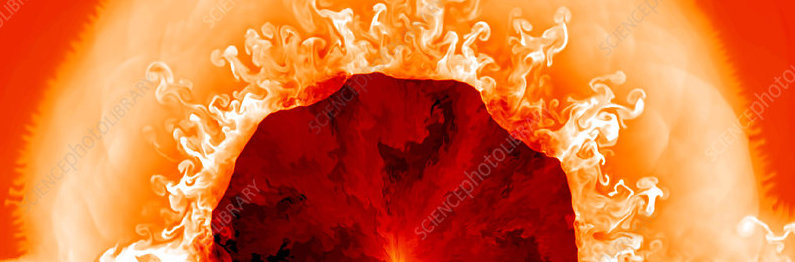} 
\chapter{Sources at the Frontier of Observations}

\begin{tcolorbox}[colback=teal!5!white,colframe=yellow!75!black,title=\sc Science Target]
{\em  Understand physical processes that underlie the most powerful astrophysical phenomena.}
\end{tcolorbox}

In addition to revealing dark astrophysical processes, gravitational waves (GWs) also offer a new window into extreme phenomena inside compact astronomical sources. While electromagnetic (EM) observations reveal the conditions in the surface regions, GWs encode information about the internal dynamics of these systems, particularly the way the distribution of matter is changing. This will allow us to address longstanding mysteries about the underlying physical mechanisms that power nature's most extreme phenomena. For example, GWs emitted along with light and neutrinos during the aftermath of the collapse of a massive star's core will yield unique insight into why the exterior of the star explodes as a supernova and the nature of the remnant that it leaves behind.  GWs from magnetar flares or glitching radio pulsars will elucidate the physical mechanism that causes these bursts of light and provide an unprecedented glimpse deep into the interior of neutron stars (NSs).  GWs produced by spinning NSs will allow us to study the structure of ultra-dense NS matter and test particle physics theories in extreme conditions of strong gravity.  Sensing these sources of GWs with a network of 3G detectors in concert with EM observatories and neutrino detectors will allow us to learn new physics that is otherwise inaccessible.  

\begin{tcolorbox}[standard jigsaw,colframe=ocre,colback=ocre!10!white,opacityback=0.6,coltext=black,title=\sc Key Science Goals]
Observations with the 3G network, further enhanced with EM and neutrino observatories, will allow us to probe new extreme environments and answer key questions about exotic astrophysical transients: 
\begin{itemize}[leftmargin=*]
\item {\bf GWs from core-collapse supernovae.} 
How do massive stars explode to form NSs and black holes? What are the physical processes involved and how can they be harnessed to advance our understanding of dense matter, neutrinos, and dark matter? 

\item {\bf Continuous GW emission from isolated or accreting NSs.} How does dense matter support elastic and magnetic stresses? Does GW emission limit the spin frequencies of NSs?

\item {\bf Bursts of GWs from magnetars and other pulsars.} What is the role of magnetic fields in bursts of EM radiation emitted by NSs? How stable is the ultra-dense matter of NSs?   

\end{itemize}
\end{tcolorbox}

At least one binary NS merger has been detected as a multi-messenger source with a GW signature and emission across the EM spectrum \cite{GBM:2017lvd}. 
Although no other multimessenger GW sources have yet been detected, other astrophysical phenomena are expected to produce detectable signatures in multiple messengers. Core-collapse supernovae in our galaxy, rare events with broad implications for astrophysics, nuclear physics and particle physics, remain the most promising multimessenger sources in the Universe. Simulations indicate that the intense GW, neutrino and EM emissions coincident with or following the collapse and explosion of a massive star would produce a large signal in terrestrial detectors, and would allow us to discern fine spectral and temporal features. Spinning asymmetric pulsars, NSs that emit a lighthouse-like beam of EM radiation as they spin, are also expected to be continuous GW sources. A sudden speed up in these otherwise regular pulses, called {\em glitches,} and episodic energetic outbursts in highly magnetized NSs, called {\em giant flares}, may also produce a coincident signal in both EM and GWs. Other fast and energetic phenomena, including fast radio bursts (FRBs) whose origin and prevalence remain most mysterious, may also emit GWs if they are associated with NSs or magnetars. With a network of 3G detectors, each of these potential multimessenger GW sources would yield new insight into key problems in modern physics and astronomy. 

 \section{Core-Collapse Supernovae}

The physical processes that drive core-collapse supernovae to violently explode and form vast nebula that can seed new stars and produce heavy elements remain mysterious \cite{Muller:2016izw}.
Simulations that take into account general relativity and the extreme nuclear and neutrino physics needed to model core-collapse supernovae predict that both neutrinos and large scale asymmetric matter flows that are necessary for explosion, also produce strong GW emission \cite{Radice:2018usf}. These studies indicate that the  formation of the proto-neutron star (PNS) halts collapse, and the shock wave generated by its formation stalls prematurely unless revived by dynamics that breaks spatial symmetries, neutrino heating, and or magnetic field interactions.  
A GW signal from a galactic core-collapse supernovae, would encode this inner dynamics, and offer vital new information. Furthermore, since neutrinos emitted from the hot and dense PNS are also detectable, GWs provide the complementarity necessary to unlocking longstanding mysteries about the explosion mechanism. Together, GWs and neutrinos would also provide detailed information about fundamental processes that can reveal nuclear and particle physics inaccessible in the laboratory.          

Simulations in two (2D) and three (3D) dimensions have shown that GWs are generated by rotational flattening, pulsations of the newly formed PNS, convection, non-radial accretion flows and instabilities, and other asymmetries  associated with the effects of strong magnetic fields. 
The dominant GW emission occurs during a phase of neutrino-driven convection and an instability in the shock wave of matter that has bounced off of the PNS's surface and stalled in place, called {\em Standing Accretion Shock Instability} (SASI) \cite{Yakunin:2015wra, Morozova:2018glm, Radice:2018usf}.
Oscillations in the star's matter in the near-surface layers of the PNS  (the $\ell = 2,$ f- and g-modes)  \cite{Murphy:2009dx,Yakunin:2015wra,Morozova:2018glm} also contribute strongly to GW emission. Indeed, 
GW emission from each phase of a core-collapse supernova provides diagnostic constraints on the explosion mechanism and the dynamics of the nascent PNS.

\noindent {\bf Core collapse and bounce:} General-relativistic studies \cite{Dimmelmeier:2007ui,Dimmelmeier:2008iq, 1997A&A...317..140M,Richers:2017joj} show that the GW burst signal from core bounce has generic characteristics for a wide range of progenitor rotation rates and profiles \cite{Zwerger:1997sq} and is a good probe of the bulk parameters of the collapsing iron core \cite{Abdikamalov:2013sta,Fuller:2015lpa}. 

\noindent {\bf Neutrino-driven turbulent convection outside and inside the PNS:} Milliseconds after core bounce, prompt convection in the region between the PNS and standing shock produces a short-period (of the order of tens of ms) burst of GWs peaking at $\sim 100$\,Hz. Subsequent stochastic mass motions that persist for tens of ms post bounce, can lead to significant broadband emission ($10$--$500$\,Hz with a peak at about $100$--$200$\,Hz) \cite{Kotake:2009rr, Muller:2011yi, Andresen:2016pdt, Mueller:2003fs, Murphy:2009dx, Yakunin:2010fn, Mueller:2014rna, Yakunin:2015wra, Morozova:2018glm}. GWs from the inner PNS convection zone is also broadband and range from $500$\,Hz to a few kHz \cite{1997A&A...317..140M, Marek:2008qi, Yakunin:2010fn, Mueller:2014rna, Morozova:2018glm, Andresen:2016pdt}. 

\noindent {\bf PNS oscillations:} The dominant fundamental oscillation modes, the quadrupolar g-modes \cite{Torres-Forne:2017xhv} and the fundamental f-mode, of the nascent PNS excited by accretion or convection and driven by gravity and pressure forces, generate GW emission  
\cite{Murphy:2008dw, Cerda-Duran:2013swa, Morozova:2018glm, Torres-Forne:2017xhv}. 
The frequency and amplitude evolution of the dominant f-mode is largely determined by the PNS mass, radius, and temperature  and thus a sensitive probe of the equation of state and of neutrino transport. Simulations indicate GW emission in the frequency band $\sim 200$--$500$ Hz at  early times (< few 100 ms) when the core is more extended 
and $\sim500$--$2000$ Hz at later stages \cite{Muller:2011yi,Torres-Forne:2017xhv}. 

\noindent {\bf SASI:} This is an instability associated with shock waves that exists in both 2D and 3D simulations. It is characterized by a nonlinear sloshing mode in 2D, and by both sloshing and spiral modes in 3D \cite{Blondin:2006yw}.   
The SASI produces characteristic time modulations both in neutrino and GW signals and provides information about the propagation of the shock wave and explosion dynamics  \cite{Mueller:2003fs,Murphy:2009dx,Yakunin:2010fn,Yakunin:2015wra,Kuroda:2016bjd,Andresen:2016pdt,Mueller:2014rna,Kuroda:2017trn}. 

\noindent {\bf Black hole formation:} When a massive rapidly spinning star collapses to form a black hole  it will be accompanied by an intense burst of GW emission, followed by a fast ringdown as the newly formed black hole settles down to a Kerr spacetime \cite{Ott:2010gv}. By contrast, black hole formation a few seconds after collapse in non-rotating or slowly rotating progenitors is likely to manifest as an abrupt cutoff of the GW emission after their characteristic frequencies (set by oscillation modes of the PNS) increase to several kHz \cite{Cerda-Duran:2013swa,Pan:2017tpk}. 
\vskip5pt


\noindent
\begin{minipage}{0.45\textwidth}
\hskip5mm A ten-fold increase in sensitivity relative to current ground-based detectors would boost the distance at which core-collapse supernova are detectable to 100 kpc, which would include the Large Magellanic Cloud and a number of smaller dwarf galaxies, increasing the chance of detection. Fig.\,\ref{fig:sn_fig1} plots the spectrum of 3D core-collapse supernova GW signals for a source placed at $100$\,kpc. They have signal-to-noise ratios for Advanced LIGO at design sensitivity in the range $0.5$--$6$, which is below reliable detectability levels. In contrast, they reach values in the range of $12$--$130$ for the 3G network,
levels that would not only allow us to detect the expected signals, but also to determine detailed properties of the progenitor star and the physics of the explosion~ \cite{Heng:2008ww, Powell:2016wke}. 

\hskip5mm In addition to providing critical clues needed to unravel the mystery of how core-collapse supernovae explode, GW observations of these systems will yield a unique insight into the state of hot and dense matter. 
\end{minipage}
\hfill
\begin{minipage}{0.52\textwidth}
\begin{tcolorbox}[standard jigsaw,colframe=gray,colback=gray!10!white,opacityback=0.6,coltext=black, title=\small\sc GW Amplitude of Supernovae]
\vskip-5pt
{\hskip-10pt\includegraphics[width=1.10\textwidth]{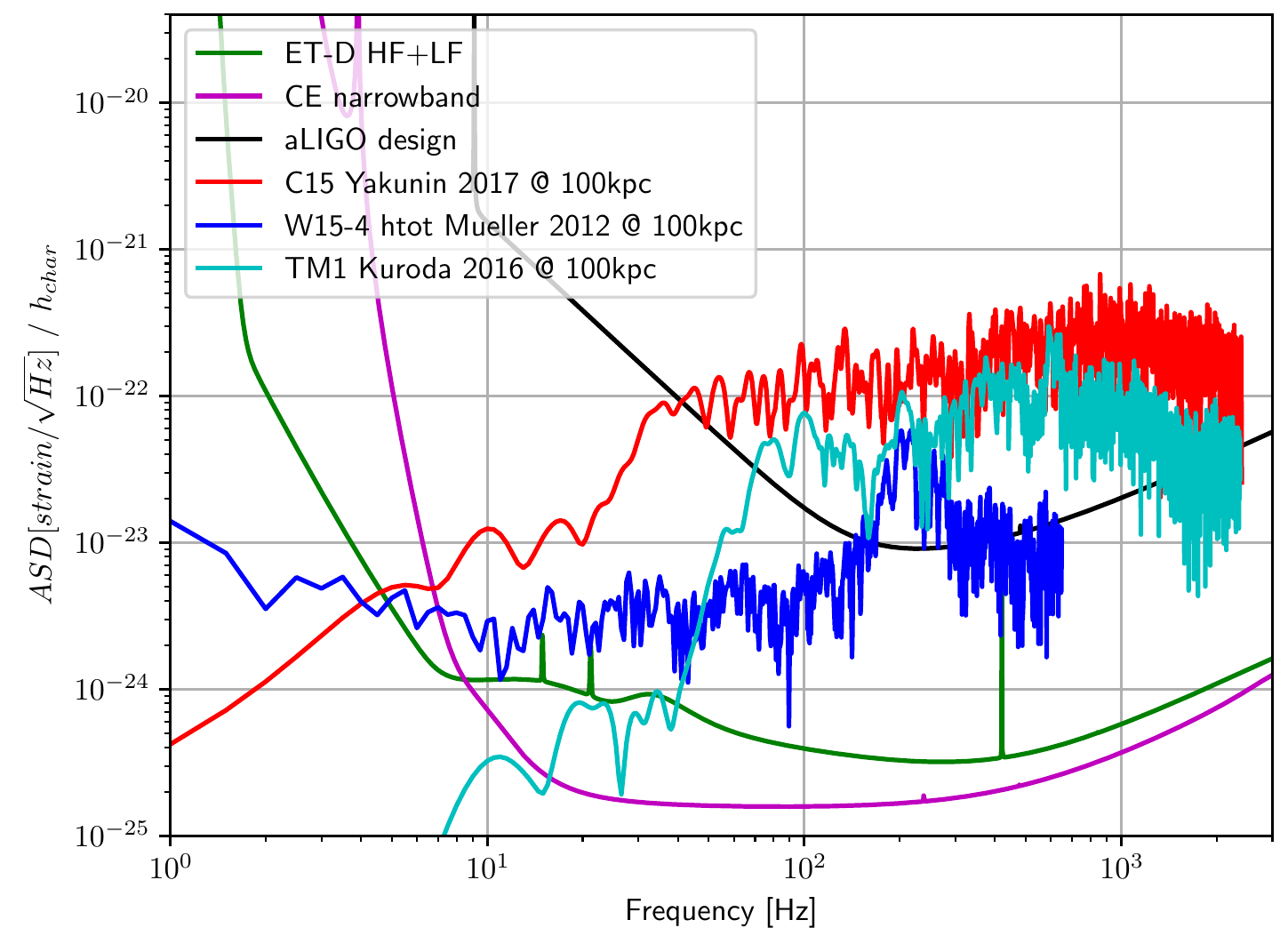}}
\captionof{figure}{\small Characteristic strain vs.\ frequency of three typical 3D 
core-collapse supernova simulations: C15 \cite{Yakunin:2017tus}, W15-4 \cite{Muller:2011yi}, 
and TM1 \cite{Kuroda:2016bjd}. The expected design sensitivity curves for Einstein Telescope (ET-D)~ \cite{Hild:2010id}, 
Cosmic Explorer (CE)~\cite{Dwyer:2014fpa}, 
and Advanced LIGO (aLIGO) \cite{Aasi:2013wya} are also shown.}
\label{fig:sn_fig1}
\end{tcolorbox}
\end{minipage}
\vskip10pt


GWs emitted by PNS modes and the SASI signals at $\sim 100$\,Hz to $250$\,Hz will allow us to constrain the physics of hot, ultra-dense nuclear matter in the newborn NS~\cite{Morozova:2018glm,Kuroda:2016bjd}. Recent work has also demonstrated that first-order phase transitions from nuclear to quark matter in the PNS core would imprint unique signatures on the detectable GW emission \cite{Zha:2020gjw}. The time evolution of the GW frequency will also allow us to chronicle the mass accretion history before and after the shock wave is reignited. The duration and the strength of the GW emission is expected to be correlated with the mass of the progenitor star \cite{Yakunin:2010fn,Mueller:2014rna,Morozova:2018glm}. With multiple detections the  correlations between GWs, neutrinos and EM radiation will help unravel the connection between stellar mass, explosion dynamics, and the nature of the remnant.    


Importantly, the onset of the neutrino emissions in core-collapse supernovae coincides with the onset of GW emission to within a few ms \cite{Andresen:2016pdt, Kuroda:2016bjd, Yakunin:2015wra, Murphy:2009dx, Kotake:2011yv}. The detection of neutrinos by Super-K/Hyper-K \cite{Abe:2016waf}, Deep Underground Neutrino Experiment \cite{Ankowski:2016lab}, Jiangmen Underground Neutrino Observatory \cite{Lu:2014zma}, IceCube \cite{Abbasi:2011ss}, Large Volume Detector \cite{Agafonova:2007hn}, Borexino \cite{Cadonati:2000kq}, KamLAND \cite{Tolich:2011zz}, and yet more sensitive neutrino detectors anticipated for the 2030's, will allow us to optimally extract the GW signal \cite{Kuroda:2017trn}. Both signals are produced at the same interior locations, resulting not only in time-coincidence, but have correlated modulations and polarizations, which aid with signal extraction and interpretation. If the progenitor core is rotating, there are additional, distinctive modulation signatures \cite{1997A&A...317..140M}. Thus, joint multimessenger analyses can not only enhance detectability but also more reliably probe physical processes, especially at the highest densities and temperatures. This aspect is critical to studies that aim to harness the extreme conditions encountered in supernovae to either discover or constrain new physics pertaining to dense matter, neutrinos, and dark matter.  
The core spin could be measured with GW and neutrino multimessenger detections, as the GW frequency is twice the modulation frequency of the neutrino signal \cite{Ott:2012kr, Yokozawa:2014tca, Kuroda:2017trn, 1997A&A...317..140M}.  
Bounce and explosion times are very difficult to localize with GWs alone, they require corresponding neutrino detections to pin point the source with accuracy. 
The 3G network will be critical to extracting all the physics from observable core-collapse supernovae and will synergistically leverage planned telescopes, satellite missions and 
neutrino detectors.



\section{Sources of Continuous GWs}

The detection of continuous GWs from NSs in the 3G network would  
provide evidence for deformations and clues about NS structure and their thermal, spin, and magnetic field evolution \cite{Sieniawska:2019hmd}.   
The emission of continuous 
GWs at detectable amplitudes requires a large mass quadrupole in a rapidly rotating compact object. A variety of mechanisms have been proposed to sustain non-axisymmetric  distribution of matter and energy, also called ellipticity, in a rotating compact object \cite{2017MPLA...3230035R} . Most prominent examples include elastic stresses in the crust, deformations due to magnetic fields, and the growth of r-modes in accreting NSs (a fluid mode of oscillation for which the restoring force is the Coriolis force) \cite{1998ApJ...502..708A,1998ApJ...502..714F}. 
Deformed NSs are nearly monochromatic sources of continuous GWs because the change in frequency 
is often smaller than $10^{-9}~\mathrm{Hz\,s}^{-1}$ over long periods of at least a few weeks and typically years. Rapidly spinning NSs are powerful sources since the intensity of the GW radiation is proportional to the sixth power of the spin frequency. Such high spin can be imparted at birth due during a core-collapse supernova \cite{Spruit:1998sg}, or from accretion of matter and angular momentum from a companion star \cite{Heuvel:2017ziq}.

\noindent {\bf Isolated NSs:} A solid NS crust can sustain (nonaxisymmetric) deformations or ellipticity, whose size depends on the composition, material properties and its evolution \cite{Ruderman:1969}. Although maximum fiducial ellipticities  $\sim 2\times 10^{-6}$ can be realized  in the crust~\cite{Douchin:2001sv}, fiducial ellipticities of $\sim 10^{-9}$ that provide a floor on the spin-down of millisecond pulsars  seem more likely \cite{Woan:2018tey}.

\noindent {\bf Accreting NSs in Binaries:} NSs in binary systems can also emit continuous GWs. They are more likely to present larger deformations than their isolated siblings, due to their internal magnetic fields\cite{2005ApJ...623.1044M}. Surface magnetic
field compressed by infalling material can produce large quadrupolar ellipticity
\cite{2005ApJ...623.1044M} and asymmetric heating in the crust due to accretion could lead to thermal deformations \cite{1998ApJ...501L..89B}. The excitation of GW-emitting unstable r-mode is also more likely in accreting NSs  
\cite{2014MNRAS.442.1786A,2014MNRAS.442.3037L} and may explain why we do not observe NSs spinning at their theoretical upper limit 
\cite{2003Natur.424...42C,2008AIPC.1068...67C,2017ApJ...850..106P}.



Current observations limit the ellipticity of a canonical NS at $10$ kpc emitting GWs above 500 Hz (150 Hz) to be $<10^{-5}$ ($10^{-4}$) \cite{Pisarski:2019vxw} and targeted searches of known nearby pulsars provide more stringent bounds. For example, observations of the pulsar J0711$-$6830 require that the ellipticity be less than $1.2\times 10^{-8}$ \cite{Authors:2019ztc}. However, to detect fiducial ellipticities of $\lesssim 10^{-9},$ new detectors with a substantially lower noise floor compared to Advanced LIGO and Virgo are necessary. Multimessenger observations of the source in GW, radio, X-rays, and/or gamma-rays can closely track the spin torques and orbital evolution of Galactic NSs to unravel the physical mechanism responsible for deformations and GW emission. 


\section{GW Bursts Associated with NS Flares and Glitches}

NSs can produce GW bursts due to the dynamics associated with giant flares of X-rays produced by highly magnetized NSs or sudden spin-up called glitches in otherwise stably rotating NSs. The coincident detection of GWs from these sources would be transformative as it will allow us to study mechanisms that operate in the deep interior and with unprecedented detail. 

\noindent \textbf{Magnetar flares:} Magnetars, highly magnetised NSs with magnetic fields exceeding
$10^{14}$\,G, are observed as anomalous X-Ray pulsars or soft gamma-ray
repeaters \cite{Kaspi:2017fwg}. Soft gamma-ray
repeaters show recurrent X-ray activity that include
frequent short-duration bursts ($10^{36}$--$10^{43}$~erg~s$^{-1}$ with durations
of $\sim0.1$~s) and, in some cases, energetic giant flares \cite{Turolla:2015mwa}
($10^{44}$--$10^{47}$~erg~s$^{-1}$ within $0.1$~s with X-ray tails that can
extend to several $100$~s). 
To date, three giant flares \cite{Israel:2005av,Strohmayer:2005ks,Strohmayer:2006py} have been detected, and several bursts \cite{Huppenkothen:2014cla,Huppenkothen:2014pba}
have been observed that showed quasi-periodic oscillations. 
Since these events are thought to involve substantial
structural changes within the NSs and due to the large involved energy,
magnetars are potential GW sources~\cite{Lasky:2015zpa,Glampedakis:2017nqy}. If a significant fraction of the X-ray energy is channelled into GWs a magnetar at $10$~kpc with a magnetic field at the pole of $B_{\rm pole}\sim10^{15}$~G, would produce a strain of $h\sim 10^{-27}$ in  the detector. The signal would consist of a high frequency component  corresponding to the f-mode around $1$--$2$~kHz, and a low frequency component associated to Alfv\'en oscillations with frequency $\sim 100$~Hz, which depends on the magnetic field strength. 


%

\noindent \textbf{Pulsar glitches:} Radio pulsars known for their very stable spin periods can occasionally
exhibit a sudden increase in their rotation frequency. These are called \textit{glitches}
and several hundred glitches have been observed in over 100 pulsars \cite{Espinoza:2011pq}. Physical models for the explanation of glitches involve a substantial rearrangement of the NS  structure on a short time scale, and are therefore expected to produce bursts of gravitational radiation. This dynamics is, however, not well understood and the predictions of the emission of GWs and their
detectability vary widely. The most optimistic scenarios suggest that a signal should be marginally detectable even by Advanced
LIGO and Virgo \cite{Melatos:2015oca,Bennett:2010tm,Prix:2011qv}, while in pessimistic scenarios even 3G instruments cannot detect the signal \cite{Sidery:2009at}. Moderately
optimistic  scenarios predict the signals to be detectable by the 3G network \cite{Keer:2015saa}. As with magnetar flares, the coincident detection of GWs with glitches would be a breakthrough, and even non-detection by 3G instruments would be able to distinguish between different scenarios. 


\section{Outlook for GW sources at the frontier of observations}
A network of 3G detectors is essential to correlate multimessenger signals from a host of extreme phenomena from compact sources. It would offer unprecedented opportunities to learn about the birth and extreme behavior of stellar remnants, 
\textit{and obtain new insight into physical processes in dense matter and those that underlie explosive phenomena such as core-collapse supernovae}. A galactic core collapse stands out as a singular source for multimessenger astrophysics, the coincident detection of GWs, neutrinos and EM radiation will be key to understanding the mechanisms that power core-collapse supernovae explosions. Temporal and spectral correlations between these three messengers will provide definitive insights into the formation of NSs and stellar mass black holes, properties of matter at extreme density and temperature.
Continuous GWs from spinning NSs, and bursts of GWs associated with magnetar flares and pulsar glitches also has the potential to unravel the mechanisms at play and the properties of dense matter and extreme magnetic fields.  Moreover, with its astounding reach, the potential of 3G detectors is difficult to overstate and serendipitous discoveries are guaranteed to take place.


\vspace{8mm}

\begin{tcolorbox}[colback=teal!5!white,colframe=red!75!black,title=\sc Science Requirements]{
Decades after their discoveries much of the physics of most explosive processes in the Universe remains hidden. Understanding these multimessenger phenomena would require:
\begin{itemize}
    \item  GW observatories that are \emph{10–100 times} more sensitive than current ground-based facilities in the 50-1000 Hz range are key to unravel the inner dynamics of supernova explosions.
    \item Current understanding of compact multimessenger sources such as magnetar quakes and pulsar glitches dictate a factor of ten greater sensitivity than advanced detectors at high frequencies.
    \item To observe continuous waves from NSs with ellipticities of one part in a billion requires sensitivity improvements by a factor of ten over the entire frequency range.
\end{itemize}
}
\end{tcolorbox}

%% file: acronyms.tex
\chapterimage{figures/bbh} 
\chapter*{Acronyms \& abbreviations}
\addcontentsline{toc}{chapter}{\color{ocre} Acronyms and Abbreviations}
\begin{table}[h!]
\parbox{1.0\linewidth}{
\begin{tcolorbox}[standard jigsaw,colframe=ocre,colback=ocre!10!white,opacityback=0.6,coltext=black,title=\sc \large Acronyms and abbreviations used in the text]
\small
\vskip-5pt
\label{tab:acronyms}
\parbox{.50\linewidth}{
	\begin{tabular}{rl}
	3G	& \emph{Third-generation}, the next generation of ground-based gravitational-wave observatories \\
	    & \emph consisting of 1 ET in Europe, 1 CE each in the US and Australia\\
	BAO	& \emph{Baryon Acoustic Oscillations} \\
	BBH	& \emph{Binary Black Hole}, a binary system of two black holes \\
	BH	& \emph{Black Hole}\\
	BNS	& \emph{Binary Neutron Star}, a binary system of two neutron stars\\
	CE	& \emph{Cosmic Explorer}, concept for a US third generation interferometer with 40 km arms \\
	CMB	& \emph{Cosmic Microwave Background} \\
	ECO	& \emph{Exotic Compact Object}, an alternative to a neutron star or a black hole \\
	EM	& \emph{Electromagnetic} \\
	ET	& \emph{Einstein Telescope}, concept for a European triangular shaped interferometer with 10 km arms\\
	GR	& \emph{General relativity} \\
	GSF	& \emph{Gravitational Self-Force} \\
	GW	& \emph{Gravitational Wave} \\
	GW150914	& Binary black hole merger event detected on 14 September 2015 \\
	GW151226	& Binary black hole merger event detected on 26 December 2015 \\
	GW170104	& Binary black hole merger event detected on 4 January 2017 \\
	GW170814	& Binary black hole merger event detected on 14 August 2017 \\
	GW170817	& Binary neutron star merger event detected on 17 August 2017 \\
	\textit{INTEGRAL} & \emph{International Gamma-Ray Astrophysics Laboratory} \\
	LIGO	& \emph{Laser Interferometer Gravitational-Wave Observatory}, 4 km arm length interferometers in \\
	     &  in the US at Hanford WA and Livingston LA\\
	LIGO-India	& LIGO interferometer being built in India \\
	\textit{LISA}	& \emph{Laser Interferometer Space Antenna} \\
	NR	& \emph{Numerical Relativity} \\
	NS	& \emph{Neutron Star}\\
	NSBH	& \emph{Neutron Star--Black Hole}, a binary system of one neutron star and one black hole \\
	PM	& \emph{Post-Minkowskian} \\
	PNS	& \emph{Proto-Neutron Star} \\
	PN	& \emph{Post-Newtonian} \\
	QCD	& \emph{Quantum Chromodynamics} \\
	SASI	& \emph{Standing Accretion Shock Instability} \\
	SM	& the \emph{Standard Model} of particle physics \\
	SN, SNe	& \emph{Supernova}, \emph{Supernovae} \\
	SNR	& \emph{Signal-to-noise ratio} \\
    Virgo & 3 km arm length interferometer located in Cascina, Italy \\
	\end{tabular}
        }
        \end{tcolorbox}
        }
        \vskip-5pt
\end{table}  
\clearpage